# Geometry Reorganizes Nonequilibrium States that Gate Topological Switching in an Active Polymer

**Shiqi Zhou***
School of Physics, Central South University, Changsha, Hunan, China
**Corresponding author: chixiayzsq@163.com*

Abstract. Even under identical external conditions, a driven system can occupy distinct long-lived states with different susceptibilities to the same rare transition. We demonstrate this in a confined partially active polymer, where independent unknotted replicas occupy distinct mechanical states with different topological futures. Matched-control and event-resolved analyses reveal a susceptible reservoir and a short threading precursor. We test this state dependence prospectively by launching 2272 Langevin-noise branches from 71 preserved unknotted phase-space states using a classifier prespecified from independent data. Most strikingly, the same classifier strongly separates finite-time knotting at one pore radius, yet reverses its susceptibility ordering at a neighboring radius. Thus geometry does not merely shift an average knotting rate; it selects among long-lived nonequilibrium states and reorganizes the mapping between a pre-existing mechanical coordinate and future topological fate.

Rare transitions in driven matter are commonly described by external control parameters together with a small set of collective variables. A central challenge is whether a low-dimensional coordinate that orders transition susceptibility within one environment retains the same dynamical meaning when the environment changes. Our simulations reveal an unexpected form of persistent state heterogeneity: independent replicas at identical pore radius and activity can remain in distinct long-lived mechanical states and exhibit very different knotting propensities. The central question is therefore not only how confinement changes an average rate, but whether a pre-existing internal state carries prospective information about a stochastic topological transition—and whether that state-to-fate ordering transfers across geometry. A change in the mean switching rate can arise because geometry redistributes the populations of long-lived states, because it changes the commitment associated with a given state coordinate, or both. Distinguishing these mechanisms requires state-resolved prospective measurements. This question connects to metastability, reaction-coordinate construction, transition-path physics, rare-event prediction, nanoconfinement, and nonequilibrium state preparation. Polymer topology provides a stringent probe because a local threading event creates a persistent nonlocal constraint. Knotting thus reports how boundary geometry reorganizes accessible conformations and topological pathways [1–7]. Confinement reshapes loops, backfolds, and escape routes, while tangential activity transports contour and injects work along the chain [8–18]. Recent studies have documented diverse forms of activity–topology coupling, including activity-driven knot generation and topology manipulation, formation–lifetime-controlled knot enhancement, active-knot dynamics, knot-modified collapse of active rings, and polarity- and placement-programmed topological gating in spatially patterned active polymers [19–23].

Here we ask whether persistent state heterogeneity at fixed controls has dynamical consequence. The evidence is deliberately separated across three levels: matched-control mechanics identifies a long-lived susceptible reservoir; event-aligned trajectories determine how that reservoir approaches threading without equating it with the transition path; and a prespecified branching experiment tests future fate from preserved phase-space states under independent Langevin noise. Protocol controls separately reject simple one-size or one-rate reductions. The resulting claim is stronger than a confinement-dependent knotting curve: geometry selects among long-lived nonequilibrium states, and neighboring geometries can reverse the predictive ordering of a fixed mechanical coordinate. The analysis therefore separates three physical levels that are often conflated in rare-event descriptions: state preparation, transition-path geometry, and stochastic commitment.

We first ask whether the mechanical differences persist when the external controls are held fixed. We therefore compare only independent replicas with identical (R,Pe), preventing trends caused simply by changing pore radius or activity from being mistaken for state differences. For the long trajectories spanning the knot-occupancy crossover in the two high-activity campaigns, larger knot occupancy $P_K$ is associated with lower mean wall load in unknotted frames ($r = -0.90$, exact stratified permutation $p = 0.0069$) and smaller unknotted transverse gyration ($r = -0.84$, $p = 0.023$) (Fig. 1). Because the shuffling is restricted within matched controls, neither relation can be generated simply by changing pore radius or nominal drive. Knot-prone replicas therefore occupy a weakly wall-coupled, transversely compact unknotted reservoir. This result is retrospective, but it supplies a mechanical signature whose relevance can be tested by temporal ordering and prospective branching.

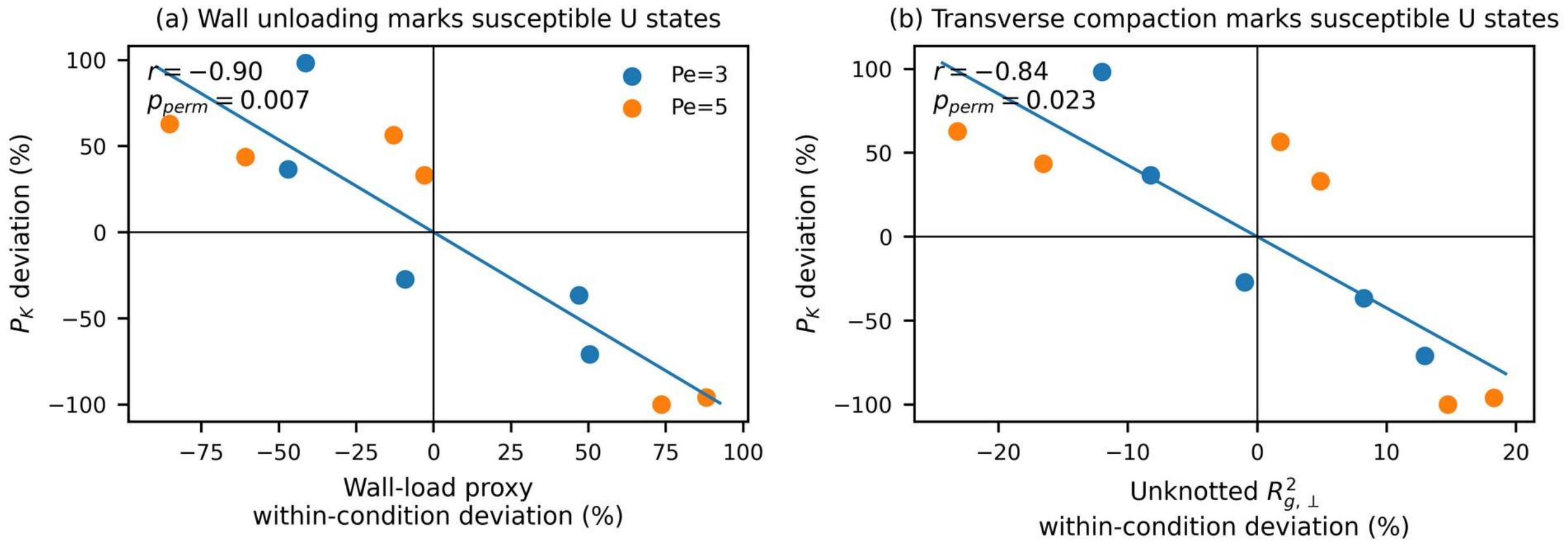

FIG. 1. Mechanical state selection at fixed external controls. (a) Within identical (R,Pe) conditions, knot-rich replicas have lower unknotted wall load. (b) The same replicas are more transversely compact. Values are deviations from the matched-condition mean; colors denote Pe. Correlations and p values are exact stratified-permutation results. Additional event-aligned controls are shown in Fig. S9 [35].

Transition-resolved trajectories connect the reservoir signature to knot entry without equating the reservoir with the transition state. Active-end nonlocal contacts rise and transverse end-to-end extent shortens before first knot detection (Fig. 2); the active-end minimum nonlocal distance likewise falls, while the active near-wall fraction does not increase (Fig. S9 [35]). Across the broad data set, 14,145 of 14,291 detected formation events (98.98%) first appear in the pore interior (Fig. S47 [35]). Thus susceptibility and transition pathway are distinct: weak wall coupling marks the long-lived reservoir, whereas transient interior compaction/threading marks the short approach to switching.

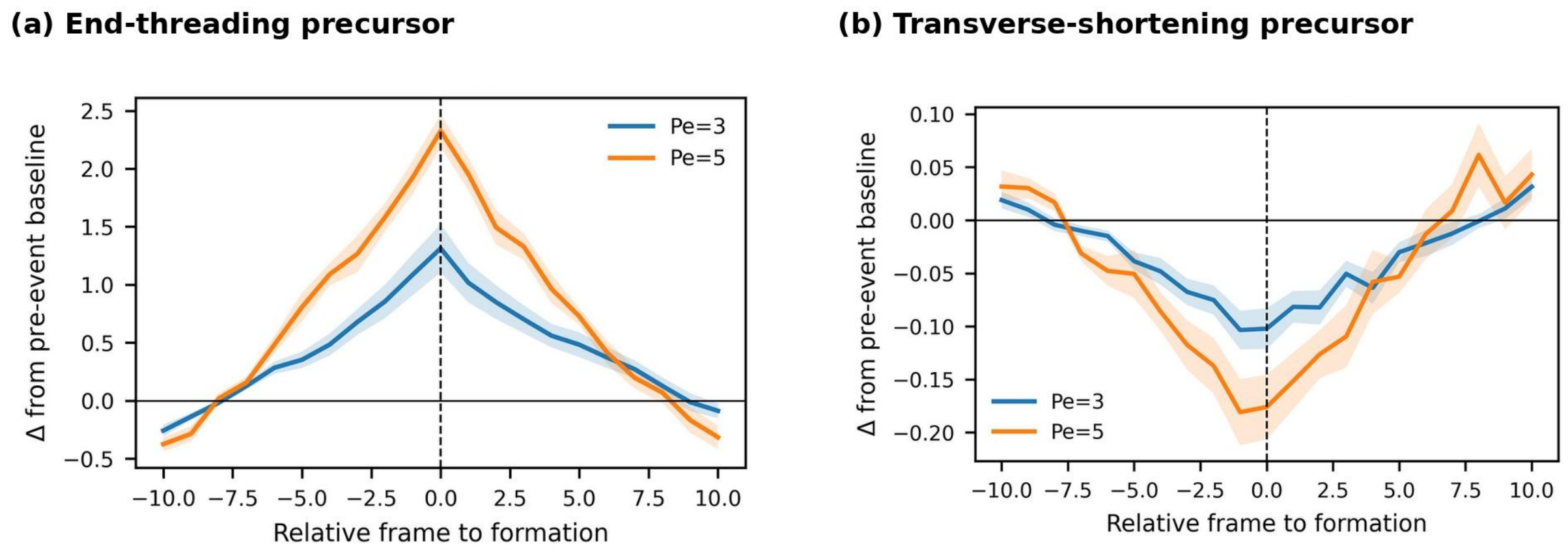

FIG. 2. Event-aligned approach to knot formation. (a) Active-end nonlocal contacts rise before the first detected knotted frame. (b) Transverse end-to-end extent decreases before formation. Curves are baseline-subtracted within each trajectory and then averaged with equal trajectory weight; bands are SEM. Additional event-aligned wall and distance controls are shown in Fig. S9 [35].

The opposite wall signatures on the two time scales are physically informative rather than contradictory: susceptible unknotted reservoirs are weakly wall-loaded on average, yet wall load can rise transiently during the short approach to an overwhelmingly interior threading event. The boundary therefore need not be the local site of topology creation to control the event; it can act upstream by selecting the long-lived states from which interior threading becomes accessible.

Additional controls rule out reductions that could otherwise mimic state dependence. Knot survival is strongly nonexponential (Fig. S13 [35]), and the chain-length response changes when active fraction rather than active-block length is held fixed, even under common-seed and common-horizon comparisons (Fig. S45 [35]). The broad and long-run scans place R=1.85 and 1.90 inside the same narrow replica-heterogeneous crossover region (Fig. S46 [35]), while formation-rate and complete-episode lifetime summaries vary separately across confinement (Fig. S47 [35]). Together with cases having similar transverse size but very different topological occupancy, these results disfavor a single memoryless rate, a universal size-only reduction, or a trivial interpretation in which the two branching radii sample unrelated regimes.

We then test future switching directly at $Pe = 3$ and $R = 1.85, 1.90$ (Fig. 3). A prespecified configuration score $s_{\mathrm{conf}}$ (operationally defined in End Matter), its condition-specific $Q20/Q80$ LOW/HIGH thresholds, parent eligibility, and the primary observation duration $H = 2 \times 10^7$ steps were fixed from independent calibration data before branch outcomes were inspected. Here $H$ is the prespecified duration of observation after branching, not a fitted relaxation time, lifetime, or half-life. Eligible parents had remained continuously unknotted for at least 2000 saved frames. Here LOW and HIGH label only the lower and upper tails of the prespecified configuration score—the lowest and highest 20% of the independent calibration distribution; they are not labels assigned from the subsequently observed knotting outcomes. The procedure yielded 71 parent states—20 LOW and 20 HIGH at $R = 1.85$, and 11 LOW and 20 HIGH at $R = 1.90$—from 13 of 16 independent parent-generation trajectories. Each parent retained coordinates, image flags, velocities, and active forcing, and seeded 32 descendants differing only in Langevin noise, for 2272 branches. The parent-level endpoint $q_H(X) = P(T_{\text{knot}} < H|X)$ therefore measures the probability of first detected knotting before this prespecified observation duration from a preserved phase-space state; because one parent-generation trajectory can contribute several parents, those parents are not treated as statistically independent; permutation, resampling, and uncertainty estimation account for this dependence at the source-trajectory level [35].

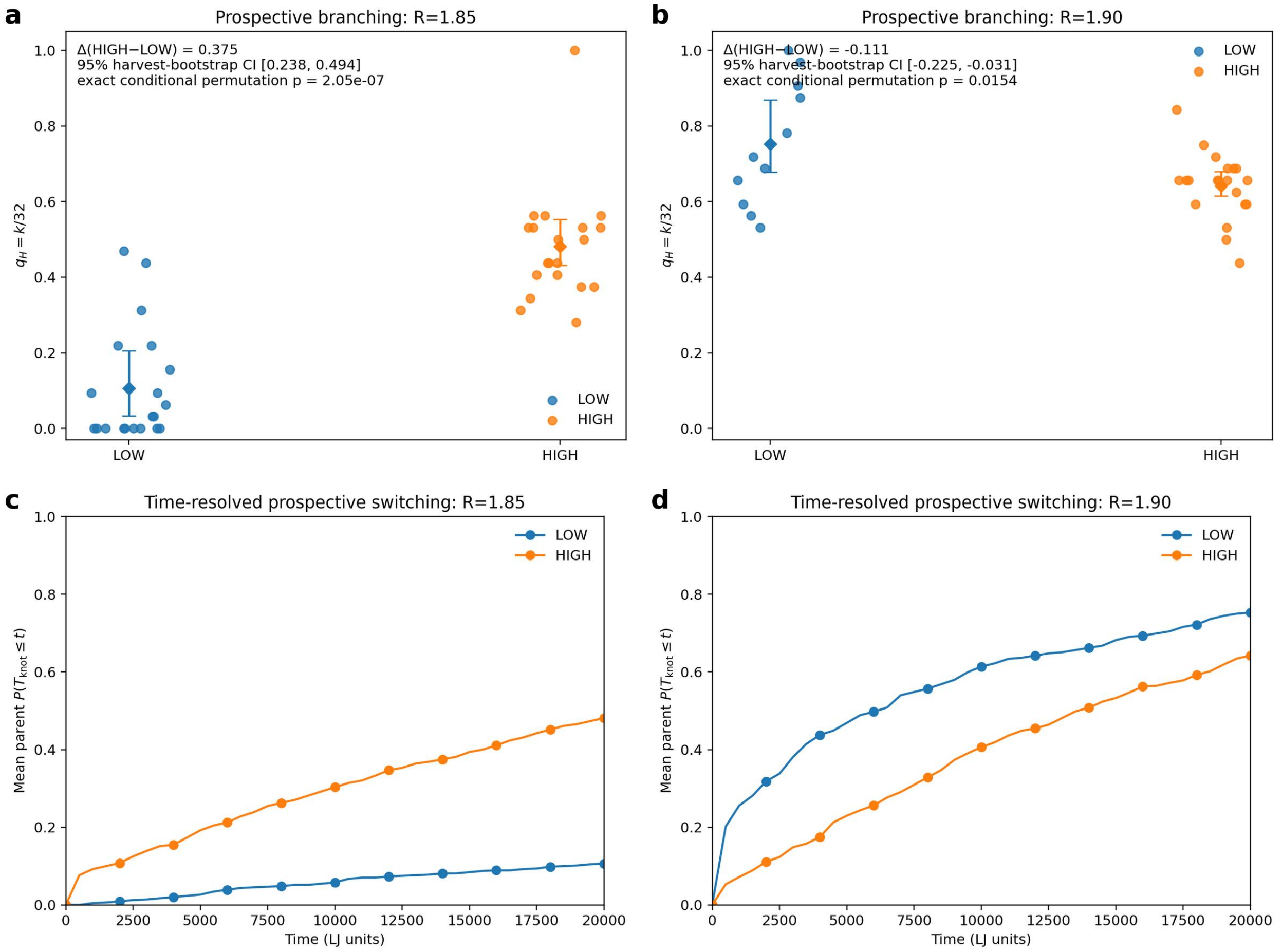

FIG. 3. Prospective branching test of a prespecified configuration classifier. (a,b) Parent-level finite-horizon knotting probability $q_H = \frac{k}{32}$ at $R = 1.85$ and 1.90. Points are parent states; diamonds show group means with 95% source-trajectory-cluster bootstrap intervals. Exact permutations preserve class counts within source trajectory. (c,d) Parent-averaged cumulative incidence of first knot detection. The prespecified HIGH>LOW ordering is strongly confirmed at $R = 1.85$ and reverses at $R = 1.90$.

At $R = 1.85$, the prespecified ordering is strongly confirmed: HIGH parents produce 308 knotting events in 640 branches versus 68 in 640 for LOW, giving mean $q_H = 0.481$ and 0.106 and HIGH − LOW $= 0.375$, with a 95% source-trajectory-cluster bootstrap interval [0.238, 0.494] and exact conditional permutation $p = 2.05 \times 10^{-7}$ (Fig. 3a,c). At $R = 1.90$, the same classifier reverses rather than merely weakens: HIGH gives $q_H = 0.642$, whereas LOW gives $q_H = 0.753$, so HIGH − LOW $=- 0.111$, with a 95% source-trajectory-cluster bootstrap interval [-0.225, -0.031] and $p = 0.0154$ (Fig. 3b,d). The difference between the radius-specific effects is 0.486, with a 95% source-trajectory-cluster bootstrap interval [0.326, 0.648] and $p = 7.88 \times 10^{-9}$. The reversal is already present at $2 \times 10^6$ and $5 \times 10^6$ steps and persists to H. The prespecified 11-versus-20 class imbalance at $R = 1.90$ was retained rather than rebalanced after outcomes were known.

The sign reversal is the decisive prospective failure of a geometry-independent one-coordinate ordering based on $s_{\mathrm{conf}}$. It is not caused by cancellation between the two ingredients of $s_{\mathrm{conf}}$: after controlling for source trajectory, both trailing transverse size and wall load associate with $q_H$ in one direction at $R = 1.85$ and the opposite direction at $R = 1.90$ [35]. Continuous-score slopes likewise change from +0.114 to -0.052, and every leave-one-source-trajectory-out HIGH−LOW estimate remains positive at $R = 1.85$ and negative at $R = 1.90$. Thresholding, one unusual source trajectory, or a post hoc class definition therefore cannot account for the result. A compact/unloaded mean-state direction carries predictive information, but its ordering is geometry dependent: no single monotonic mapping $q_H = f(s_{\mathrm{conf}})$ preserves the ordering across these neighboring radii. The reversal is therefore not merely a loss of classifier accuracy; it is a change in the ordering relation itself. This distinction matters beyond classifier performance. A coordinate intended to transfer across control conditions must preserve the ordering of commitment probabilities, not merely correlate with outcomes locally. Here an independently calibrated coordinate succeeds at one radius and reverses at the neighboring radius, while both radii lie in the same heterogeneous crossover region (Fig. S46 [35]). Geometry therefore changes not only which states are occupied, but also the dynamical meaning of a projection through that state space.

Because $s_{\mathrm{conf}}$ alone does not preserve susceptibility ordering across the two radii, we examined the 1000-frame pre-branch histories ending at each parent frame for a possible second direction. This search is explicitly exploratory because it was performed after the $R = 1.90$ outcome was known. After accounting for which parent-generation trajectory each parent came from, two fluctuation measures remain positively associated with future switching probability at both radii: the covariance of transverse size with wall load and the temporal variability of axial hairpin count. The axial hairpin count is the number of robust reversals of bond direction projected along the pore axis; bonds that are nearly transverse to the axis are omitted from the reversal test (precise definition in the Supplemental Material), so its temporal variability measures reorganization of forward- and backward-pointing contour segments. For visualization we combine them into an equal-weight, within-radius standardized lability coordinate $s_{\mathrm{lab}}$. The coupling between the prespecified static score and $s_{\mathrm{lab}}$ reverses with geometry, from $r =+ 0.718$ at $R = 1.85$ to $r =- 0.661$ at $R = 1.90$ (Fig. 4), while the score-adjusted $s_{\mathrm{lab}}$ coefficient remains positive at both radii (Figs. S43 and S44; Table S18 [35]). This post-outcome analysis does not validate a new committor; it provides a concrete two-coordinate hypothesis for an independent test.

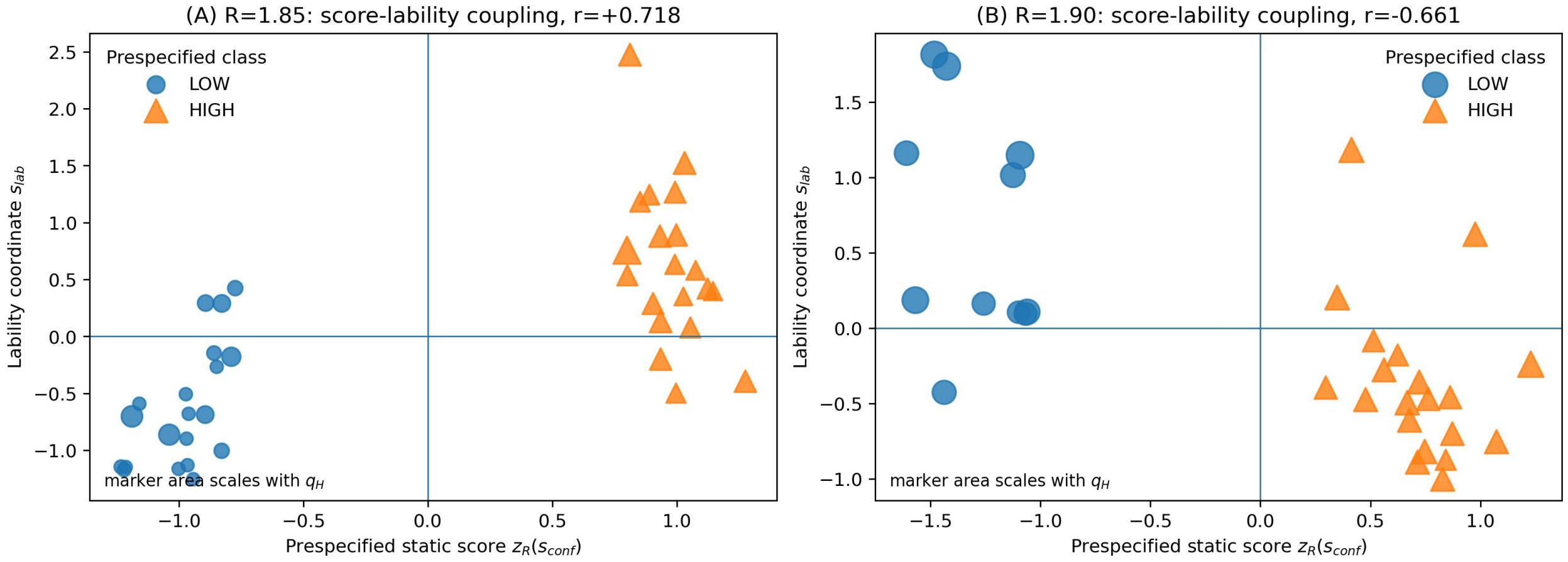

FIG. 4. Geometry reverses the coupling between a prespecified static state score and exploratory dynamical lability. (a) At $R = 1.85$, $s_{\mathrm{conf}}$ and $s_{\mathrm{lab}}$ are positively coupled ($r =+ 0.718$). (b) At $R = 1.90$, the coupling is negative ($r =- 0.661$). Marker area encodes the measured parent-level $q_H$. Larger symbols indicate a larger fraction of the 32 descendants knotting before the fixed observation duration. Because $s_{\mathrm{lab}}$ was defined only after the prospective reversal was known, this figure is a mechanistic hypothesis rather than an independent confirmatory test; direct $s_{\mathrm{lab}} - q_H$ relations are shown in Figs. S43 and S44 [35].

Additional pre-branch diagnostics show that the residual structure itself changes with radius: active-end contact and power fluctuations are most prominent at R=1.85, whereas wall/transverse lability, backfolding, and passive-end accessibility dominate at R=1.90 (Figs. S41 and S42 [35]). This is not evidence for a universal second scalar. Instead, it suggests a dynamical-lability direction whose coupling to the static compaction/unloading direction is reorganized by geometry.

Complementing recent active-knotting studies centered on conformation, knot motion, formation rates, and lifetimes [19,20], nonequilibrium energetic diagnostics provide an independent consistency check without invoking an equilibrium free-energy landscape. Event-centered active-power bursts increase strongly with activity while internal- and potential-energy bumps vary much less (Fig. S10 [35]). Across the Pe=1 confinement crossover, the occupancy-derived ΔF_eff changes sign while the confined internal-energy contrast ΔU_conf remains positive; the residual

TΔS_eff is therefore only a driven-state diagnostic, not a thermodynamic entropy (Fig. S11 [35]). The final schematic in Fig. S48 [35] synthesizes the prospective reversal and exploratory lability result: neighboring radii can occupy differently oriented regions of a multidimensional nonequilibrium state manifold, so projection onto one static score can reverse even when an exploratory dynamical-lability direction remains positively associated with finite-horizon switching.

The branching experiment also fixes the causal scope of the claim. Parent states were selected, not mechanically imposed, so $q_H(X)$ is a finite-horizon conditional probability for a preserved phase-space state $X$, not an equilibrium probability or a proof that manipulating $s_{\text{conf}}$ alone changes the true committor. Nor does the sign reversal exclude every possible scalar coordinate; it shows specifically that the independently fixed $s_{\text{conf}}$ is not a geometry-independent ordering variable across the tested radii. It is nevertheless stronger than a retrospective correlation: classification is completed before any branch outcomes are generated, all descendants of a parent share the same phase-space state, branch noise is independent, and the classifier, thresholds, and endpoint remain fixed after the reversal is observed. The experiment therefore establishes prospective state dependence while leaving direct intervention on state coordinates for future work.

Confinement therefore acts at two logically distinct levels: it redistributes long-lived nonequilibrium states and can change the mapping from a coordinate within those states to rare-event fate. The evidence was built in discovery order: broad scans first exposed persistent heterogeneity under identical controls; matched-control and event-resolved analyses separated a susceptible reservoir from the short transition-path signature; and a classifier frozen from independent data was then tested on 2272 branches generated with independent Langevin noise. Together, these results establish three broader physical principles for rare-event dynamics. First, a reactive reservoir need not resemble the configuration immediately preceding transition. Second, a local boundary can control a predominantly interior topological event through global state selection. Third, a coordinate that predicts future susceptibility within one environment need not retain its ordering after the environment changes. More generally, a change in an observed rare-event rate need not correspond to a uniform change in transition dynamics: it can arise from reweighting long-lived states, from changing state-specific commitment, or from both. This decomposition makes collective-variable discovery and collective-variable transferability distinct problems. A coordinate can be locally predictive yet fail—or reverse—as a common ruler when geometry reorganizes the occupied manifold. The implication is not that a low-dimensional reaction coordinate cannot exist, but that its portability across environments is itself a falsifiable physical property rather than an assumption. This perspective links confined active polymers to general problems of metastability, rare-event commitment, reaction-coordinate discovery, nanoconfinement, biomolecular topology, and nonequilibrium state preparation.

Acknowledgments.— This work is supported in part by the High Performance Computing Center of Central South University.

Data Availability.— The processed numerical data underlying the figures and tables, together with the custom analysis scripts used for the prospective branching and robustness analyses, are not deposited in a public repository because a dedicated project repository is not currently maintained; these materials are available from the corresponding author upon reasonable request. The raw molecular-dynamics trajectories are retained on institutional storage because of their size and are available for verification from the corresponding author upon reasonable request.

Conflicts of Interest.— The author declares no conflicts of interest.

AI Use Disclosure.— OpenAI ChatGPT (GPT-5.6 Sol) was used to assist with manuscript drafting and revision and with drafting some analysis and visualization code. The author defined the scientific questions and analyses, reviewed and edited the AI-assisted code as needed, executed all code on author-controlled data, independently verified the numerical outputs and figures, and takes full responsibility for the manuscript and results.

## End Matter

### A. Simulation model and data sets

We study a linear Kremer-Grest/WCA bead-spring chain at reduced temperature T=1 in a repulsive cylindrical pore, with tangential active forcing applied to a terminal contour block [24–27]. The baseline data set uses $N = 300$ and active beads 181–300. Knot states and localized knot arcs are identified with the KymoKnot procedure [28]. The broad survey contains 120 trajectories over 66 sampled (R,Pe) points; a long-trajectory data set densely samples the crossover region $R = 1.825$–1.925 at Pe = $3$ and 5. A separate Pe = $3$ chain-length study compares $N = 200$,300,450 under fixed active fraction and fixed active-block length. The prospective experiment uses eight new parent-generation trajectories at each of $R = 1.85$ and 1.90. Further simulation parameters, validation details, and additional analyses are given in the Supplemental Material [35].

The cylindrical wall acts through the truncated-and-shifted 9–3 Lennard-Jones external potential $U_{\mathrm{wall}}(g) = U_{93}(g) - U_{93}(g_c)$ for $0 < g < g_c$, and $U_{\mathrm{wall}}(g) = 0$ for $g \geq g_c$, with $U_{93}(g) = \epsilon_w \left[\frac{2}{15}\left(\frac{\sigma_w}{g}\right)^9 - \left(\frac{\sigma_w}{g}\right)^3\right]$. For bead i, $g_i = R - \rho_i$, where $\rho_i$ is its perpendicular distance from the cylinder axis and $g_c = \left(\frac{2}{5}\right)^{\frac{1}{6}}\sigma_w$ is the location of the minimum of $U_{93}(g)$, which defines the wall cutoff.

**B. Statistical analysis**

Independent statistical units are trajectories for matched-control and chain-length analyses, trajectories after within-trajectory averaging for event-aligned analyses, and parent configurations for the branching experiment. Dense saved frames are never treated as independent replicas. Parent-level $q_H$ is computed from 32 stochastic descendants. HIGH/LOW inference is performed separately at each radius using exact class-label permutation constrained within source trajectories; confidence intervals resample complete source-trajectory clusters. The $R = 1.90$ imbalance (11 LOW, 20 HIGH) is retained because the prespecified Q20/Q80 rule was not relaxed after outcomes were available. Exploratory second-coordinate analyses use only pre-branch trailing histories ending at the parent frame, apply multiplicity correction to the candidate panel, and do not alter the primary classifier or endpoint. OpenAI ChatGPT (GPT-5.6 Sol) assisted in drafting some analysis and visualization code; the author specified the analyses, executed all code on author-controlled data, and independently verified the outputs. Standard bootstrap/blocking/survival procedures follow Refs. [29–31].

**C. Notation and operational definitions**

Geometric and wall observables.—Let $\mathbf{e}_a$ be the cylinder-axis unit vector and $\mathbf{r}_{\mathrm{cm}}$ the chain center of mass. For the equal-mass beads used here, the transverse mean-square radius of gyration is $R_{g,\perp}^2 = N^{-1}\sum_i \{|\mathbf{r}_i - \mathbf{r}_{\mathrm{cm}}|^2 - [(\mathbf{r}_i - \mathbf{r}_{\mathrm{cm}}) \cdot \mathbf{e}_a]^2\}$. For bead i with radial gap $g_i = R - \rho_i$, the frozen custom analyzer (version 2.3) defines the signed wall force $f_\rho$ with the positive radial direction pointing away from the cylinder axis (increasing $\rho_i$); inward repulsion is negative. The wall-load proxy is $W = -\sum_i f_{\rho,i}$, so larger positive W denotes a larger net inward normal load. The gyration terminology and conventions follow standard polymer definitions [32–34].

Prospective score and classes.—A bar denotes the 1000-saved-frame trailing mean ending at the parent frame (the parent frame plus the preceding 999 saved frames). At each radius R, $z_R(x) = \frac{x-\mu_R}{s_R}$ uses the mean and sample standard deviation computed with denominator $n_R - 1$, where $n_R$ is the number of calibration landmarks at that radius of an independent discovery calibration set constructed with exactly equal landmark counts from each of five discovery seeds. The prespecified score is $s_{\mathrm{conf}} = \frac{-z_R(\overline{W}) - z_R\left(\overline{R_{g,\perp}^2}\right)}{2}$. Q20$_R$ and Q80$_R$ are the condition-specific 20th and 80th percentiles of the discovery score distribution; LOW means $s_{\mathrm{conf}} \leq Q20_R$, HIGH means $s_{\mathrm{conf}} \geq Q80_R$, and the middle 60% is not selected.

Finite-time susceptibility and correlations.—For a preserved parent phase-space state $X_i$, $q_H(X_i) = P(T_{\mathrm{knot}} < H|X_i)$. The horizon $H$ is the fixed elapsed simulation time measured from the parent frame to the prospective endpoint: $H = 2 \times 10^7$ MD steps, equal to 2000 saved intervals and $2 \times 10^4$ reduced time units for $\Delta t = 0.001$. It is a prespecified observation window, not an estimated characteristic time. Operationally, the plotted parent-level value is its 32-branch estimate $\hat{q}_H = k_i/32$, where $k_i$ is the number of descendants whose first detected knot occurs before $H$. $P_K$ denotes the fraction of topology-valid saved frames classified as knotted. Unless otherwise stated, $r$ denotes the Pearson product-moment correlation coefficient; exploratory Spearman rank correlations are denoted $\rho_S$.

# Supplemental Material for Geometry Reorganizes Nonequilibrium States that Gate Topological Switching in an Active Polymer

Shiqi Zhou*
School of Physics, Central South University, Changsha, Hunan, China
*Corresponding author: chixiayzsq@163.com

Scope and inferential map. The Letter is designed to stand on the main-text evidence. Sections S1–S12 document the model, matched-control signatures, temporal precursors, survival behavior, protocol controls, and statistical conventions. Section S13 contains the prespecified prospective branching experiment and the primary state-dependence inference. Section S14 is explicitly post-outcome and exploratory; none of its candidate variables changes the parent labels, endpoint, or primary significance tests. Section S15 synthesizes existing controls and context; it introduces no separate simulation data set or primary inferential test. Section S16 closes the argument by separating quantitative energetic consistency checks from a final schematic state-manifold synthesis; neither is used to define the prospective classes or primary inference.

## S1. Simulation model and data sets

The baseline $N = 300$ data set contains 120 trajectories over 66 sampled (R,Pe) points for a bead-spring polymer with active beads 181–300. The long-trajectory data set samples R={1.825,1.850,1.875,1.900,1.925} across the crossover region for Pe = $3$ and 5 for nominal seeds 4–8; one Pe = $5, R = 1.925$ trajectory is absent, leaving 49 populated long trajectories. The interaction model is the Kremer–Grest/WCA chain implemented in LAMMPS with a repulsive cylindrical 9–3 wall [1–4]. Knot states and localized arcs are obtained using our custom Python post-processing analyzer (version 2.3), which uses the KymoKnot procedure for knot identification and localization [5]. Autocorrelation-aware blocking, bootstrap uncertainty, Kaplan–Meier survival analysis that retains censoring information, and independent mass-weighted gyration calculations follow Refs. [6–11]. Related active-polymer confinement in cylindrical channels is discussed in Ref. [14].

The cylindrical wall external potential is $U_{\mathrm{wall}}(g) = U_{93}(g) - U_{93}(g_c)$ for $0 < g < g_c$, and $U_{\mathrm{wall}}(g) = 0$ for $g \geq g_c$, with $U_{93}(g) = \epsilon_w \left[\frac{2}{15}\left(\frac{\sigma_w}{g}\right)^9 - \left(\frac{\sigma_w}{g}\right)^3\right]$. For bead i, $g_i = R - \rho_i$; $\rho_i$ is the perpendicular distance from the cylinder axis and $g_c = \left(\frac{2}{5}\right)^{\frac{1}{6}}\sigma_w$ is the location of the minimum of $U_{93}(g)$, which defines the wall cutoff.

The chain-length study is performed at Pe = $3$. A preliminary set of $N = 200$ and $N = 450$ trajectories was used to identify the parameter region for long production. Forty-two selected conditions were then independently restarted with three long replicas per condition. The fixed-$p_{\mathrm{active}} = 0.4$ arm contains $N = 200$ at $R = 1.80$–2.05 and $N = 450$ at $R = 1.80, 1.85$,1.90. The fixed-$N_{\mathrm{active}} = 120$ arm contains $N = 200$ at $R = 1.80, 1.85$,1.90 and $N = 450$ at $R = 1.85, 1.90$. Thirty-two long trajectories reached $1.2 \times 10^{10}$ steps and the remaining ten complete trajectories ended between $1.03413 \times 10^{10}$ and $1.189709 \times 10^{10}$ steps. The existing $N = 300$ Pe = $3$ long data set supplies the intermediate chain length. Broad occupancy, kinetic, and size diagnostics for these data sets are shown in Figs. S1-S8.

## S2. Event-aligned selection and formation-location control

The event-aligned analysis uses one selection rule for Pe = $3$ and Pe = $5$ long trajectories from the crossover-region campaign: $0.01 < P_K < 0.99$ and at least 30 formation events. This leaves 19 Pe = $3$ and 9 Pe = $5$ trajectories. Each observable is baseline-subtracted using that trajectory's mean over relative frames -10 to -6, and trajectories are then averaged with equal weight. Error bands are SEM across trajectories, not across individual events. Relative frame zero is the first knotted frame in the gap-safe topology timeline; 'precursor' therefore denotes temporal ordering at the saved-frame resolution rather than microscopic causality below that interval.

| Pe | Observable | n | $\Delta(-1)$ | SEM | p |
|---|---|---|---|---|---|
| 3 | endN_nonlocal_contact_count | 19 | 1.0886 | 0.176 | 8e-06 |
| 3 | endN_min_nonlocal_distance | 19 | -0.01251 | 0.00249 | 8.83e-05 |
| 3 | end_to_end_transverse | 19 | -0.10346 | 0.0181 | 2.04e-05 |
| 3 | wall_normal_load_proxy | 19 | 11.991 | 3.8 | 0.00551 |
| 3 | active_near_wall_fraction | 19 | -0.0019536 | 0.000703 | 0.0123 |
| 3 | near_wall_fraction | 19 | -0.00046062 | 0.000468 | 0.338 |
| 5 | endN_nonlocal_contact_count | 9 | 1.9277 | 0.144 | 9.11e-07 |
| 5 | endN_min_nonlocal_distance | 9 | -0.018897 | 0.00277 | 0.000136 |
| 5 | end_to_end_transverse | 9 | -0.18083 | 0.0311 | 0.000395 |
| 5 | wall_normal_load_proxy | 9 | 69.376 | 26.5 | 0.0308 |
| 5 | active_near_wall_fraction | 9 | -0.0033842 | 0.000954 | 0.00754 |
| 5 | near_wall_fraction | 9 | -0.0016297 | 0.000471 | 0.00854 |

Across the broad data set, 14,145 of 14,291 detected formation events are classified as interior (98.98%), leaving 146 near-wall events (1.02%). The trajectory-level and event-level wall signals answer different questions: knot-rich unknotted replicas have lower time-averaged wall load, whereas wall load rises transiently around formation while the active near-wall fraction decreases. This combination is consistent with an interior threading event launched from a weakly wall-coupled reservoir and accompanied by short-lived compression. Additional event-aligned wall, distance, and near-wall controls are shown in Fig. S9.

## S3. Within-condition state selection and operational susceptibility coordinate

Raw wall load and transverse size vary deterministically with pore radius, so the fixed-control test compares only independent replicas sharing the same (R,Pe). Exact stratified permutation tests give $r = -0.8955$, $p_{\mathrm{perm}} = 0.00693$ for the unknotted wall-load proxy and $r = -0.84$, $p_{\mathrm{perm}} = 0.023$ for unknotted transverse gyration. Axial active power has the complementary positive association. These signals define an operational susceptibility proxy $m^* = \frac{-z(W_U)-z\left(R^2_{g,\perp,U}\right)+z(P_{\parallel,U})}{3}$, where each z score is computed only within a shared control condition. Positive $m^*$ denotes a less wall-loaded, more transversely compact, more axially powered unknotted replica.

For occupancy, we regress the log-odds logit($P_K$)=ln[$P_K$/(1-$P_K$)] against the operational susceptibility coordinate.

| Response | Interpretation | Slope | Robust SE | p | n |
|---|---|---|---|---|---|
| logit($P_K$)=ln[$P_K$/(1-$P_K$)] | $\gamma = \alpha + \beta$ | 2.289 | 1.026 | 0.0257 | 10 |
| ln $\alpha_K$ | $\alpha$ | 1.607 | 0.462 | 0.000511 | 10 |
| ln $k_{\mathrm{un}}$ complete proxy | direct $\beta$ test | 1.245 | 0.836 | 0.136 | 9 |
| difference | $\beta$ heuristic | 0.683 | 1.126 | — | 10 |

The direct reverse-rate proxy is not stable enough to identify a scalar $\beta$. Leave-one-observable-out variants of $m^*$ leave the occupancy slope between 2.26 and 2.32 and the formation-hazard slope positive and significant. PCA gives nearly equal signed loadings and a first component explaining 98.6% of the variance, with correlation 0.999999 to $m^*$. Both analyses use the same selected replica observables and therefore assess internal representation stability, not held-out predictive validity. The equal-weight score is a transparent operational summary of the dominant replicated mechanical mode, not a validated committor or unique microscopic reaction coordinate.

## S4. Energetic diagnostics and state-selection integral

Event-centered rich-dump observables provide a drive-sensitive check. The case-level median active-power peak above its pre-event mean rises strongly with Pe, whereas the internal- and potential-energy bumps vary much less. The occupancy-derived quantity $\Delta F_{\mathrm{eff}}$=-$k_B$ T ln($P_K/P_U$) is used only as a nonequilibrium log-occupancy diagnostic, not as an equilibrium free energy; likewise $\Delta U_{\mathrm{conf}}$-$\Delta F_{\mathrm{eff}}$ is not interpreted as a thermodynamic entropy. The event-centered work/energy comparison and the occupancy-based energetic contrast are shown in Figs. S10 and S11, respectively.

For a minimal quantitative organization, m is taken to have unit variance with control-dependent mean $\mu(R, \mathrm{Pe}) = \chi_{\mathrm{Pe}}[R - R_0(\mathrm{Pe})]$, while the microscopic logistic slope $\gamma$ is fixed by the independent within-condition regression. Only $R_0$ and $\chi_{\mathrm{Pe}}$ are adjusted for each activity. The model reproduces the broad Pe = $1$ crossover and the sharp Pe = $5$ long-run crossover well, and captures the rising Pe = $3$ flank. Its failure at the largest Pe = $3$ radii coincides with the regime of independently largest replica heterogeneity, motivating an additional slow coordinate rather than a claim of a universal one-dimensional free-energy landscape. The resulting state-selection integral and fixed-effect residual diagnostics are shown in Fig. S12.

## S5. Lifetime-shape test of heterogeneous escape rates

A scalar $\beta$ cannot be identified reliably from complete-episode means, but the reverse-rate hypothesis has a separate distribution-level consequence. Left-censored episodes are excluded, right-censored episodes—those still knotted when the trajectory ends, so that only a lower bound on lifetime is known—are included in per-trajectory Kaplan-Meier estimates, and trajectories are then averaged with equal weight. Requiring at least five usable episodes gives 23 Pe = $3$ and 14 Pe = $5$ trajectories. A single-rate exponential has log-survival RMSE 2.66 for Pe = $3$ and 2.49 for Pe = $5$, whereas the minimal two-rate mixture gives 0.213 and 0.214, improvements by factors 12.5 and 11.6. The mixture is descriptive: it demonstrates strong escape-rate heterogeneity but is not used to identify literal microscopic states or to re-estimate $\beta$. The censored survival comparison is shown in Fig. S13. Robustness of the operational susceptibility coordinate is summarized separately in Fig. S14.

## S6. Common-seed and common-time robustness of the chain-length test

The primary chain-length panels use the common seeds 4-6 for $N = 200$,300,450. This matters most at $R = 1.85$: the $N = 300$ mean is 0.760±0.176 using all five reference seeds but 0.977±0.007 using the common three. At $R = 1.90$ the corresponding values are 0.869±0.056 and 0.902±0.030. We therefore emphasize $R = 1.90$ for the clean protocol contrast and treat $R = 1.85$ as evidence of nonmonotonicity and metastability rather than of a universal length trend.

| Protocol | R | $N = 200$ | $N = 300$ | $N = 450$ |
|---|---|---|---|---|
| fixed $p_{\mathrm{active}} = 0.4$ | 1.85 | 0.215±0.097 | 0.977±0.007 | 0.513±0.279 |
| fixed $p_{\mathrm{active}} = 0.4$ | 1.90 | 0.327±0.052 | 0.902±0.030 | 0.966±0.020 |
| fixed $N_{\mathrm{active}} = 120$ | 1.85 | 0.491±0.271 | 0.977±0.007 | 0.815±0.175 |
| fixed $N_{\mathrm{active}} = 120$ | 1.90 | 0.598±0.048 | 0.902±0.030 | 0.640±0.320 |

A second robustness test equalizes the observation time of all 42 independently restarted trajectories. $P_K$ is recomputed from each validated topology timeseries through step $1.03413 \times 10^{10}$, the shortest complete production endpoint. The common-horizon values correlate with the full-run values at $r = 0.9966$; the largest absolute case-level change is 0.0868 and the mean change is -0.0162. At $R = 1.90$, fixed-$p_{\text{active}}$ gives Delta $P_K$(450-200)=0.6388 in the full runs and 0.6477 at the common horizon, whereas fixed-$N_{\text{active}}$ gives 0.0424 and 0.1028. Thus the qualitative protocol contrast is not produced by unequal run lengths. The common-seed and common-horizon robustness checks are summarized in Fig. S15.

## S7. Exploratory chain-length mechanics and kinetic accounting

Absolute unknotted transverse size does not collapse the chain-length response. At fixed $p_{\text{active}}$ and $R = 1.90$, for example, $P_K$ increases strongly with N while the mean unknotted transverse RMS size changes only modestly and not in the direction of a simple compaction law. When the $N = 200$/450 long trajectories are compared only within identical (protocol,N,R) conditions, the raw-residual transverse-size association is negative ($r =- 0.396$ in the unfiltered set), consistent in sign with the $N = 300$ state-selection result. However, its magnitude weakens as a larger minimum number of unknotted frames is required; it is therefore treated as exploratory consistency evidence rather than an independent significant replication.

The kinetic decomposition uses $\alpha_K$=$N_{\text{form}}/T_U$ for access and the occupancy-budget quantity $\Theta_K$=$T_K/N_{\text{form}}$ for residence per detected formation. These satisfy the exact identity $P_K/(1\text{-}P_K)$=$\alpha_K\ \Theta_K$ whenever the quantities are defined. $\Theta_K$ is therefore an accounting decomposition, not an independent causal estimator of an unknotting rate. At fixed $p_{\text{active}}$ , $R = 1.90$, the median $\alpha_K$ changes little from $N = 200$ to450 while median $\Theta_K$ rises by about 57-fold; at fixed $N_{\text{active}}$ , $R = 1.85$ the corresponding changes are about1.4-fold and82-fold. These ratios identify where the occupancy odds are allocated between access and residence, but the independent evidence for heterogeneous escape remains the censored lifetime distribution in Fig. S13. The corresponding structural and kinetic decomposition is shown in Fig. S16.

The $N = 450$ fixed-$N_{\text{active}}$ control illustrates the metastability directly. At $R = 1.90$ the three long replicas have $P_K \approx 0.919$,1.000, and0.0028. The two replicas with defined unknotted geometry have nearly identical transverse RMS size but markedly different axial organization and topological residence; the saturated $P_K = 1$ trajectory has no unknotted exposure. This is why a second slow coordinate such as backfold organization or knot-core position is a more plausible next reduced description than a universal single-size scaling variable.

## S8. Inferential hierarchy, relation to previous work, and novelty

The state-selection mechanism emerged from the data in stages rather than being assumed at the outset. Broad confinement scans and replicated long trajectories first revealed unusually persistent replica-to-replica heterogeneity under identical controls. Matched-control analysis then showed that the heterogeneity had a mechanical signature; event-aligned trajectories and the interior-formation control constrained how that reservoir connects to knot entry; survival and chain-length analyses ruled out simple single-rate and single-size reductions. Only after this discovery sequence was a separate prospective branching experiment designed to test whether pre-existing mechanical states correspond to different subsequent topological switching probabilities.

This focus distinguishes the Letter from recent active-knotting studies. Li et al. [13] studied anchored actively reptating polymers, where intermittent giant conformational fluctuations and outward reptation generate continual self-knotting, knot migration, transfer, braiding, and topology engineering. Vatin et al. [12] studied polar active diblocks, where changing the active fraction produces a nonmonotonic knotting upsurge governed by competition between formation rate and knot lifetime. Zhou [15] further showed that spatial placement and signed propulsion polarity of active blocks can gate knotting, with polarity reversal changing knot survival and steady-state knotting by orders of magnitude. Together these studies establish that activity architecture can generate, manipulate, enhance, or suppress polymer knots. Here the control problem is different: cylindrical confinement reveals long-lived mechanical-state heterogeneity at fixed external controls, and the prespecified branching experiment tests whether those pre-existing states predict future topology. The geometry-dependent sign reversal then shows that a locally predictive collective variable need not provide a universal state-to-fate ordering.

The evidence therefore has deliberately different inferential levels. Retrospective matched-control signatures motivate the state description; event alignment establishes temporal ordering at the saved-frame resolution; lifetime and protocol controls constrain alternative reductions; and the prospective experiment is the confirmatory state-dependence test because the classifier, thresholds, eligibility rules, and finite observation horizon are fixed from independent calibration data before branch outcomes are examined. The subsequent second-coordinate analysis is post-outcome and explicitly exploratory. This hierarchy is essential to the broader claim: the work identifies a transferable geometry-state-rare-event framework without asserting a unique microscopic reaction coordinate or a completed causal intervention.

## S9. Statistical independence and limits of coordinate interpretation

Statistical independence is assigned at the trajectory or matched-control condition level, not to individual topology frames. Matched-replica permutation tests shuffle only within identical (R,Pe) groups. Event-centered averages are constructed per trajectory before combining trajectories with equal weight. The $N = 200/N = 450$ preliminary $2 \times 10^9$-step data set selected the parameter region of the 42 independent long productions and must not be pooled as equally long confirmatory evidence. The $N = 300$ five-seed reference is shown both in its full ensemble and restricted to common seeds 4–6. The common-horizon truncation addresses unequal endpoints among the 42 independently restarted trajectories, not all possible initial-condition or nonergodicity effects.

The operational score $m^*$ is assembled from three signed unknotted mechanical observables whose correlations with topology motivated their inclusion. PCA and leave-one-variable-out constructions therefore test representation stability rather than external predictivity. The prospective experiment addresses this limitation with a distinct prespecified configuration classifier, $s_{\mathrm{conf}} = \frac{-z(\overline{W})-z\left(\overline{R_{g,\perp}^{2}}\right)}{2}$, where the bars denote pre-branch trailing averages ending at the parent frame; normalization and Q20/Q80 thresholds are determined from independent calibration data. The classifier intentionally omits axial active power so that parent selection is configuration based. The two scores are related mechanical summaries with different inferential roles and are not treated as interchangeable reaction coordinates.

The 98.98% interior-event statistic is the fraction of detected formation locations, not an exposure-normalized hazard ratio for interior versus near-wall contour. It rules out the most literal picture in which nearly all detected events are direct wall contacts, but does not establish that a wall-assisted route is dynamically impossible. Likewise, event-aligned precursor changes preceding the first K-labeled frame can still contain unresolved within-frame motion. Complete-episode mean lifetimes underestimate the influence of strongly persistent right-censored episodes; the independent survival-shape comparison is therefore preferred to any single complete-episode mean as evidence about escape heterogeneity.

## S10. Primary prediction, experimental implications, and scope

The primary prospective prediction is that mechanically distinct unknotted states prepared at the same $(R, Pe)$ can have different finite-time knotting probabilities when the classifier, thresholds, and observation duration are specified before any branch outcomes are generated. Here the horizon $H$ denotes the prespecified duration of observation after branching—the fixed elapsed simulation time over which first knotting is monitored—not a fitted kinetic time scale, lifetime, or half-life. Section S13 reports this test. The prespecified $s_{\mathrm{conf}}$ classifier gives strong HIGH>LOW separation at $R = 1.85$ but LOW>HIGH at $R = 1.90$. Accordingly, the experiment confirms state-dependent future switching while rejecting a geometry-independent scalar susceptibility ordering. This state-resolved formulation is relevant beyond polymer knotting whenever external controls coexist with metastable internal states and rare transitions.

The reversal also defines the limit of the confirmatory claim. The earlier $\mathrm{Pe} = 3$ high-radius deviation from the one-coordinate state-selection model and the $N = 450$ fixed-$N_{\mathrm{active}}$ splitting had suggested a missing slow degree of freedom, but they did not identify it. Section S14 examines candidate variables in the same pre-branch trailing histories ending at the parent frame only after the $R = 1.90$ reversal was known. Those analyses are therefore explicitly exploratory: they do not relabel parents, change the primary endpoint, or contribute to the prospective p values.

The experimentally relevant statement is broader than a shift in mean knot occupancy. Under comparable channel confinement and nonequilibrium contour drive, pre-transition mechanical organization should stratify future switching even when external controls are identical, and geometry can change the ordering of that susceptibility. Changing the driven-block length at fixed total contour should also not be equivalent to adding passive contour at fixed driven-block length. The present results do not imply universal exponents, a unique critical radius, a thermodynamic free-energy landscape, or a quantitative mapping from simulation Pe and R to a particular experiment.

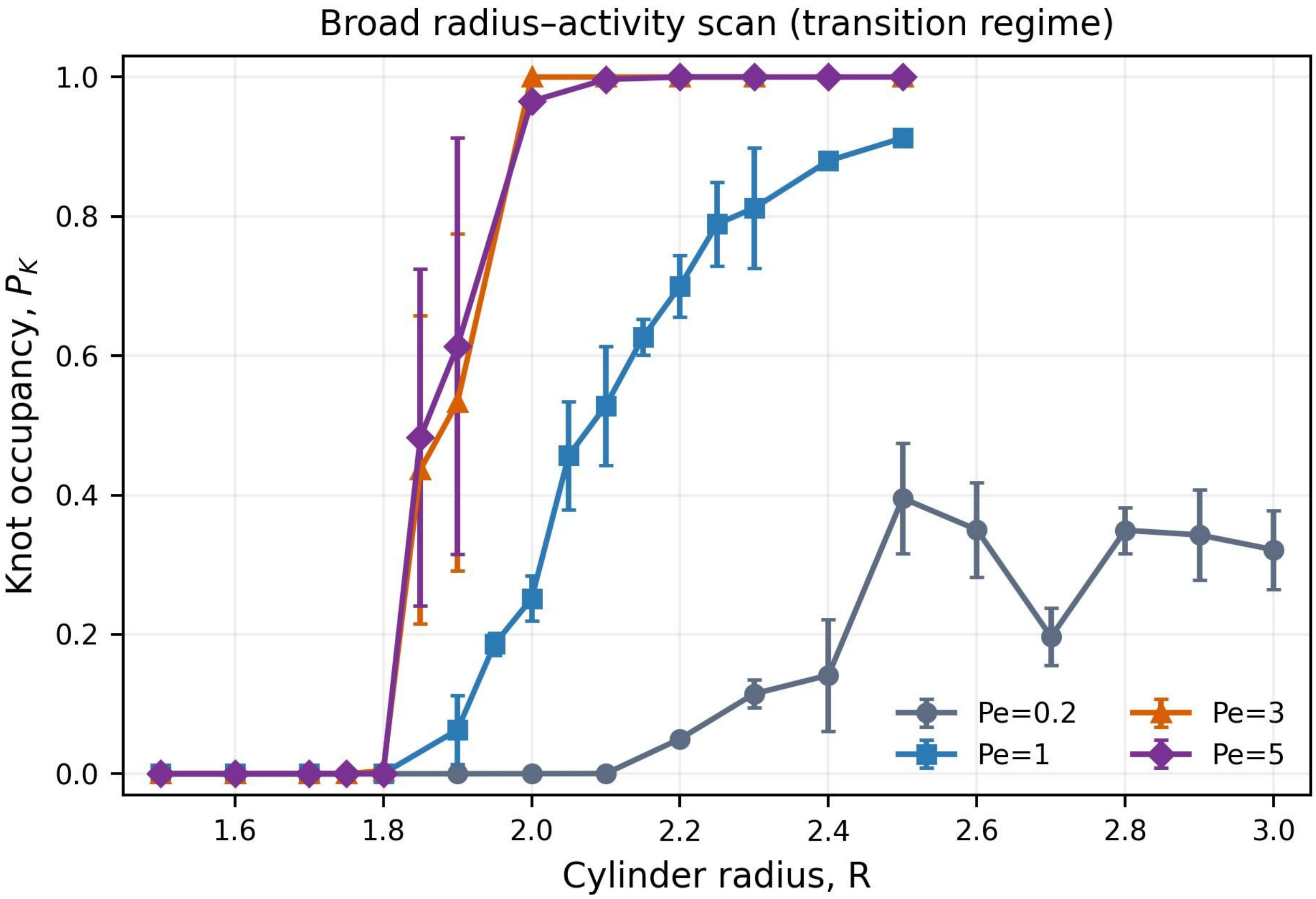

FIG. S1. Broad knot occupancy $P_K$ versus cylinder radius R for Pe = *0.2*,1,3,5. Points are parameter-level means; error bars are seed-level SEM where independent replicas are available. The scan is not a complete Cartesian grid.

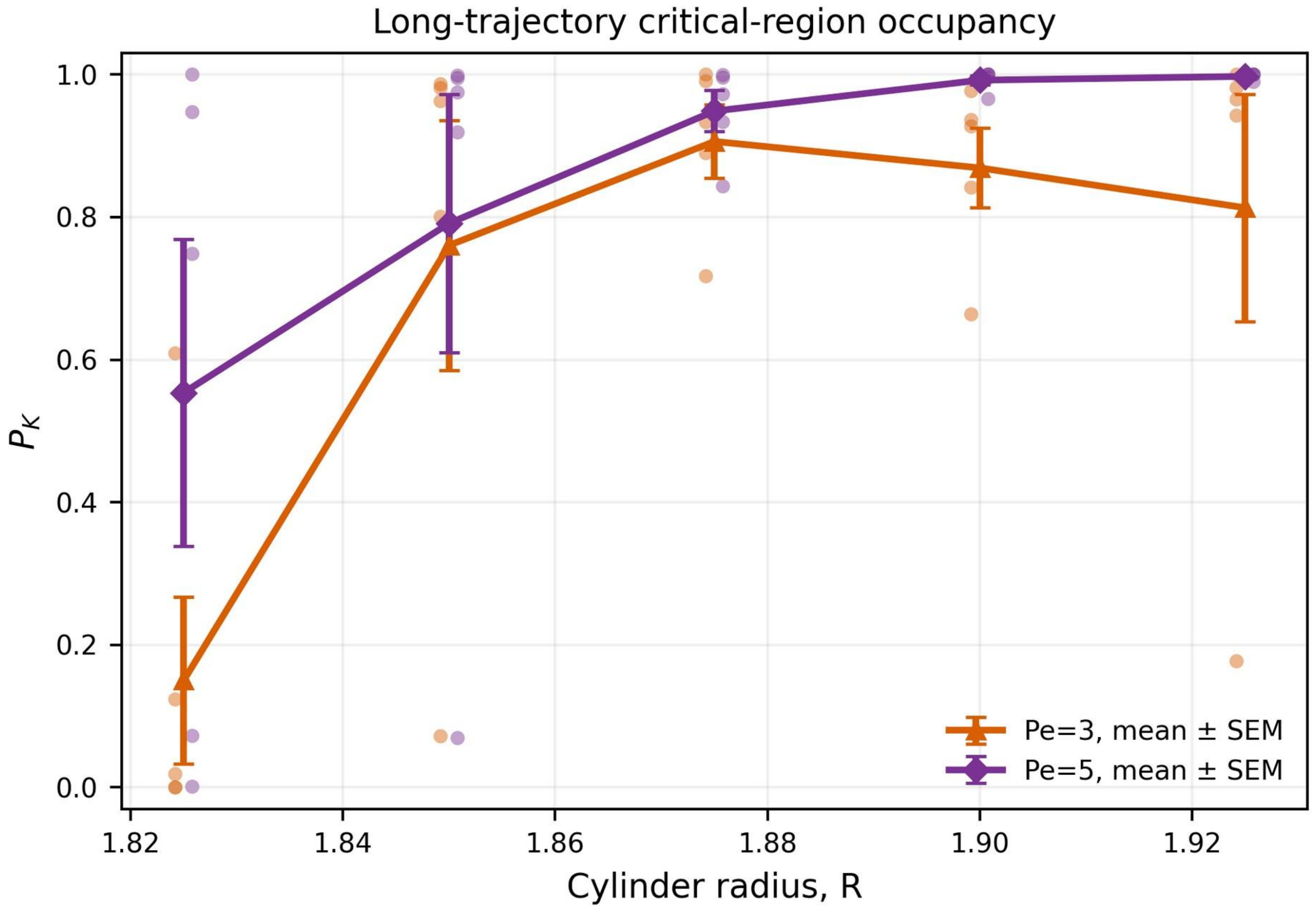

FIG. S2. Knot occupancy in the long-trajectory crossover-region campaign. Individual seed values are shown with transparent symbols; connected points are seed means±SEM. These data resolve the high-activity crossover and its replica heterogeneity.

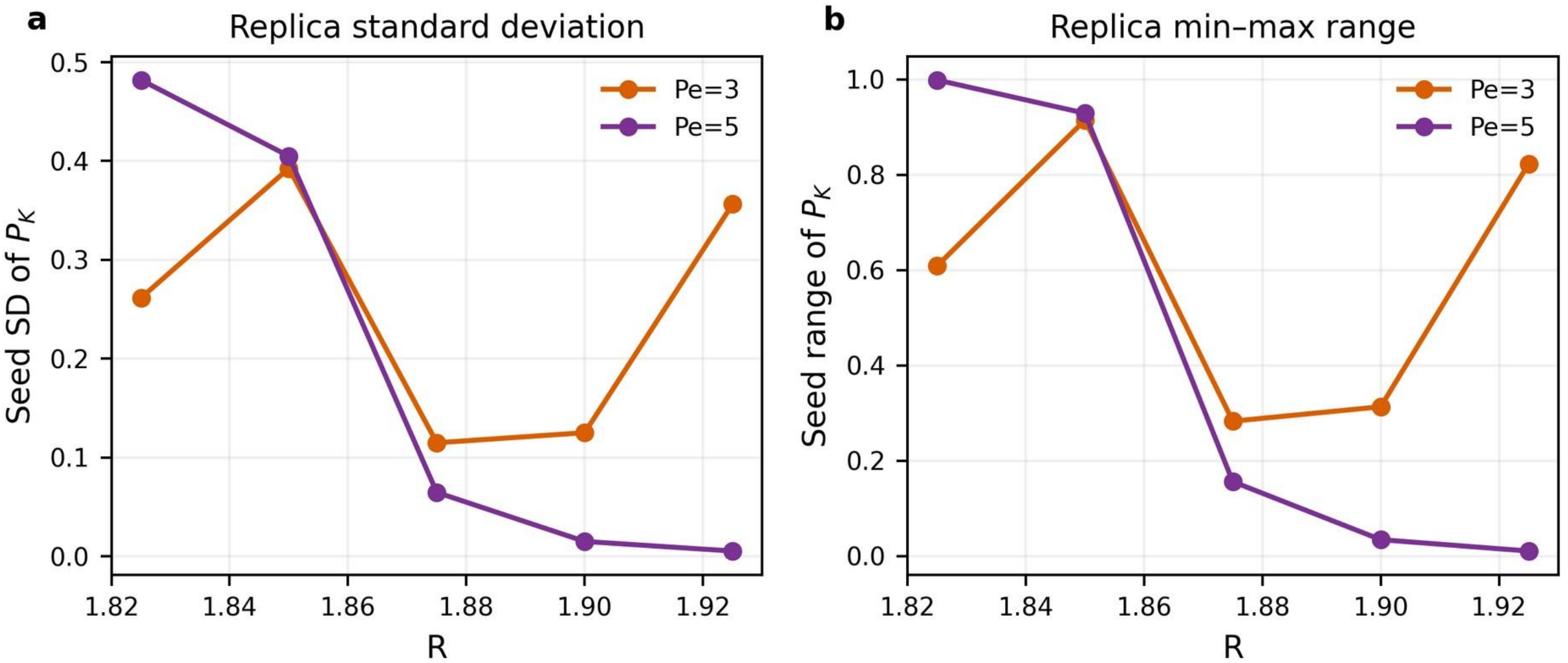

FIG. S3. Crossover-region replica heterogeneity quantified by the seed standard deviation and seed min-max range of $P_K$. Large dispersion near the crossover demonstrates correlation-limited finite-time sampling.

Figures S1–S3 separate two questions that are easily conflated in an average occupancy curve. The broad scan establishes where confinement and activity change the mean topological response, whereas the replicated long trajectories show that the transition-relevant window contains large, persistent seed-to-seed differences at the same nominal controls. The latter observation is the reason the Letter treats the internal state as a conditioning variable rather than assigning a unique susceptibility to each (R,Pe) point. We use this finite-time replica splitting operationally as evidence of long-lived

nonequilibrium-state heterogeneity; it is not interpreted as an equilibrium phase coexistence or as proof of a thermodynamic critical point.

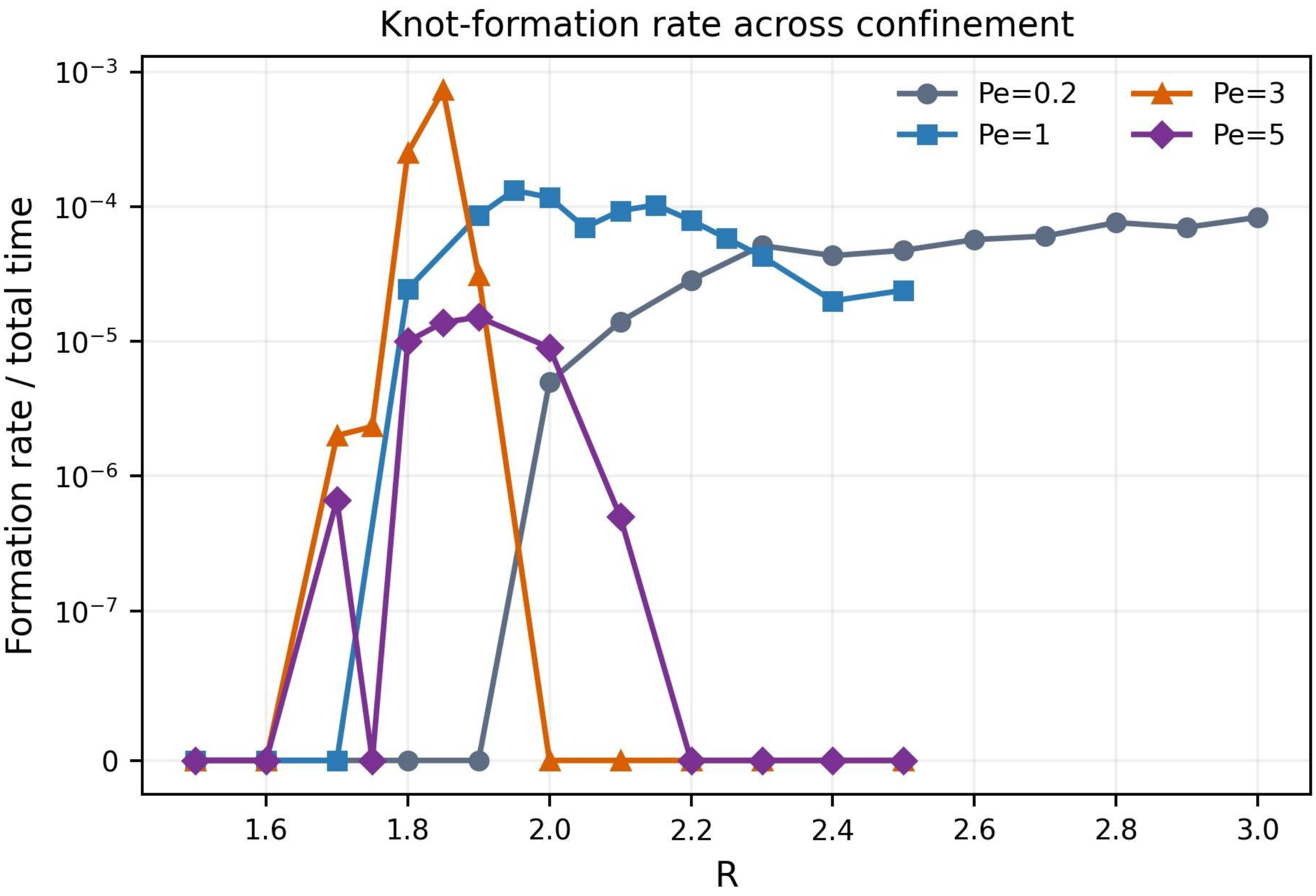

FIG. S4. Knot-formation rate per unit total valid time across the broad scan. A symmetric-log ordinate accommodates rates spanning orders of magnitude and zeros.

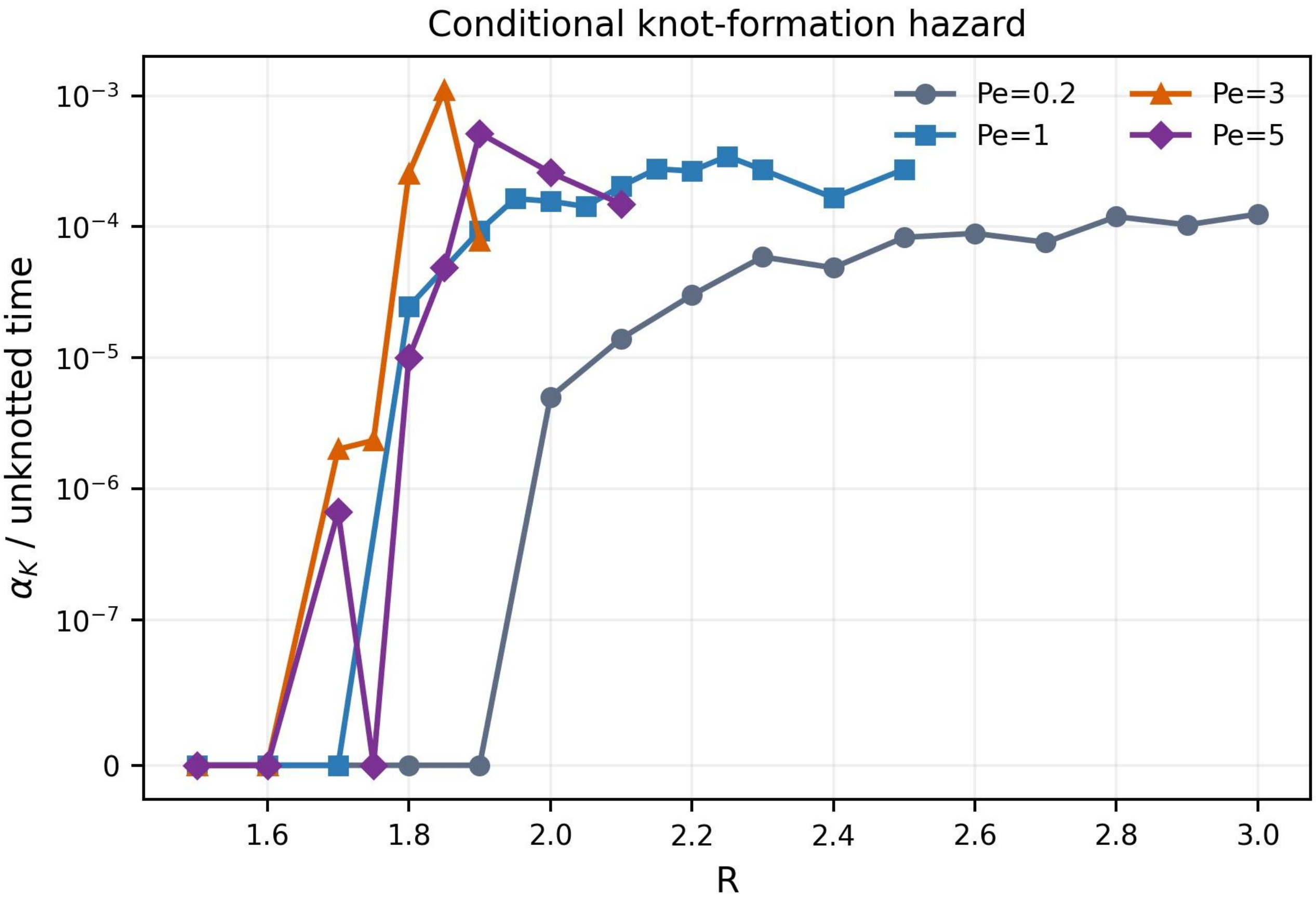

FIG. S5. Conditional formation hazard $\alpha_K = N_{\mathrm{form}}/T_{\mathrm{unknotted}}$. Conditioning on unknotted exposure separates formation propensity from the reduced opportunity for new events in trajectories that remain knotted most of the time.

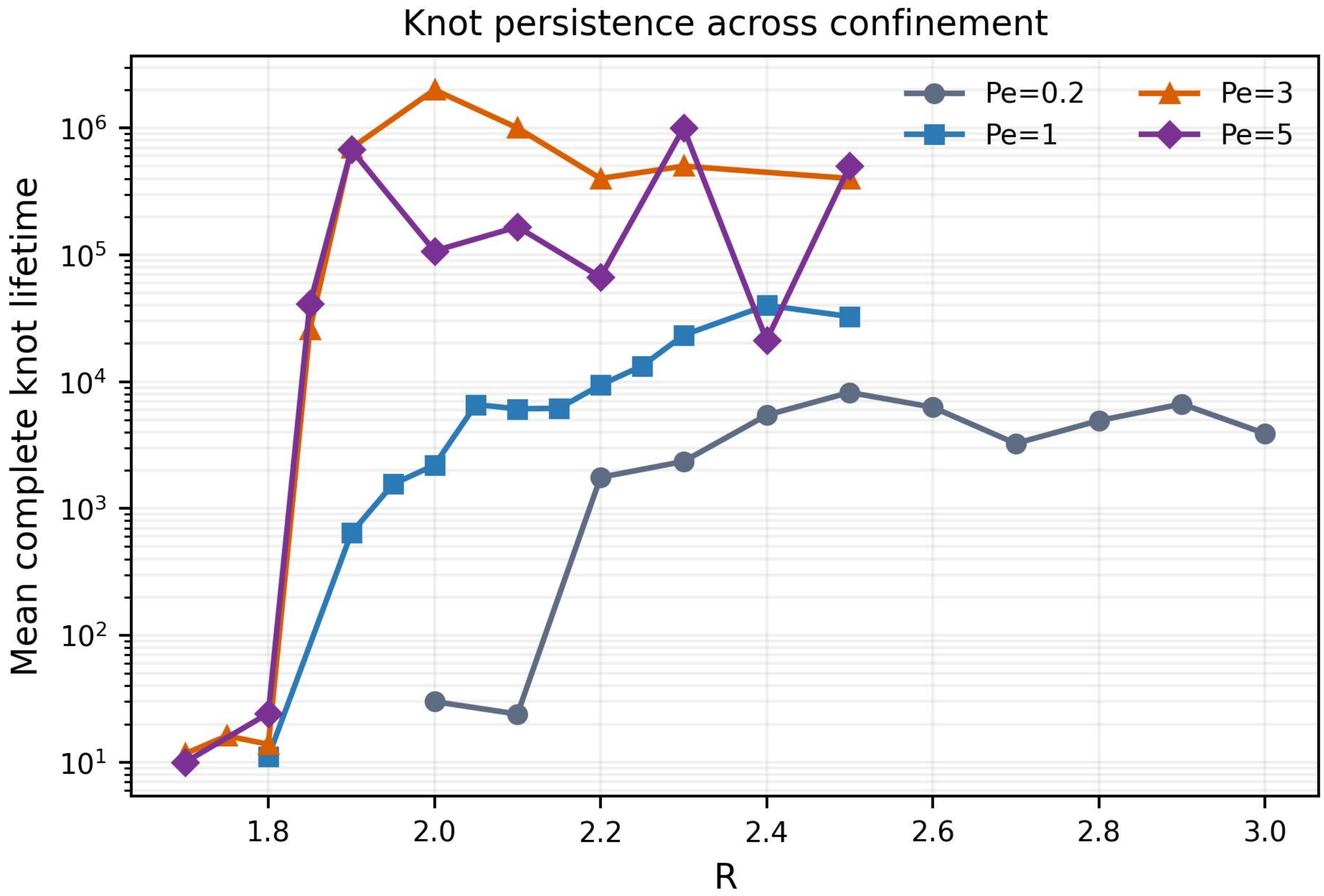

FIG. S6. Mean lifetime of complete knot episodes across the broad scan. The logarithmic ordinate highlights the increase in persistence; long-lived and censored trajectories require interpretation together with correlation times and episode counts.

Figures S4–S6 also show why occupancy cannot be identified with a single kinetic rate. The total-time formation rate mixes intrinsic access to threading with the amount of time a trajectory is available to form a new knot, whereas the conditional hazard in Fig. S5 divides by unknotted exposure. Conversely, complete-episode lifetimes in Fig. S6 omit the full duration of right-censored knots that survive to the end of a trajectory. Formation access and topological residence therefore carry distinct information. The censored survival analysis in Fig. S13 is the stronger test of persistence heterogeneity, and it is deliberately kept separate from the formation-hazard analysis rather than collapsed into a one-rate description.

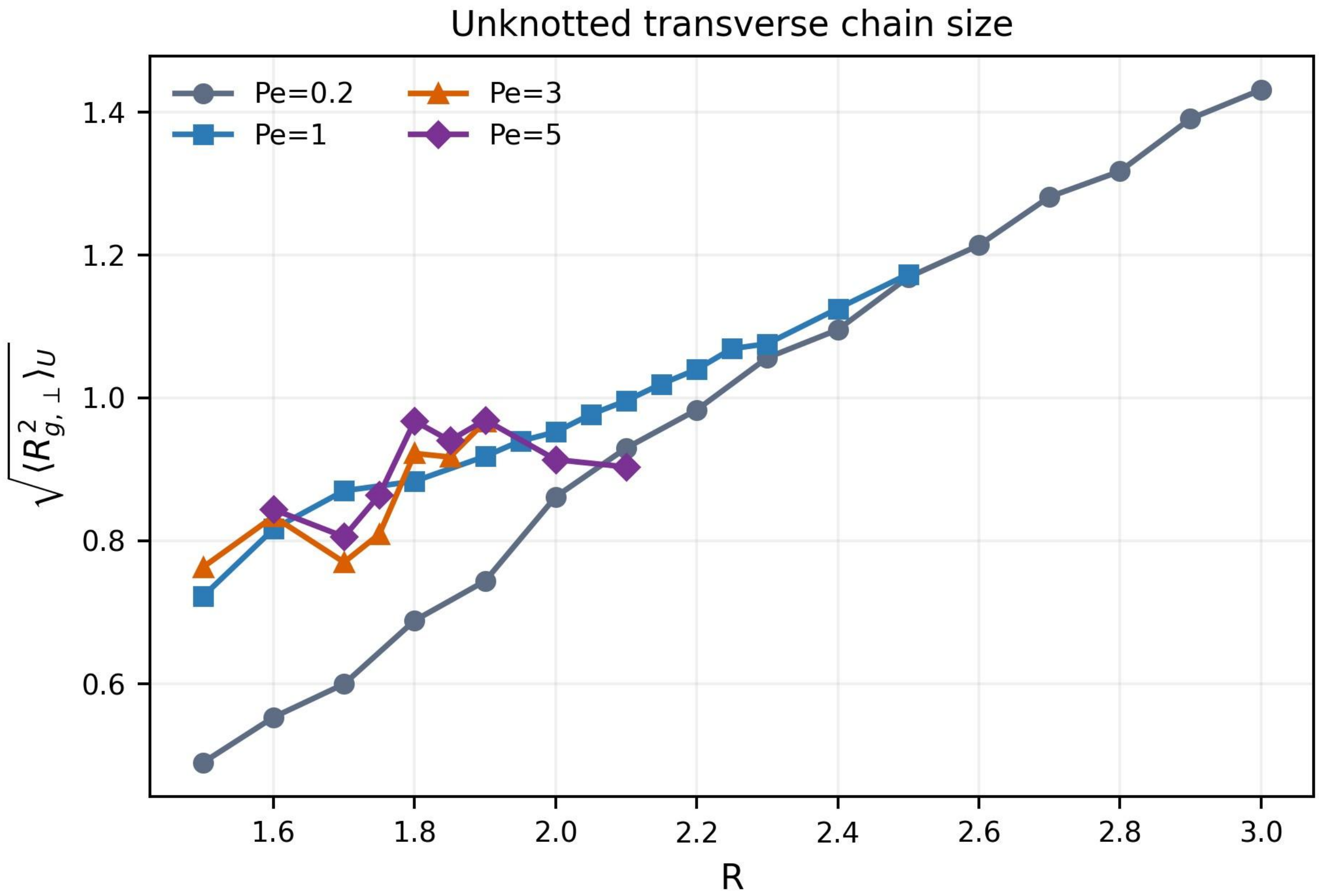

FIG. S7. Unknotted transverse RMS gyration size across the broad scan from the mass-weighted gyration calculation.

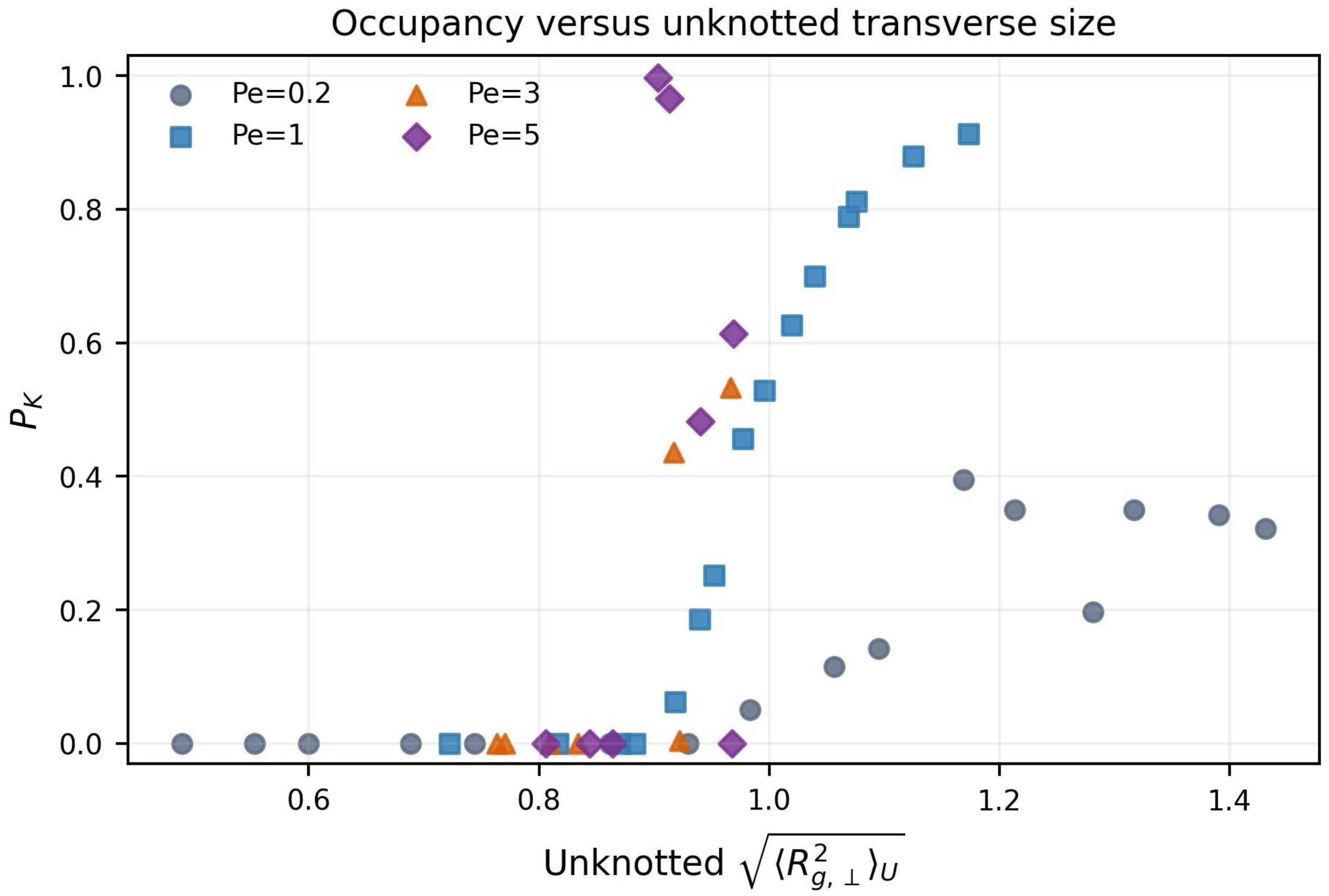

FIG. S8. Knot occupancy versus unknotted transverse RMS gyration size for the broad scan. Activity-dependent separation shows that mean transverse size alone does not collapse the active-knotting data. Several high-activity scan conditions cluster near the top boundary because distinct sampled radii map to nearly identical transverse-size and occupancy values; overlapping or near-overlapping markers are therefore retained rather than jittered.

The size diagnostics delimit rather than eliminate the role of compaction. Transverse size changes systematically with radius and is associated with susceptibility within matched controls, but the broad-scan points do not collapse onto a single activity-independent P_K–size curve. Thus absolute size contains both trivial geometric dependence and state information. This is why the main analysis removes the matched-condition mean before asking whether more compact unknotted replicas are more knot-prone: the within-condition contrast tests state selection, whereas the raw broad-scan trend alone would not distinguish state dependence from the direct effect of changing the pore.

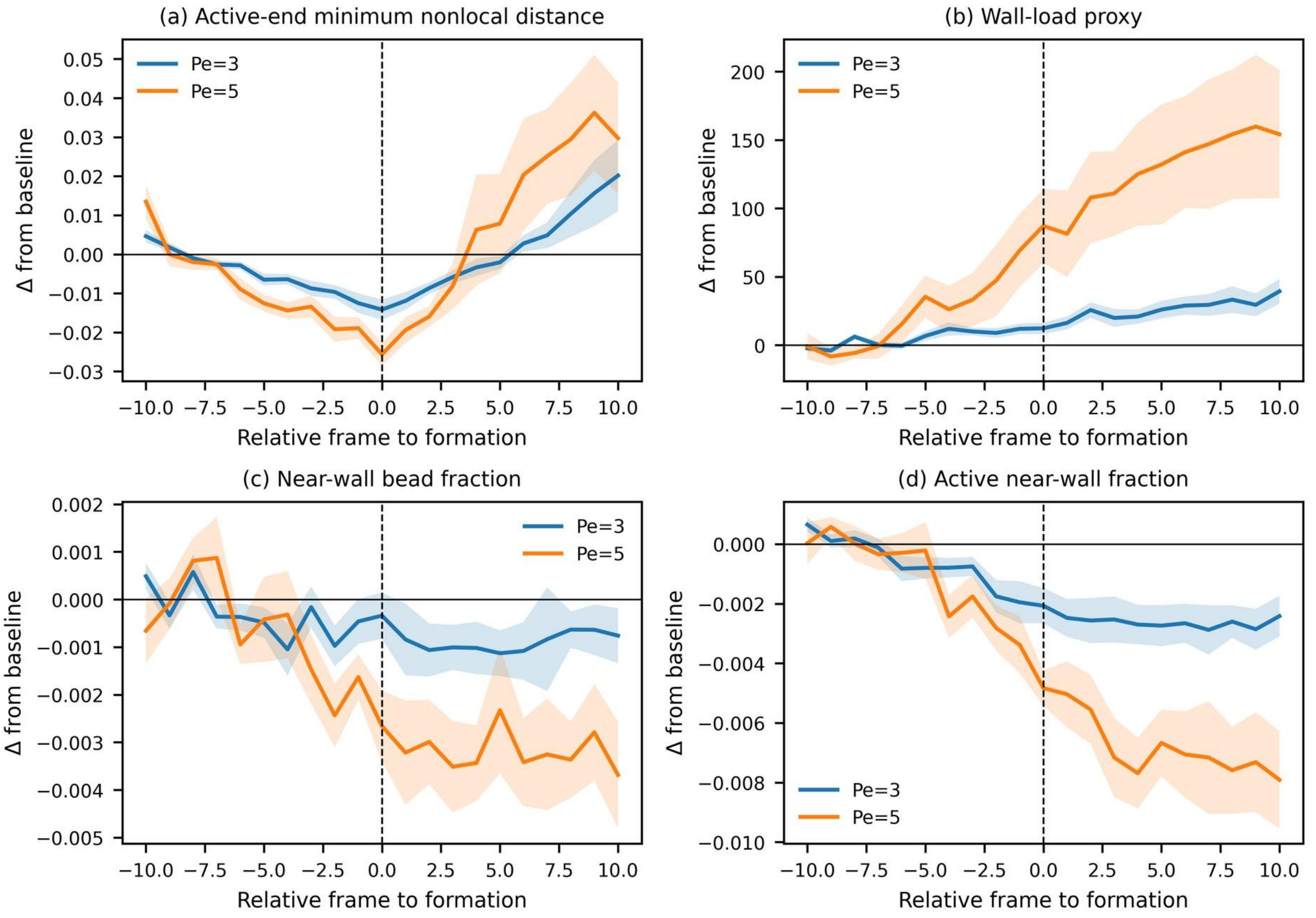

FIG. S9. Additional event-aligned controls for the same Pe = *3* and Pe = *5* critical trajectories used in the main text. The active-end minimum nonlocal distance decreases before formation; wall load rises transiently; and near-wall fractions decrease slightly, arguing against a direct wall-contact formation event.

Taken together, main-text Fig. 2 and Fig. S9 reveal a useful separation of time scales. Before the first detected knot, nonlocal approach of the active end and transverse shortening develop coherently, while the minimum nonlocal distance decreases. At the same time, the wall-load proxy rises transiently but the near-wall fractions do not rise and the active near-wall fraction decreases. This is compatible with a short compression/threading excursion inside the pore, not with a picture in which the knot is produced by sustained adsorption or a direct wall-contact event. The transient wall-load increase is also not inconsistent with the lower time-averaged wall load of susceptible reservoirs: one describes the approach to an event, the other the long-lived state from which the event is entered.

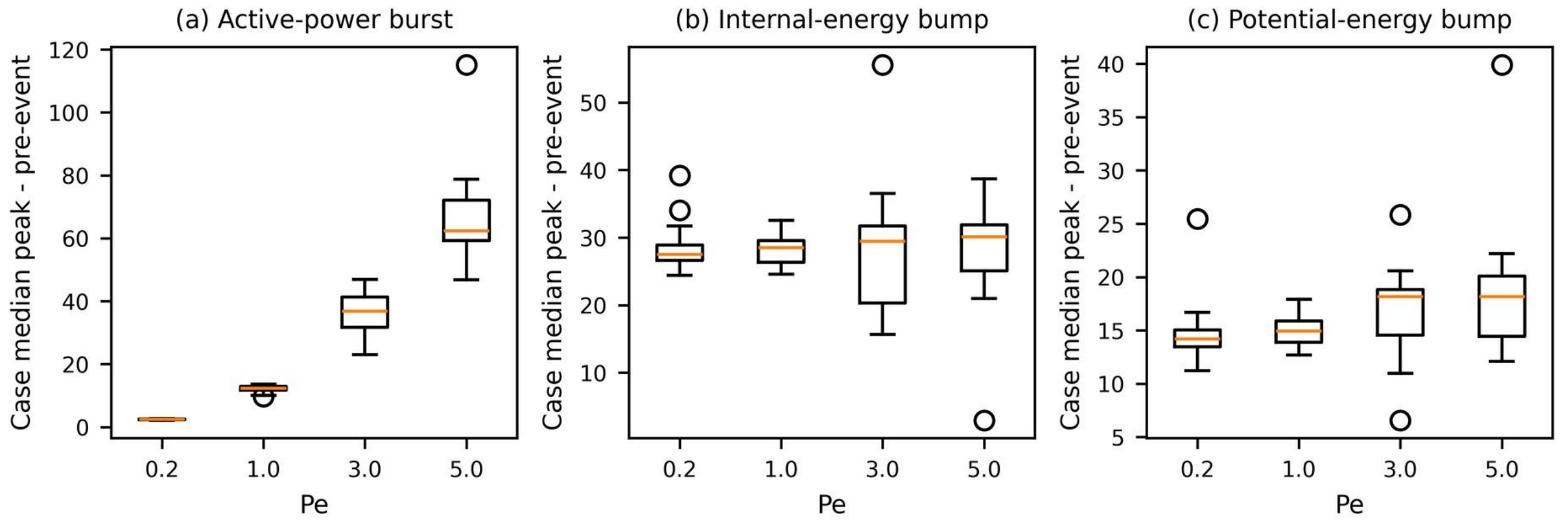

FIG. S10. Event-centered rich-dump proxies. The active-power burst grows strongly with activity, whereas internal- and potential-energy bumps vary much less.

The rich-dump observables give an independent nonequilibrium consistency check on this transition-path interpretation. The strong growth of the active-power burst with Pe, compared with the much weaker variation of internal- and potential-energy bumps, identifies injected active work as the quantity most strongly amplified near the detected transition. We therefore avoid translating these event-centered signals into an equilibrium barrier picture. Their role is narrower: they show that the approach to knotting becomes increasingly drive-dominated as activity increases, while the reservoir and prospective analyses remain defined by mechanical state rather than by an assumed free-energy coordinate.

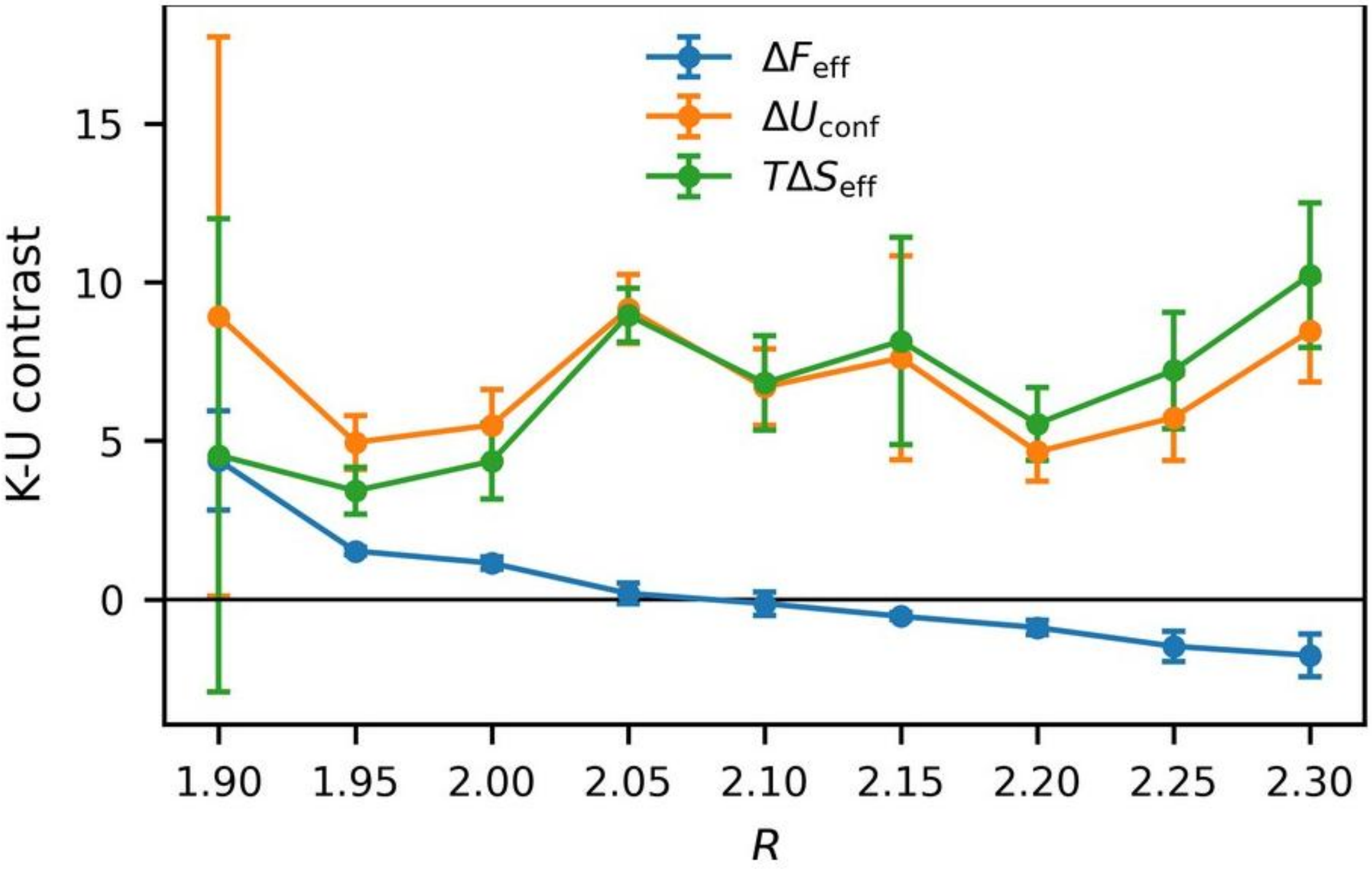

FIG. S11. Nonequilibrium occupancy/energy diagnostics across the Pe=1 confinement scan. The occupancy-derived ΔF_eff=-kBT ln(PK/PU) changes sign through the crossover while the confined internal-energy contrast ΔU_conf remains positive. The residual TΔS_eff=ΔU_conf-ΔF_eff is used only as a nonequilibrium bookkeeping diagnostic, not as a thermodynamic entropy.

This comparison makes the energetic point more restrictive than a purely structural contrast. Across the Pe=1 crossover, topology carries a positive confined internal-energy cost even where the log-occupancy contrast favors the knotted sector. The increased knot population therefore cannot be interpreted as the knotted state simply lying lower in an equilibrium potential-energy landscape. Because the polymer is actively driven and detailed balance is absent, ΔF_eff and TΔS_eff are diagnostic labels for occupancy/energy mismatch rather than thermodynamic free energy and entropy. Together with the event-centered power and energy responses in Fig. S10, the result separates injected work from conservative-energy response without using either quantity to define the branching classifier.

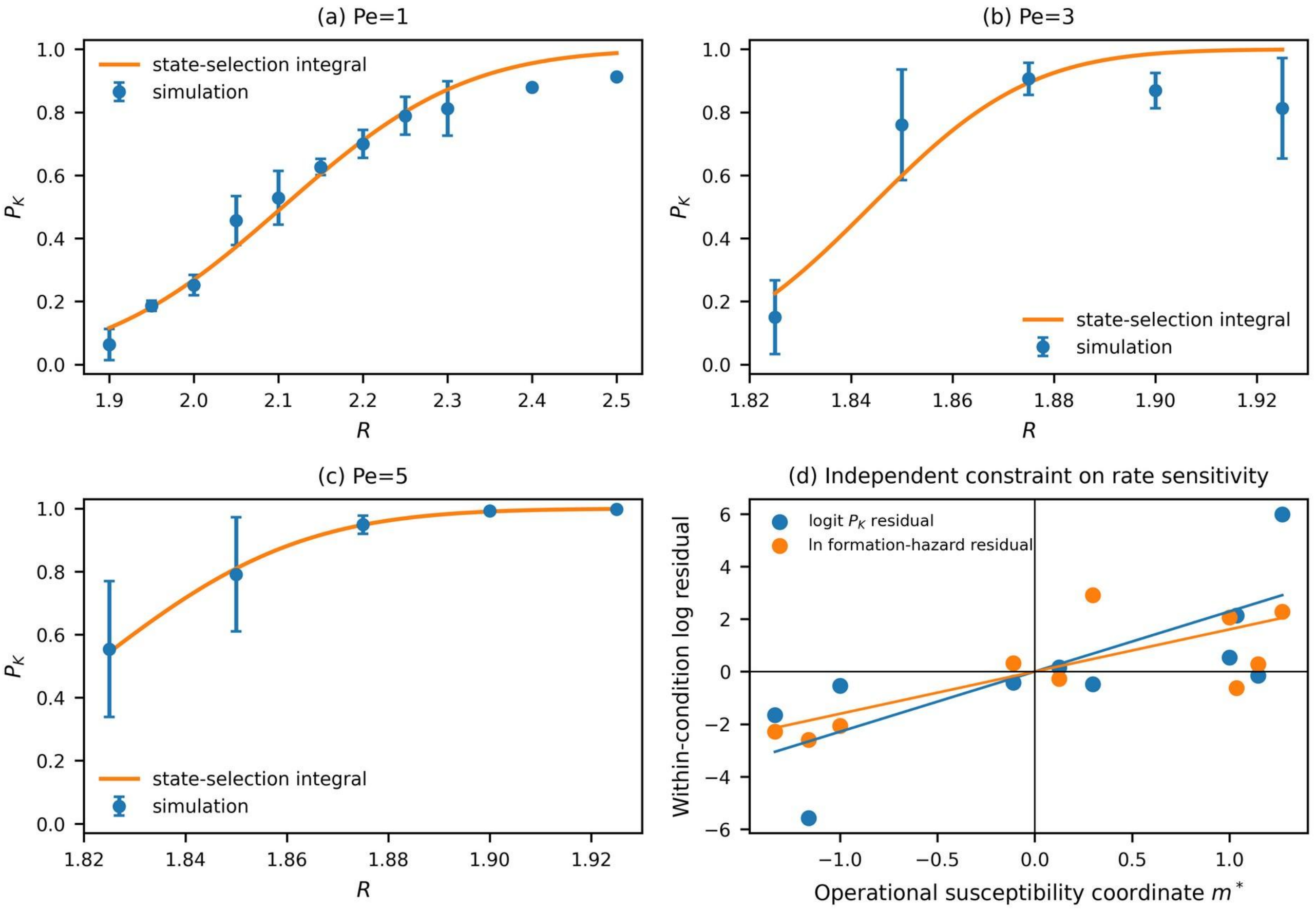

FIG. S12. Quantitative state-selection model. (a-c) Simulation $P_K$(R) and the numerical state-selection integral for Pe = *1*,3,5 with $\gamma$ fixed by the within-condition regression and only $R_0$ and $\chi$ adjusted per activity. The Pe = *3* high-R deviation is included. (d) Fixed-effect residuals constraining occupancy and formation sensitivity.

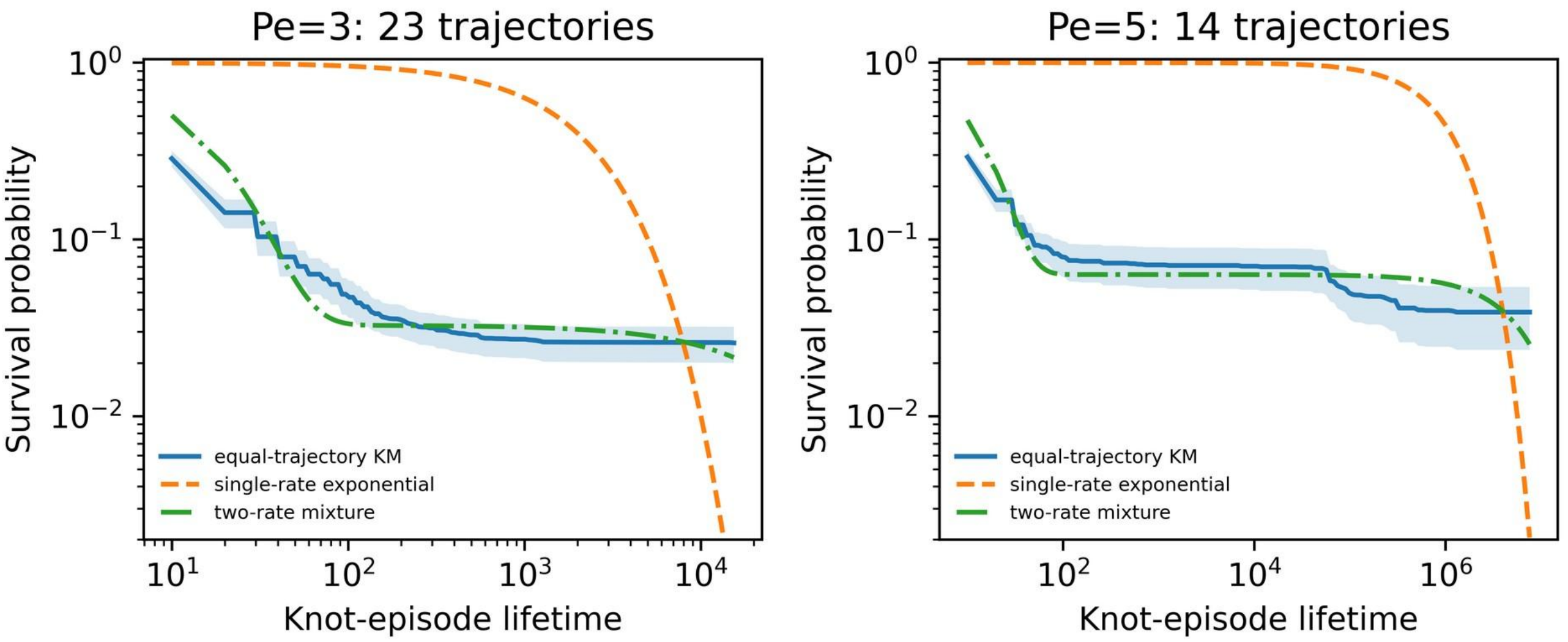

FIG. S13. Independent lifetime-shape test of escape-rate heterogeneity. Equal-trajectory Kaplan-Meier survival is compared with a single-rate exponential and a minimal two-rate mixture; left-censored episodes are excluded and episodes still knotted at the trajectory endpoint are included as right-censored observations.

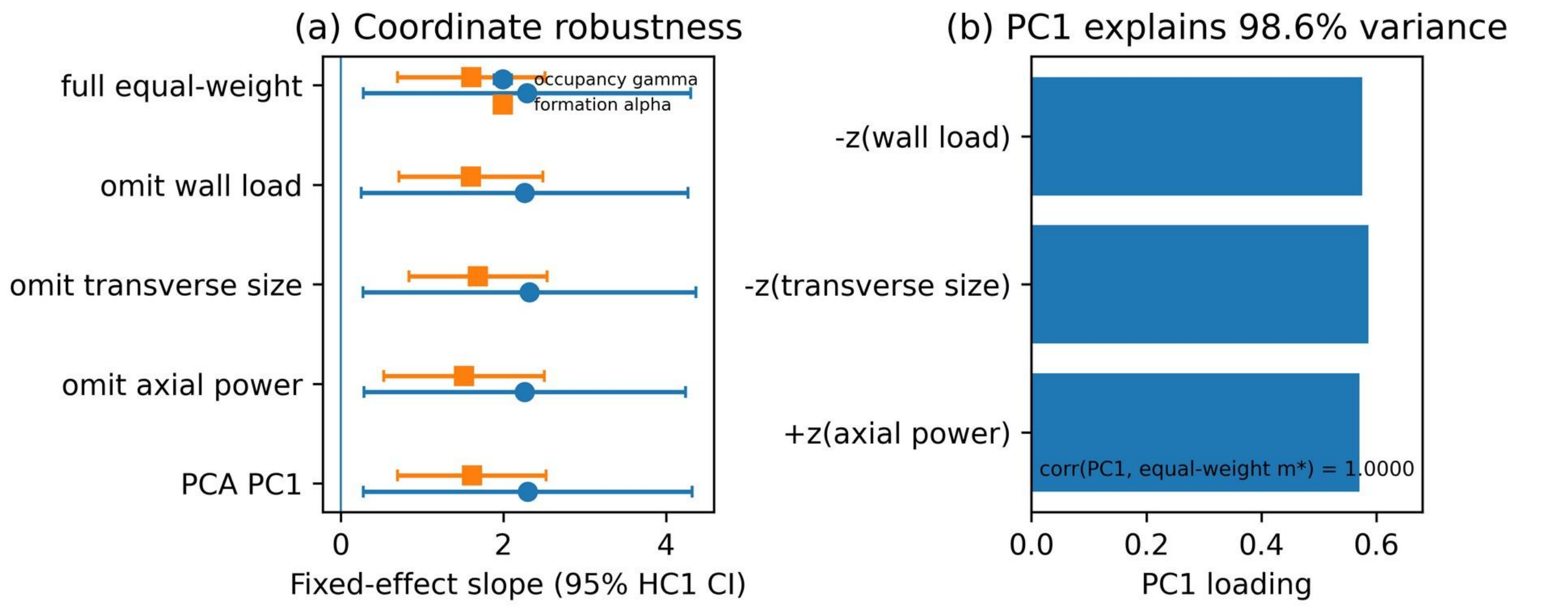

FIG. S14. Robustness of the operational susceptibility coordinate. Leave-one-observable-out variants and PCA PC1 give similar fixed-effect slopes; PC1 explains 98.6% of the signed three-observable variance and is nearly identical to the equal-weight coordinate.

Figures S12–S14 clarify what the reduced state description can and cannot claim. The state-selection integral organizes much of the radius dependence using a mechanical coordinate fixed by matched-control regressions, but its high-R Pe=3 deviation occurs precisely where replica heterogeneity is largest. At the same time, leave-one-observable-out variants and PCA preserve the dominant matched-control direction, so the signal is not carried by one arbitrarily chosen observable. These two facts point in the same direction: the compact/unloaded mode is reproducible as a local state coordinate, yet a one-dimensional projection is insufficient globally. The later prospective reversal tests this limitation without refitting the score to the branch outcomes.

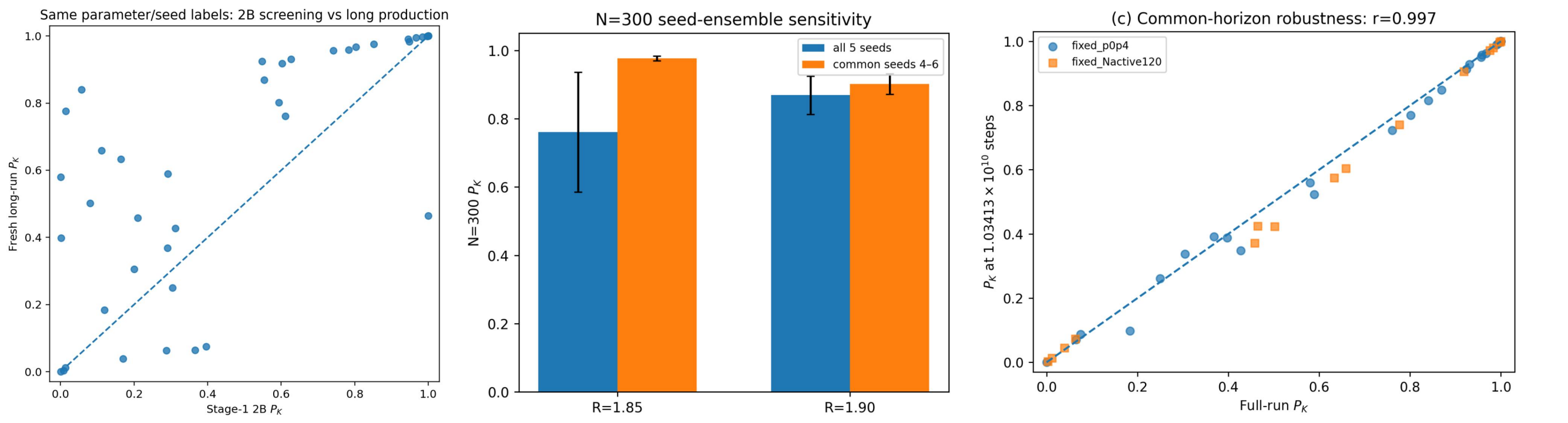

FIG. S15. Robustness of the chain-length data set. (a) The preliminary $2 \times 10^9$-step data set preserves broad rank information but differs substantially from independently restarted long production, justifying its use for selection rather than final inference. (b) $N = 300$ seed-ensemble sensitivity is substantial at $R = 1.85$ but small at $R = 1.90$. (c) Full-run $P_K$ and the common-horizon estimate at $1.03413 \times 10^{10}$ steps agree closely across all 42 independently restarted trajectories.

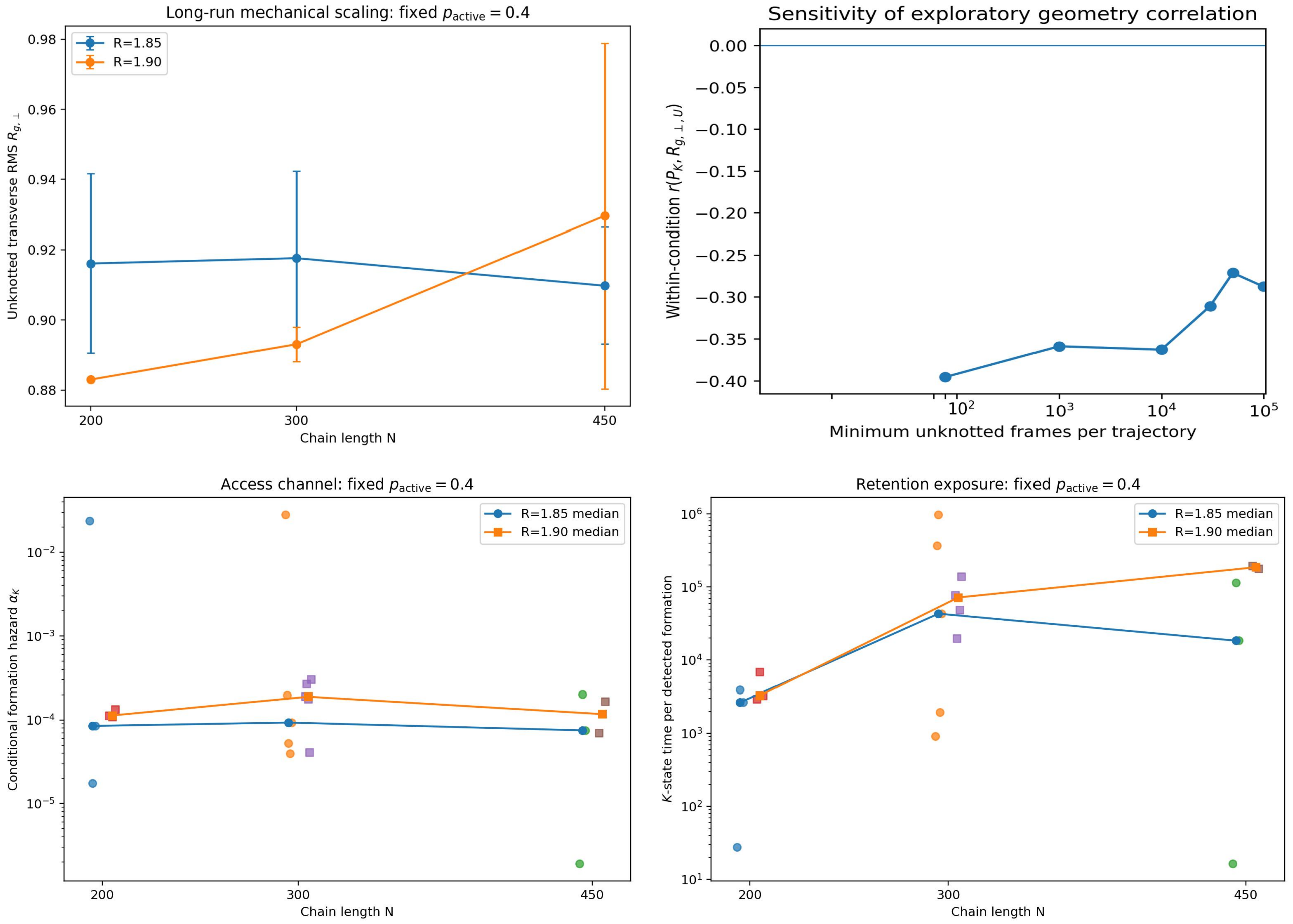

FIG. S16. Exploratory chain-length mechanical and kinetic decomposition. (a) Unknotted transverse RMS size under fixed does not yield a simple monotonic geometric collapse. (b) The within-condition transverse-size correlation remains negative but weakens when increasingly long unknotted exposure is required. (c) Conditional formation hazard and (d) occupancy-budget K-state time per detected formation for matched radii. The latter is an accounting quantity linked algebraically to , not an independent escape-rate measurement.$p_{\text{active}}$ = *0.4*$P_K$

The chain-length controls provide a second route to the same conclusion. Changing N at fixed active fraction changes both passive and active contour, whereas changing N at fixed active-block length adds passive contour without increasing the driven block. The very different occupancy responses under these two protocols therefore rule out interpreting chain length as a universal scalar control. The access/residence accounting in Fig. S16 further shows that similar occupancy changes can be allocated differently between formation opportunity and time spent knotted. Because the residence quantity is algebraically tied to occupancy, we use it only for bookkeeping; the independent evidence for heterogeneous escape comes from the censored survival shapes.

## S11. Replica heterogeneity, topology-resolved geometry, and correlation-limited sampling

This section collects the additional diagnostics needed to assess replica heterogeneity, finite-time sampling, topology-resolved geometry, and the protocol-controlled chain-length interpretation. Null and sensitivity results are retained when they delimit the range of validity of a reduced mechanical description.

Replica-resolved occupancy curves show how the character of the confinement crossover changes with activity. Pe = *1* displays a broad, comparatively smooth crossover, whereas the Pe = *3* transition window is strongly replica heterogeneous and Pe = *5* shifts toward tighter confinement before rapidly saturating (Figs. S17–S19). These comparisons separate activity-dependent crossover location from the finite-time replica splitting that motivates replica-level inference. Seed-by-radius heterogeneity and the descriptive logistic crossover context are shown in Figs. S20-S22.

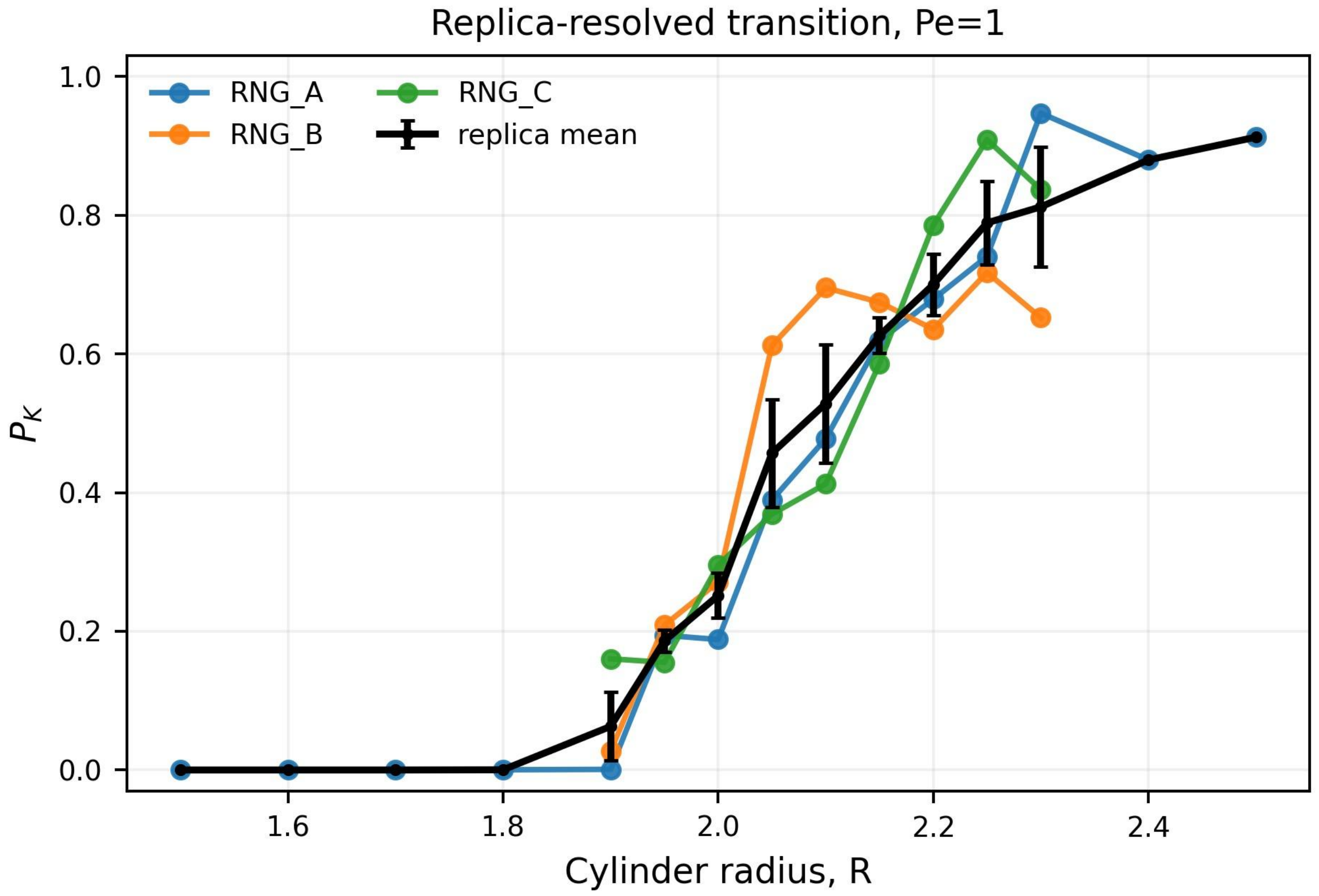

FIG. S17. Replica-resolved $P_K$ for Pe = *1*. The black curve is the replica mean and colored curves are individual trajectories; the crossover is comparatively broad and smooth.

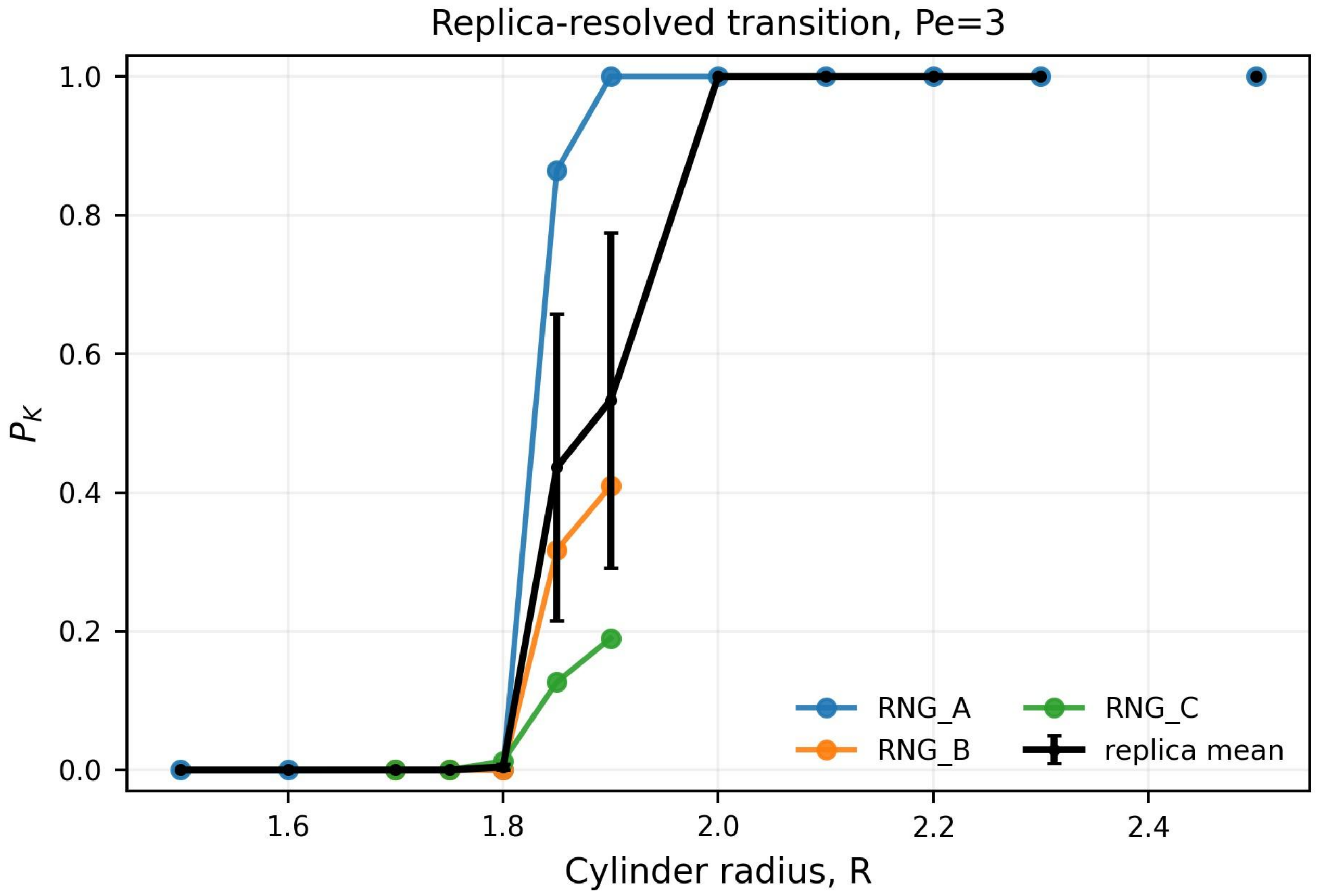

FIG. S18. Replica-resolved $P_K$ for Pe = *3*. Strong seed-to-seed dispersion appears in the crossover confinement window, motivating the explicit metastability and common-horizon checks used in the revised chain-length analysis. Lines connect only contiguous sampled conditions for each replica; isolated markers are intentionally not bridged across gaps in that replica's sampled radius sequence.

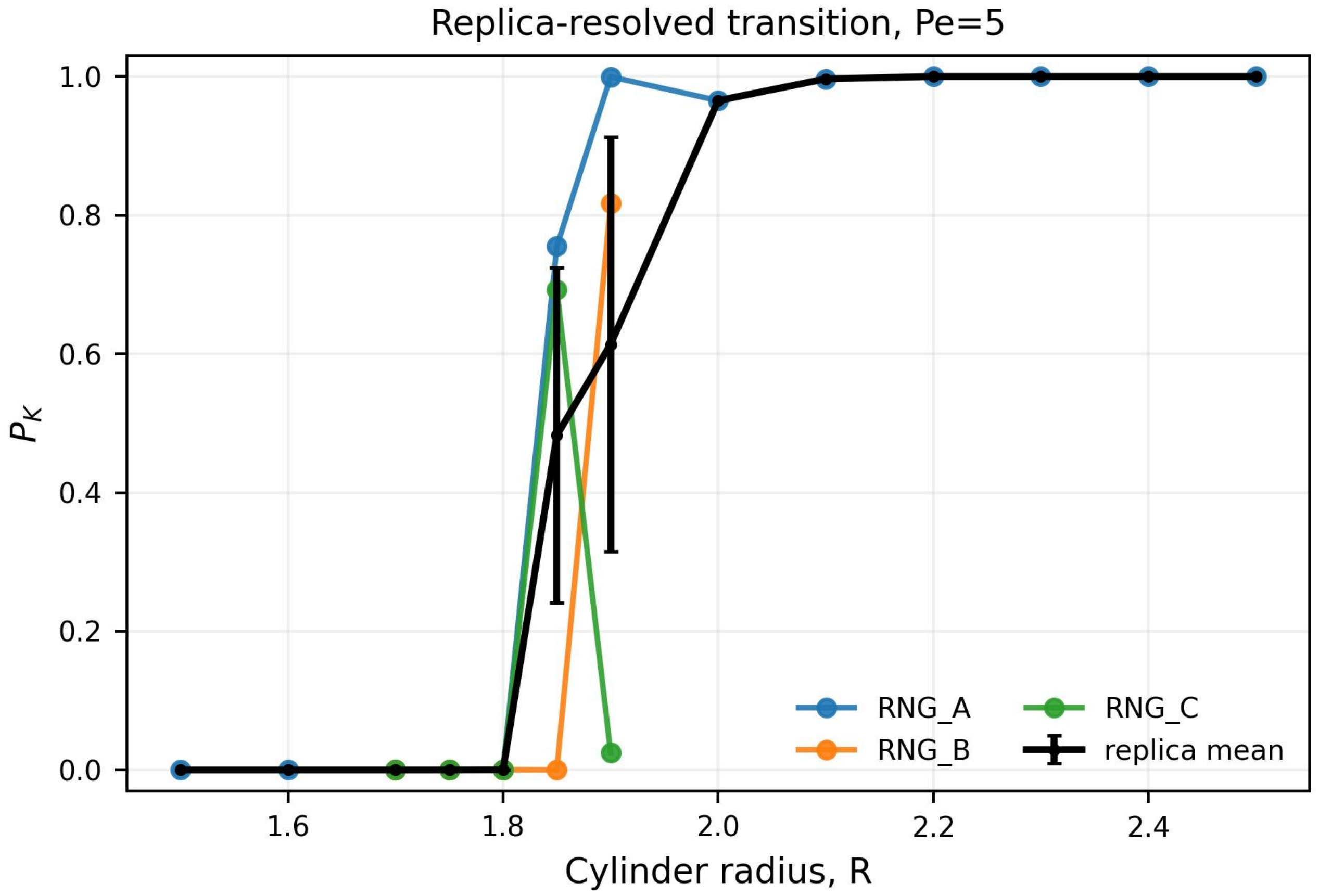

FIG. S19. Replica-resolved $P_K$ for Pe = *5*. The finite-time crossover is shifted to tighter confinement relative to Pe = *1* and becomes nearly saturated beyond the narrow transition window.

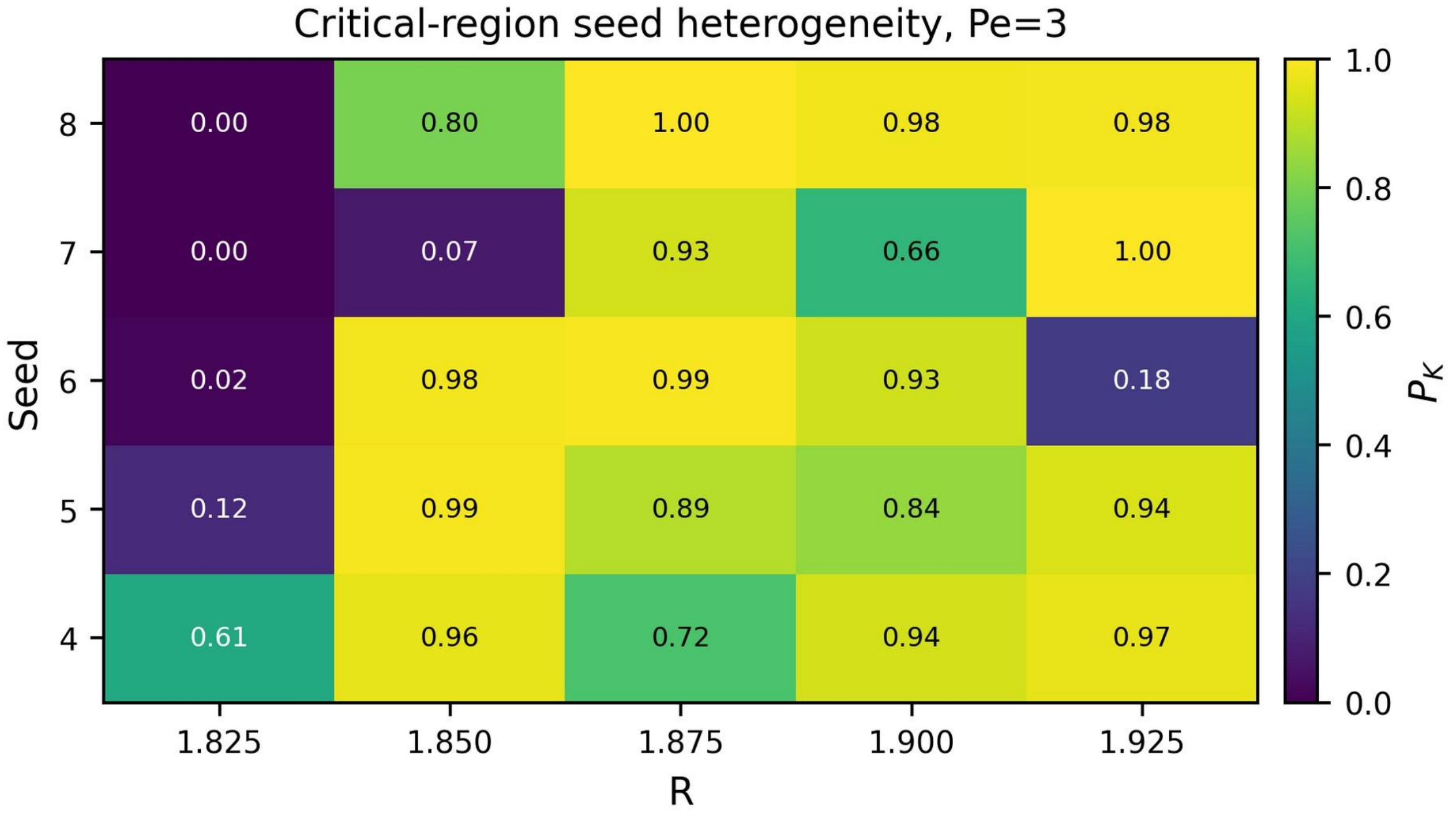

FIG. S20. Seed-by-radius heat map of $P_K$ for the Pe = *3* crossover-region long-trajectory data set. The patchy pattern shows that neighboring radii can populate distinct long-lived topological states over the available observation time.

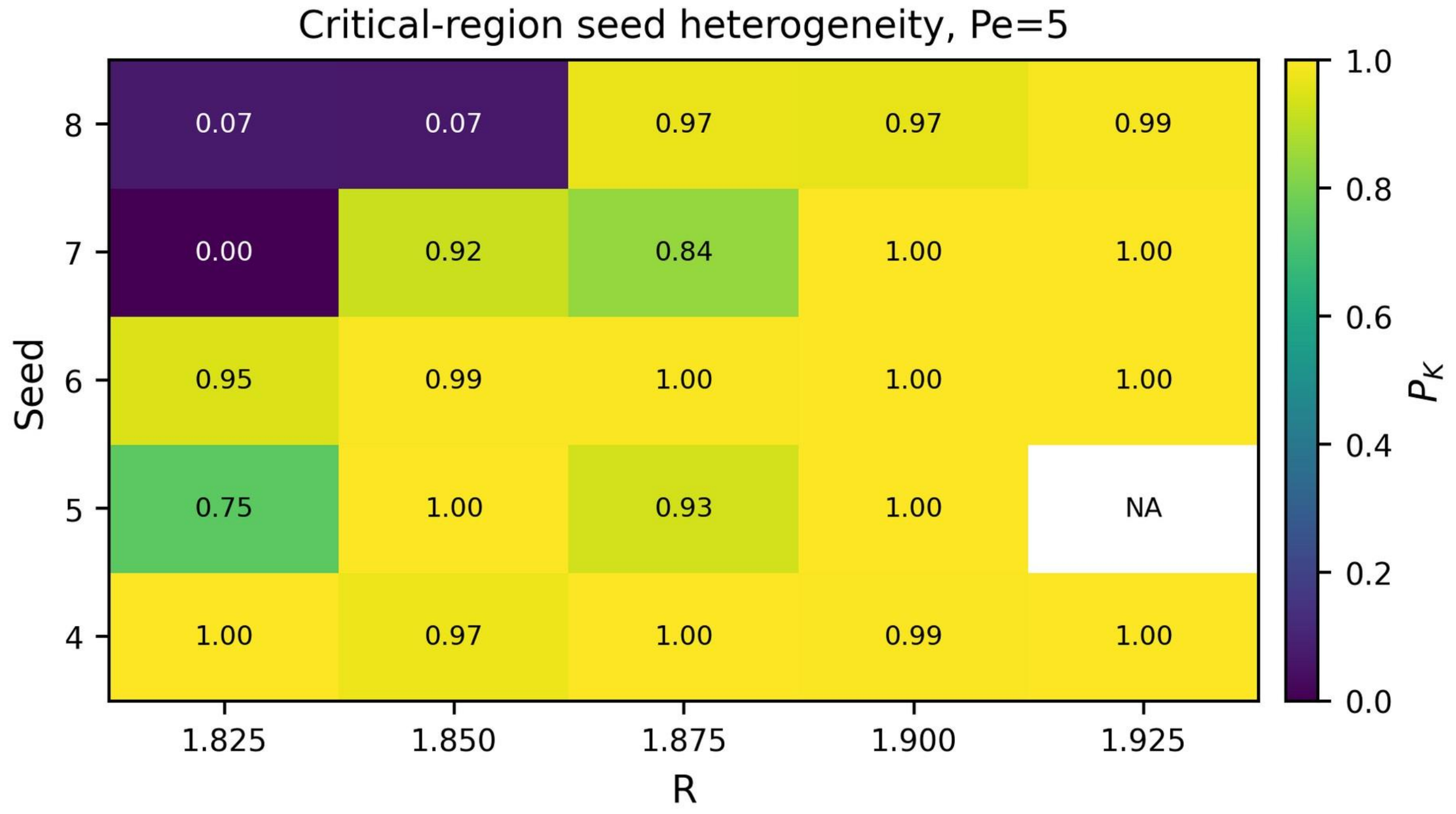

FIG. S21. Seed-by-radius heat map of $P_K$ for the Pe $= 5$ crossover-region long-trajectory data set. The unavailable Pe $= 5$, $R = 1.925$ seed5 entry is shown as missing.

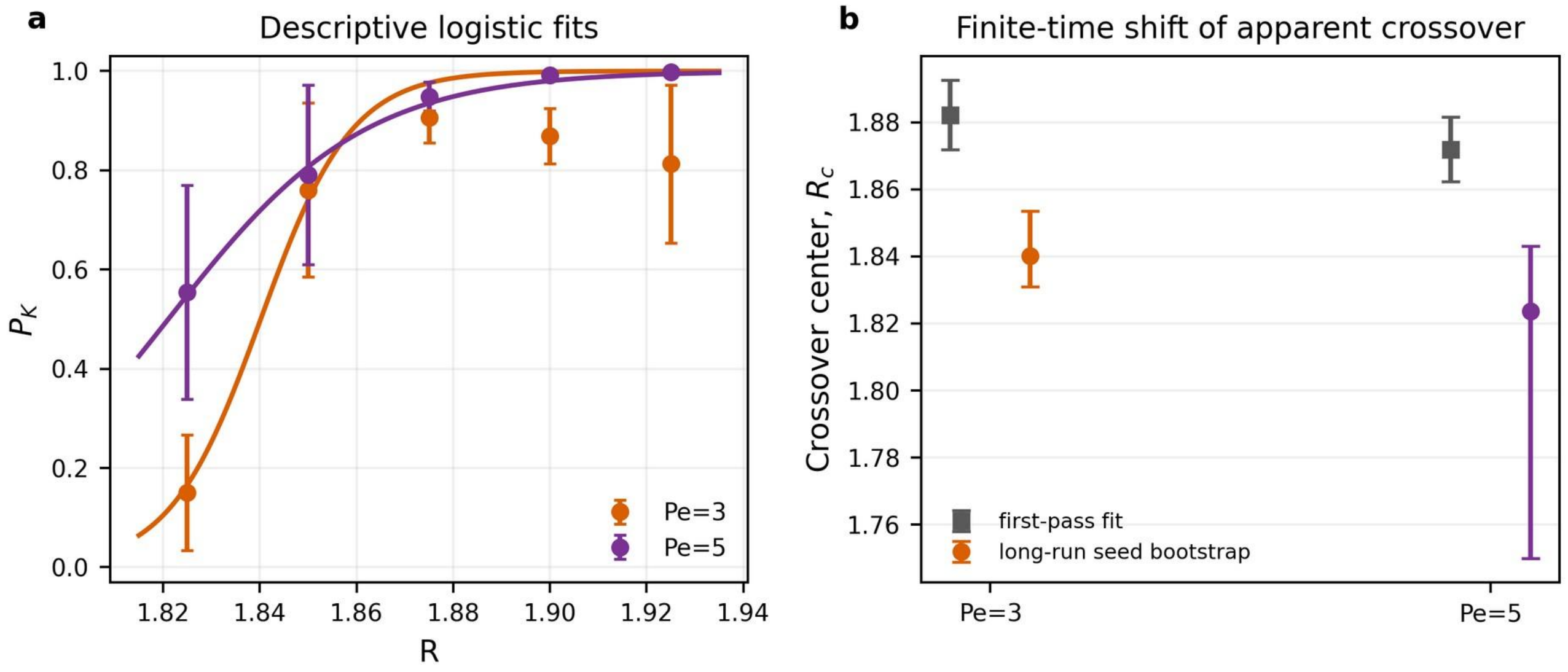

FIG. S22. Descriptive logistic crossover analysis for the $N = 300$ long data set. Fits summarize finite-time occupancy curves and are not interpreted as thermodynamic critical points. Panel (b) contains only Pe=3 and Pe=5 because the long-trajectory crossover bootstrap is defined for those two high-activity campaigns.

The replica-resolved crossover plots and descriptive logistic fits should therefore be read as context for state selection, not as a search for a unique critical radius. At Pe=3 and 5 the sharp mean crossover is assembled from trajectories that can occupy very different finite-time topological states at nearby radii. A smooth fit to the mean can summarize where the crossover lies while hiding precisely the heterogeneity that matters for prediction. The prospective radii were chosen within this state-rich region so that the branching test asks whether pre-existing state information matters under comparable external conditions, rather than comparing two qualitatively disconnected parameter regimes.

Topology-resolved structural observables provide complementary checks on the reduced state description. Localized knot length, the transverse size of knotted conformations, the K–U transverse-size contrast, and axial/transverse anisotropy all vary systematically across the crossover region (Figs. S23–S26). They establish that knotting is accompanied by state-dependent reorganization, but none supplies a geometry-independent scalar ordering that can replace the matched-control mechanical-state or prospective branching tests. Correlation-limited effective sampling is quantified separately in Fig. S27.

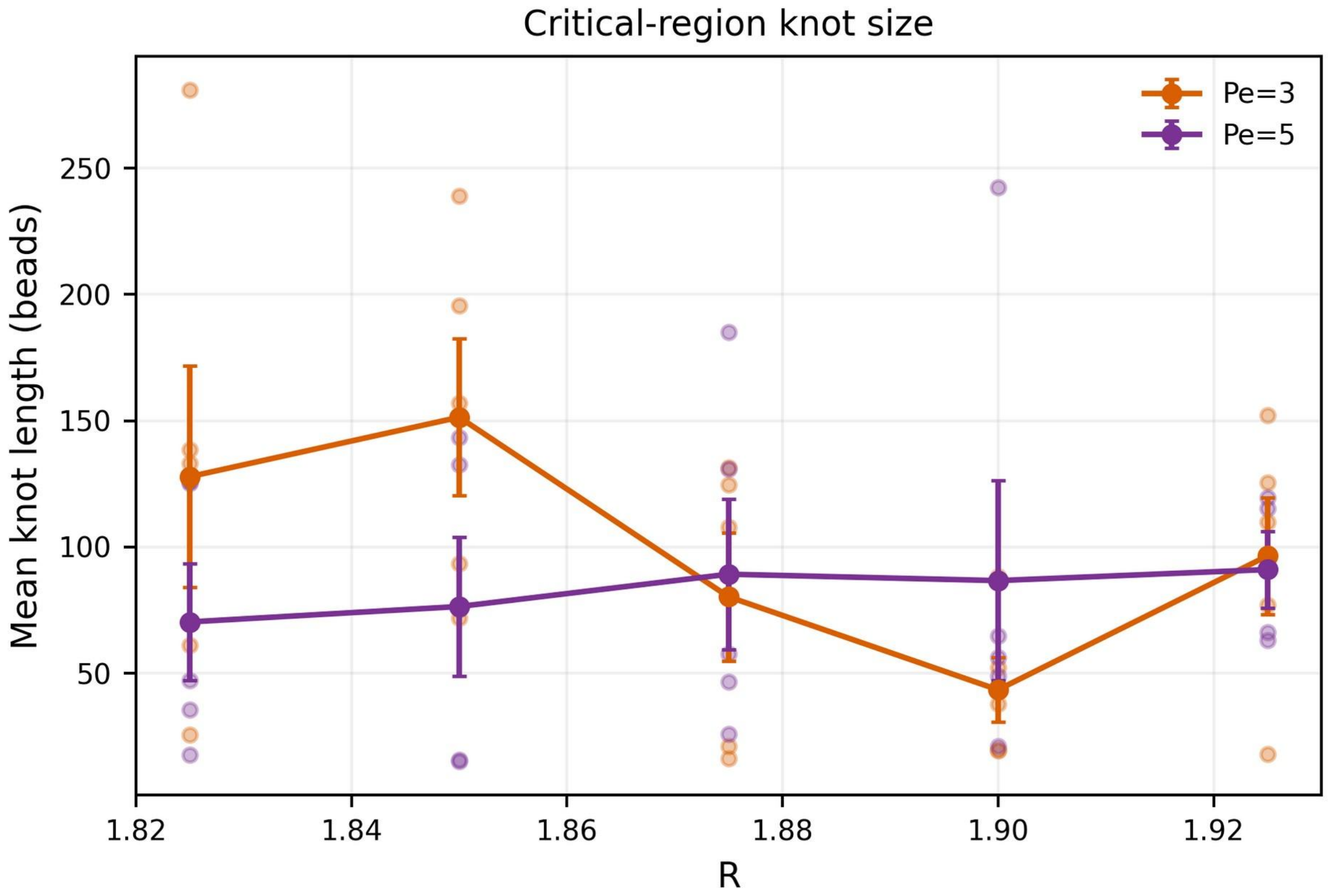

FIG. S23. Mean localized knot length in the $N = 300$ crossover-region long-trajectory campaign. Transparent points are independent trajectories; connected points summarize seed means with SEM. All trajectory-level points are retained; the broad scatter is genuine replica-to-replica heterogeneity and is not removed by outlier filtering.

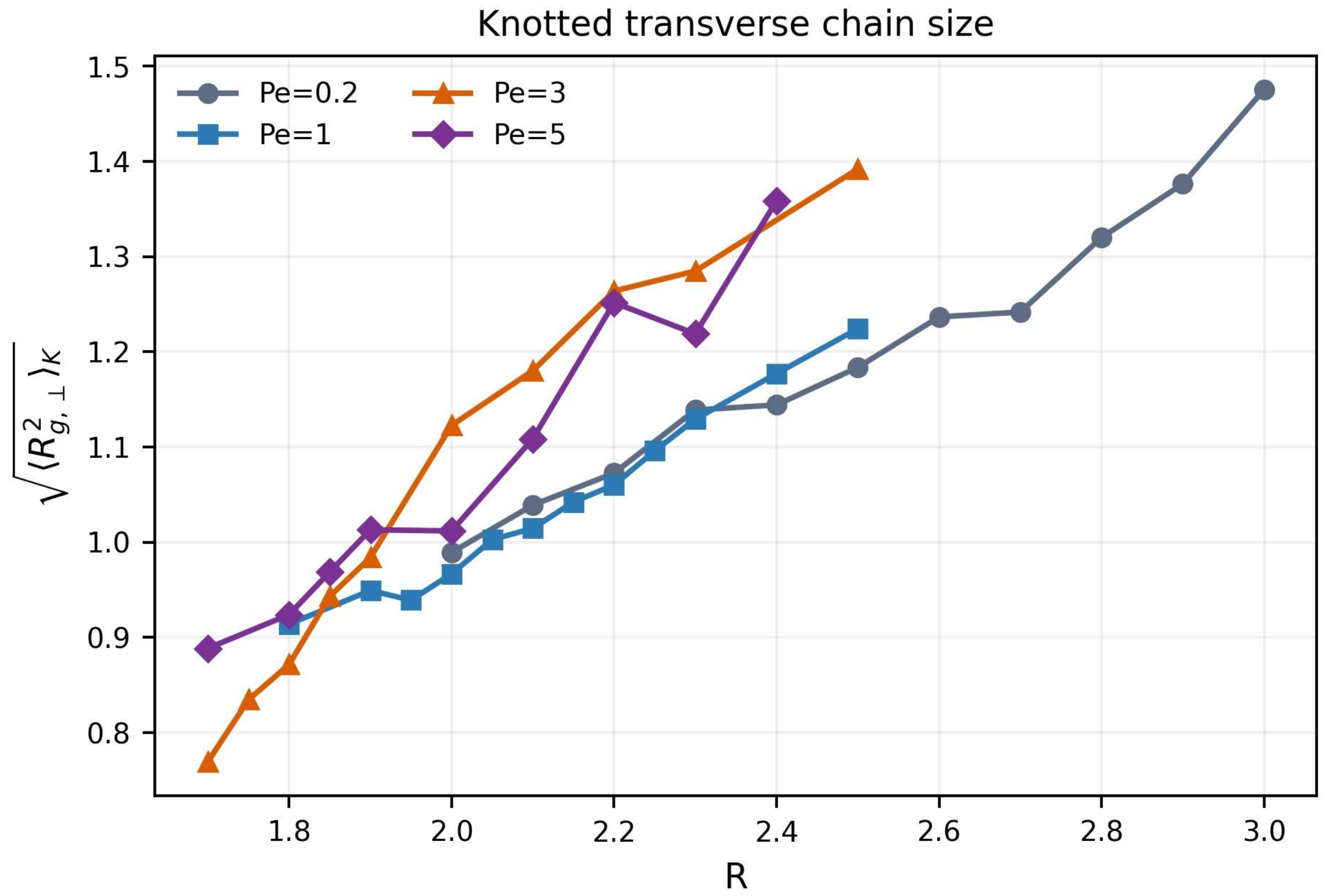

FIG. S24. Knotted transverse RMS gyration size across the broad $N = 300$ scan from the independently validated mass-weighted gyration workflow.

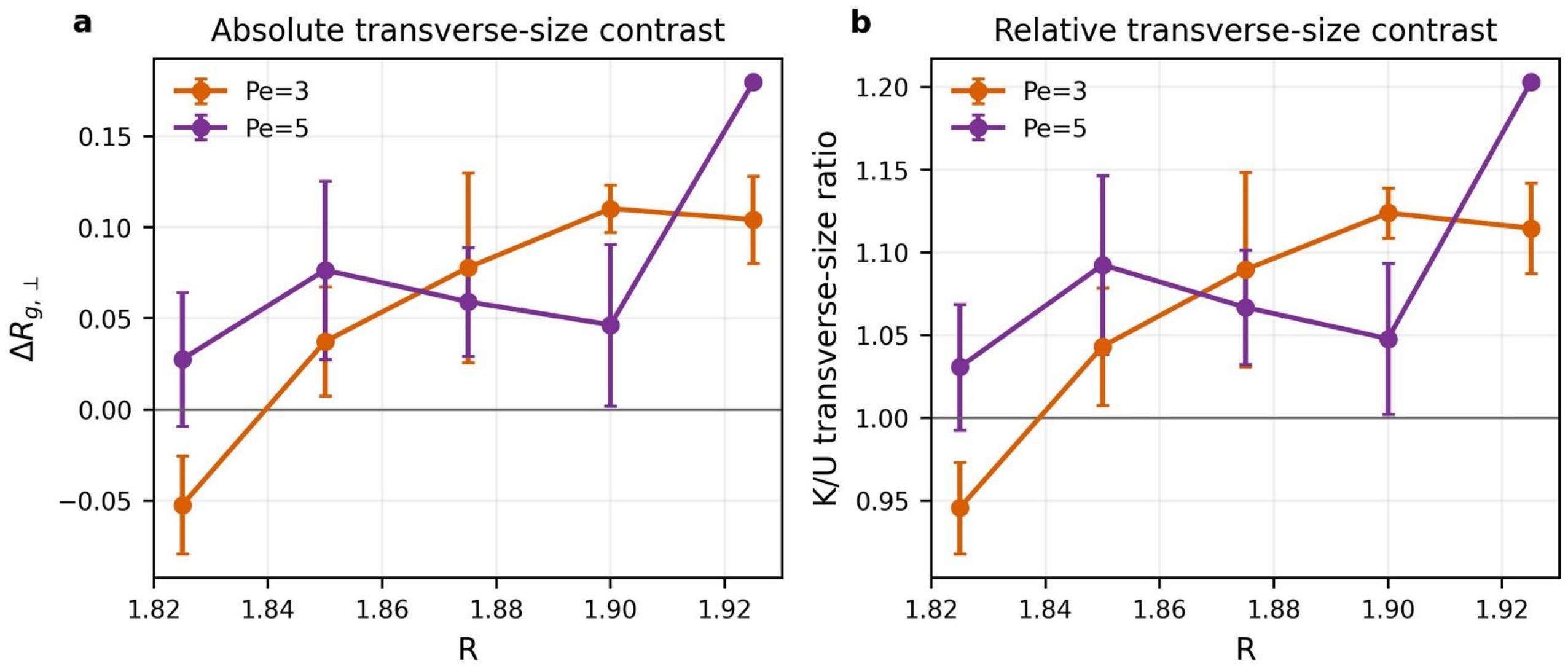

FIG. S25. Topology-resolved transverse-size contrast in the $N = 300$ crossover-region long trajectories: absolute K–U difference and ratio of knotted to unknotted transverse RMS gyration size.

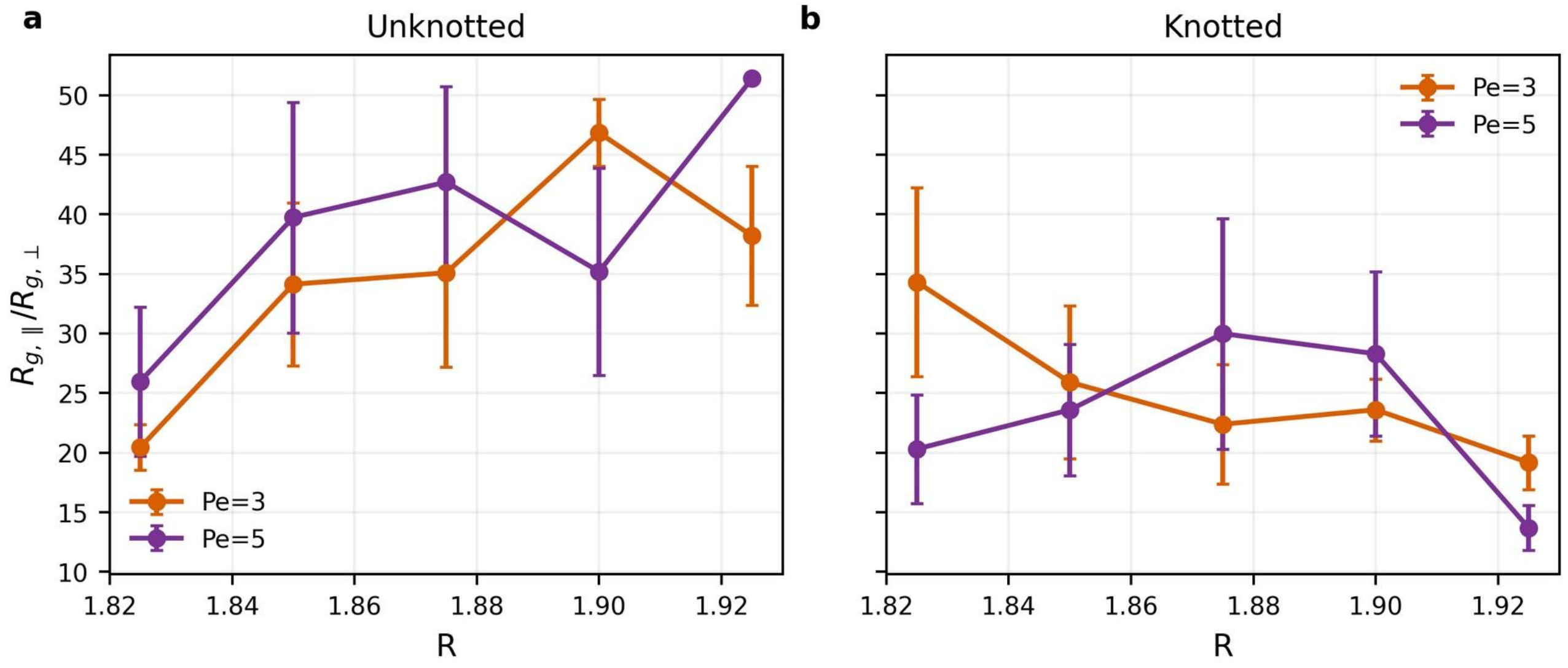

FIG. S26. Axial/transverse gyration anisotropy in the $N = 300$ crossover-region long trajectories, conditioned on unknotted and knotted frames. These observables document state-dependent shape changes but are not used as a universal scalar reaction coordinate.

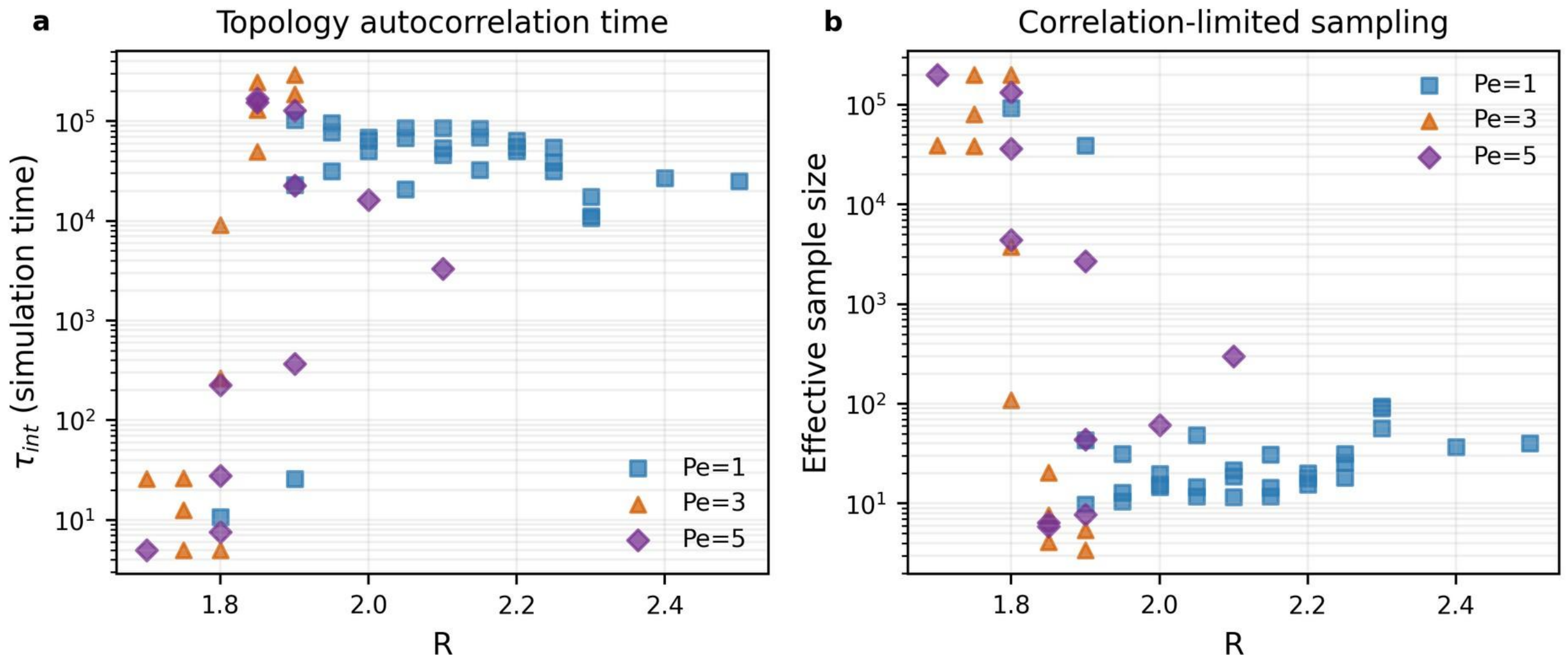

FIG. S27. Topology autocorrelation and correlation-limited effective sample size for broad-scan cases in the transition-relevant range. The long correlation times motivate replica-level rather than frame-level statistical inference.

The autocorrelation analysis explains why the unit of inference is the independent trajectory or source trajectory rather than the saved frame. Near the crossover, topology can remain correlated over long intervals, so millions of stored configurations do not imply millions of independent samples. Treating frames as independent would sharply overstate precision and would also obscure the physical point: the long correlation is itself evidence that replicas can remain trapped in distinct dynamical organizations for substantial times. The same logic motivates source-trajectory clustering in the prospective experiment, where several eligible parents can originate from one parent-generation trajectory. Formation-location and wide-pore context are summarized in Fig. S28; radius-resolved chain-length structural responses are shown in Figs. S29-S31; and within-condition formation-hazard residuals and fixed-fraction replica splitting are shown in Figs. S32 and S33.

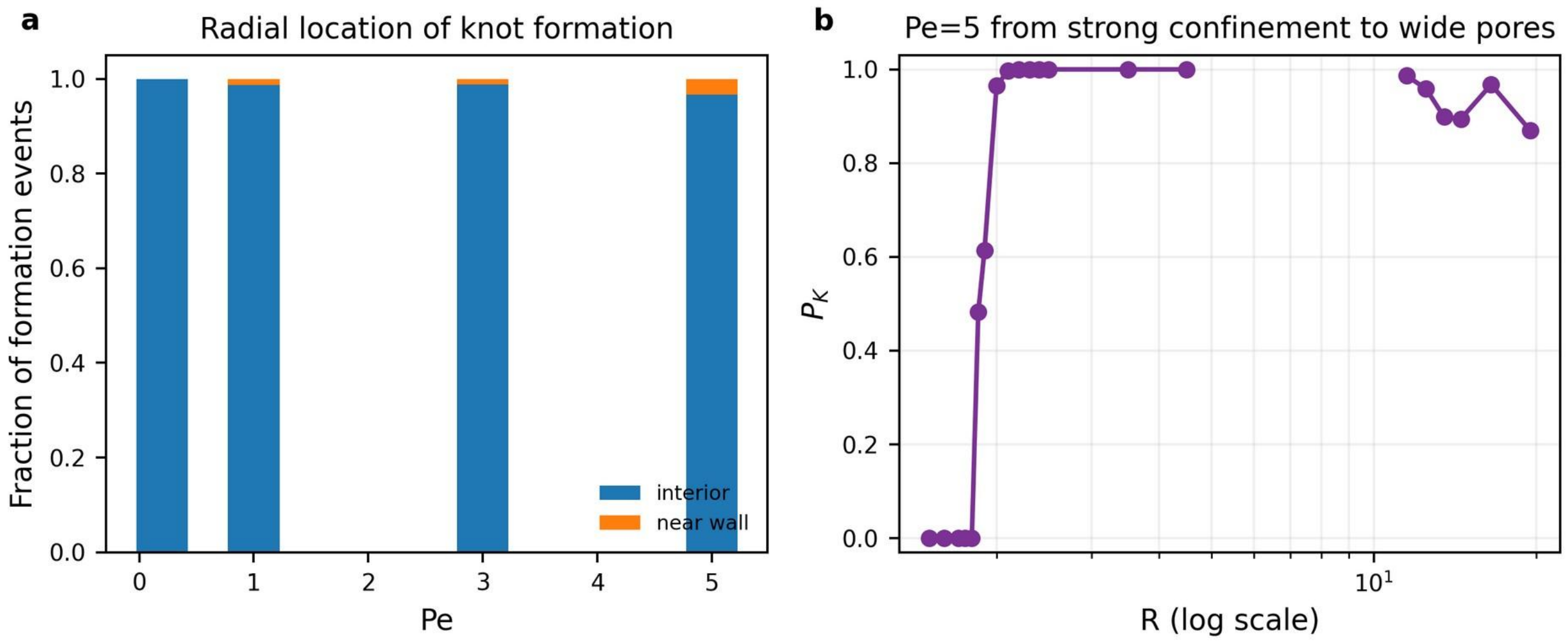

FIG. S28. Formation-location and wide-pore context. Formation is overwhelmingly interior rather than near-wall, while the wide-R Pe = *5* scan shows how the strong-confinement crossover connects to more weakly confined behavior.

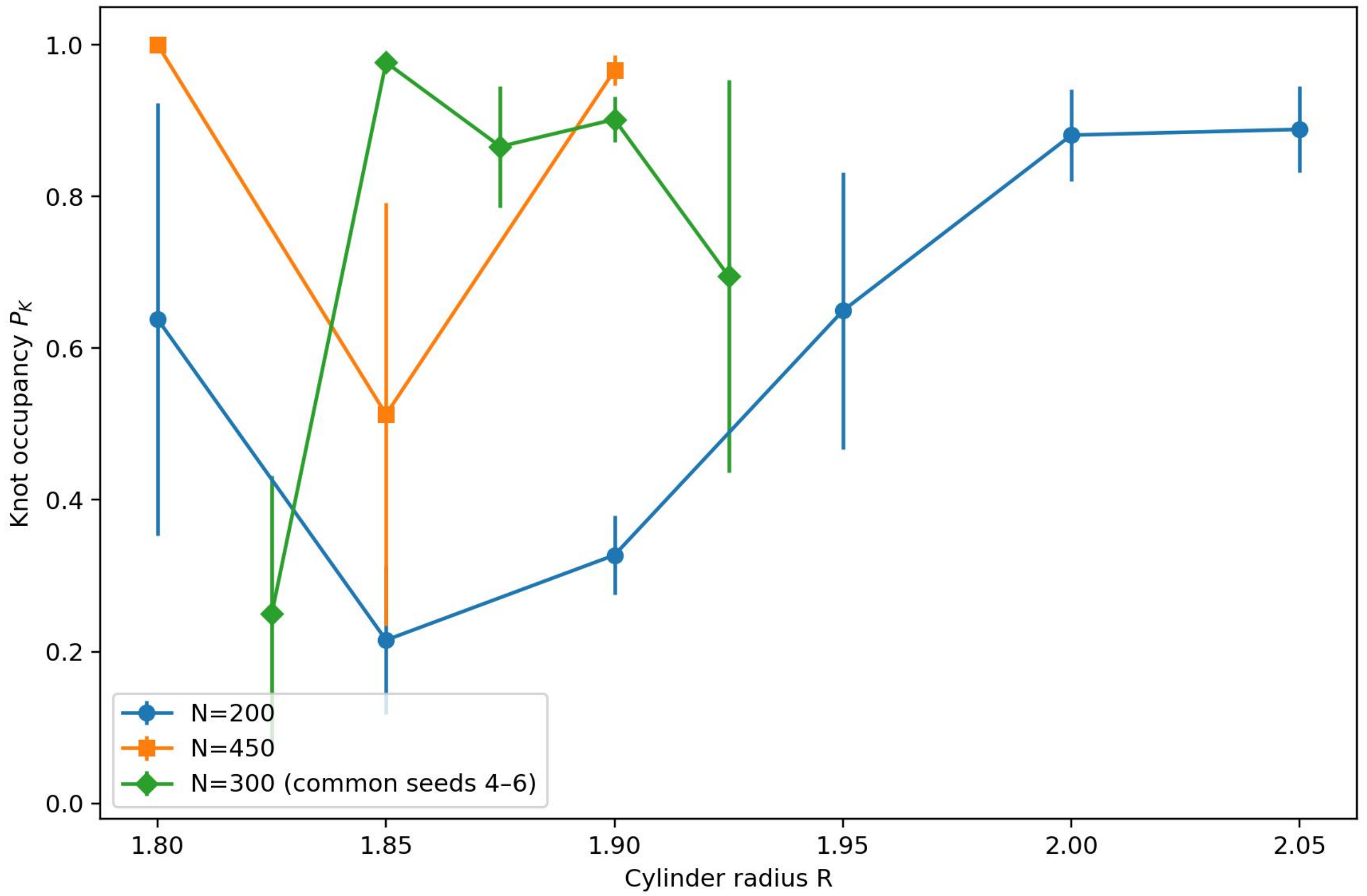

FIG. S29. Radius-resolved chain-length response at fixed $p_{active} = 0.4$. The $N = 300$ curve uses the matched seed subset 4–6 to align the seed ensemble with $N = 200$ and 450; coverage in R differs by N, so connected lines are descriptive rather than a global scaling fit.

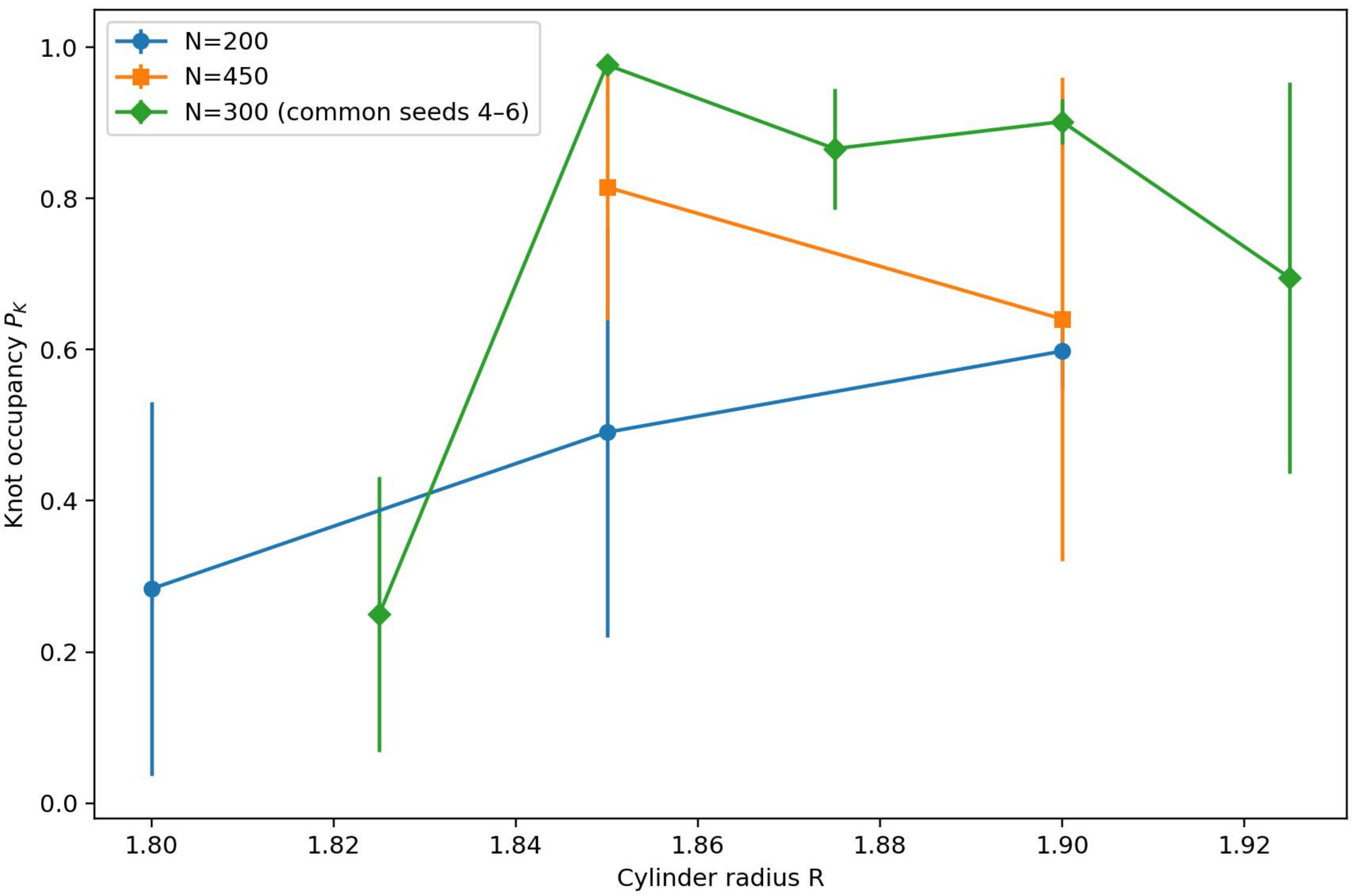

FIG. S30. Radius-resolved chain-length response at fixed $N_{\text{active}} = 120$, using the matched $N = 300$ seed subset 4–6. The response is nonmonotonic and strongly protocol dependent, especially at $N = 450$.

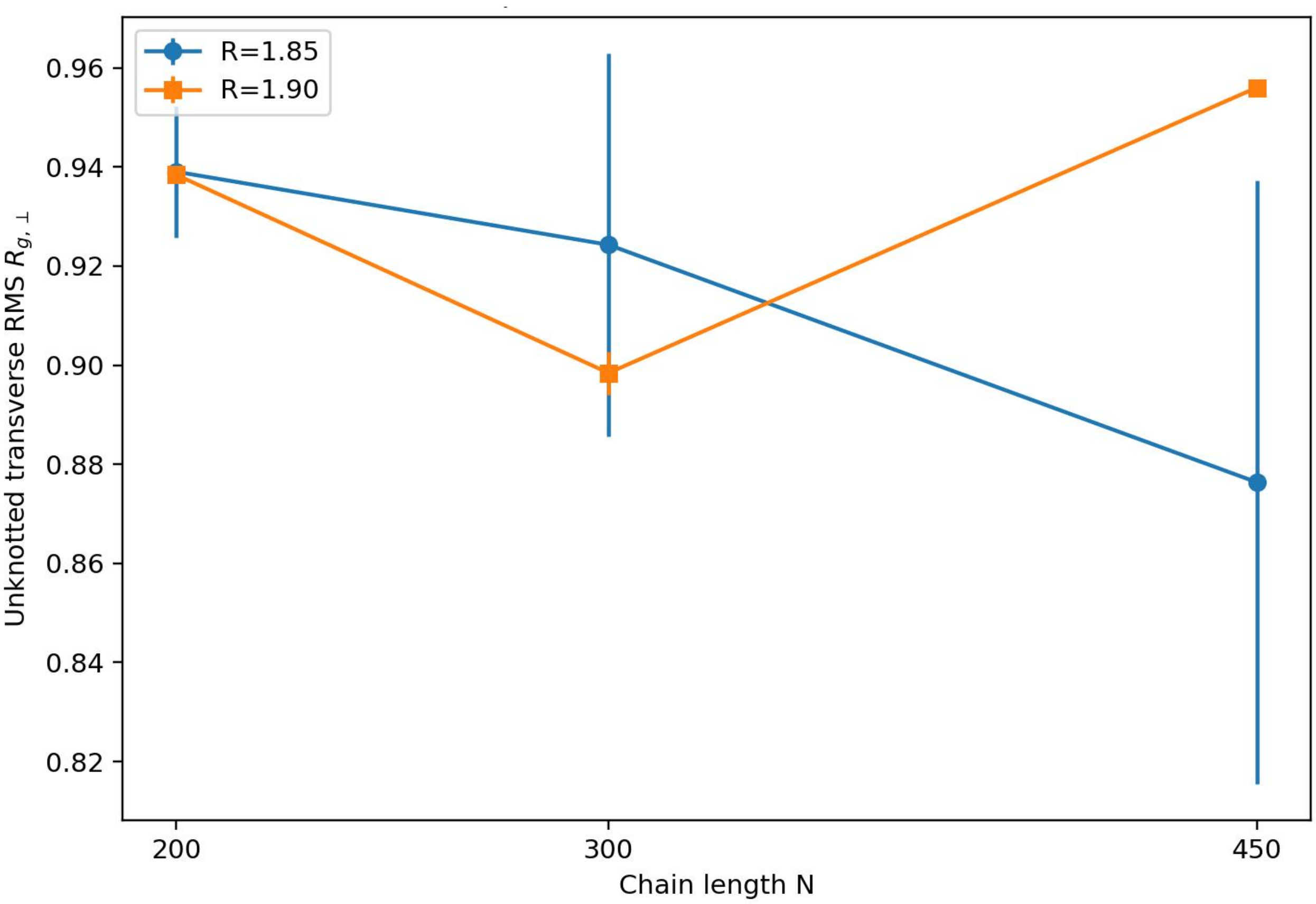

FIG. S31. Unknotted transverse RMS size under the fixed-$N_{\text{active}} = 120$ control using the common seed ensemble. The $R = 1.85$ and $R = 1.90$ trends differ, reinforcing that absolute transverse size does not provide a universal chain-length collapse.

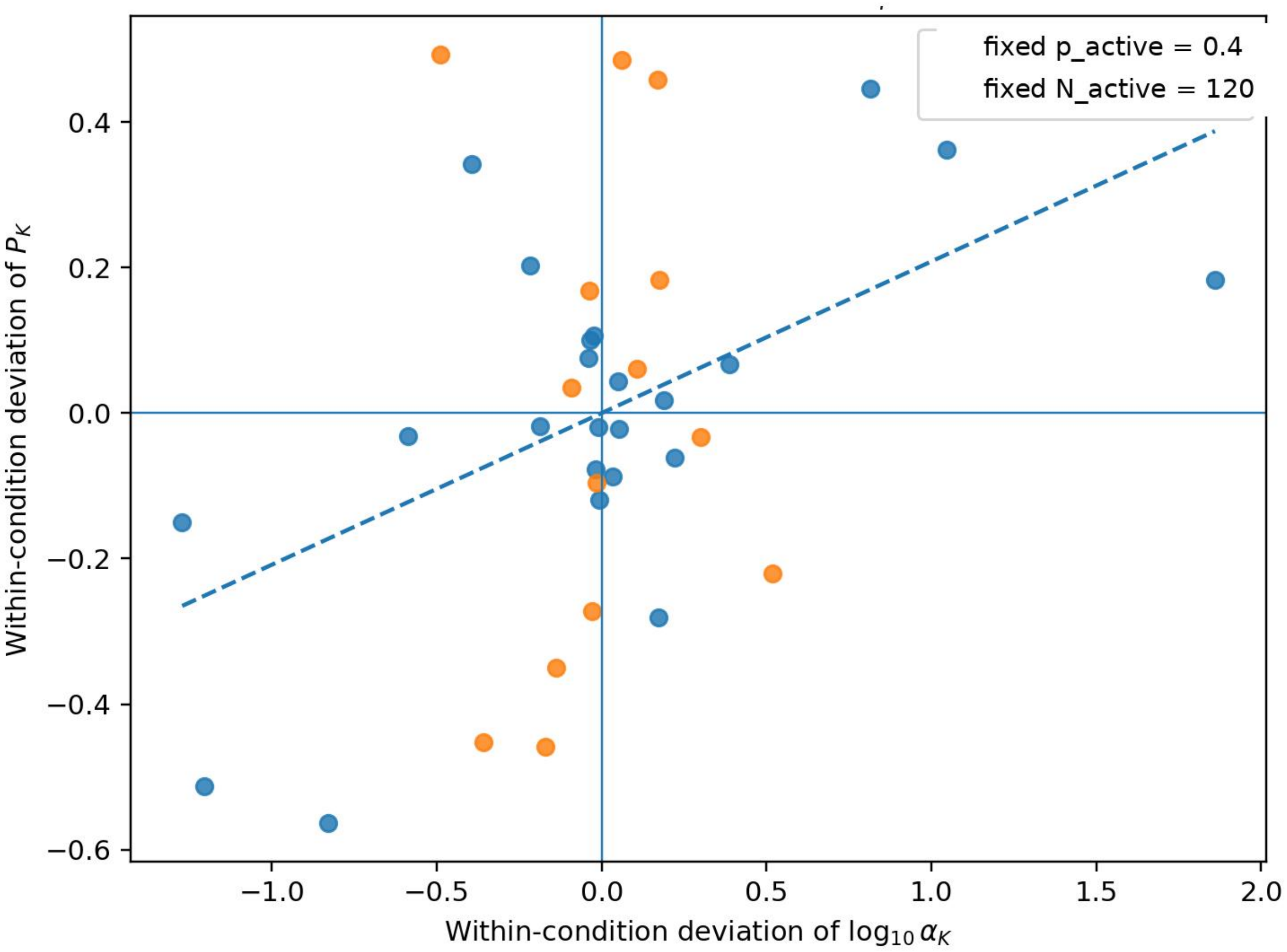

FIG. S32. Within-condition relation between knot occupancy and conditional formation hazard in the 42 independently restarted trajectories. The combined residual association is modest ($r \approx 0.415$, stratified $p_{\mathrm{perm}} \approx 0.047$); restricting to complete three-replica conditions weakens the two-sided permutation significance to about 0.069, so this is treated as supportive rather than decisive evidence.

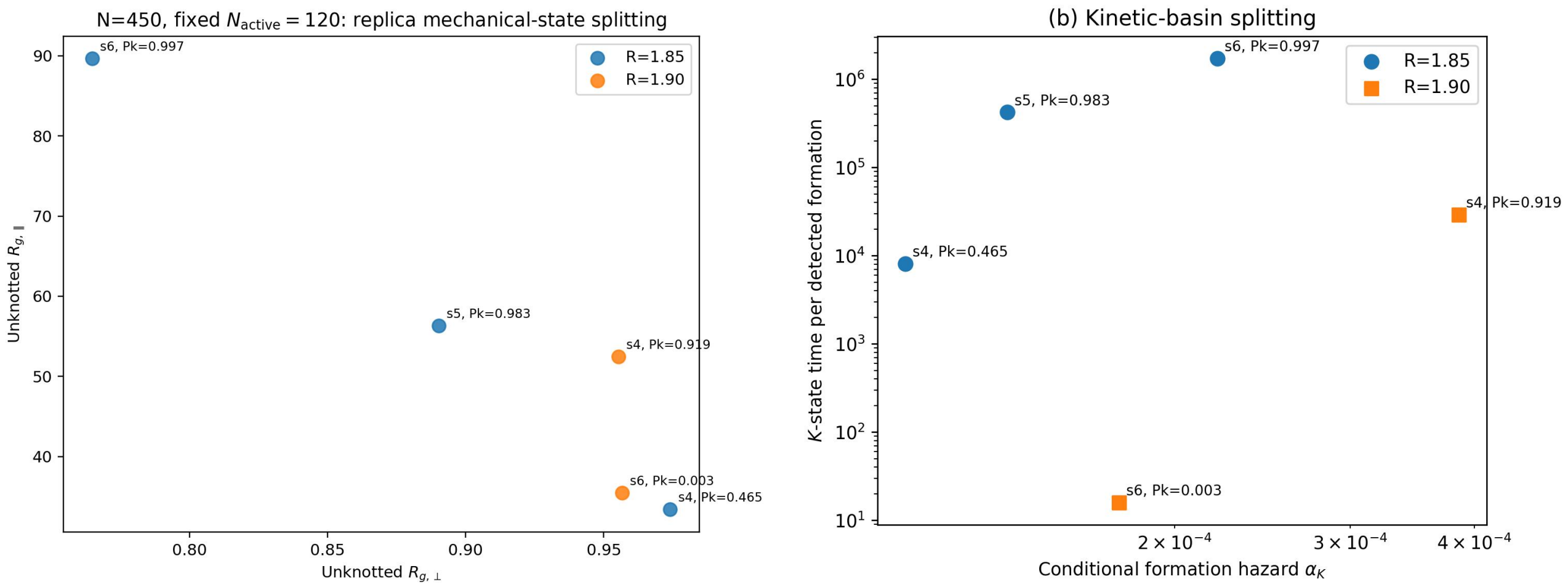

FIG. S33. $N = 450$ fixed-$N_{\mathrm{active}} = 120$ replica splitting. Mechanical organization and kinetic residence separate trajectories with nearly identical external controls. At $R = 1.90$, one trajectory is nearly always unknotted while another is nearly always knotted, demonstrating a slow-state distinction beyond transverse size alone.

The nearly binary N=450 replica splitting is particularly informative because it exposes what a mean structural measure can miss. At the same external controls, trajectories with comparable transverse size can spend almost all of their time in opposite topological sectors, while axial organization and residence differ strongly. This is an extreme version of the state-dependence problem seen in the N=300 crossover. It motivates searching for slow organization and lability variables—such as backfold accessibility and axial hairpin dynamics—rather than refining a universal size law. The subsequent branching experiment is needed because these retrospective examples alone cannot establish prospective commitment.

The replica splitting at $N = 450$ also motivates a direct test of anisotropy as a reduced coordinate. After removing within-condition means, the anisotropy residual is positively correlated with occupancy but is not statistically robust ($r \approx 0.309$, $p_{\mathrm{perm}} \approx 0.119$; Fig. S34). We therefore retain anisotropy as a useful structural diagnostic and negative-control candidate rather than promote it to a universal susceptibility coordinate. Detailed protocol-resolved access/retention accounting is shown in Fig. S35.

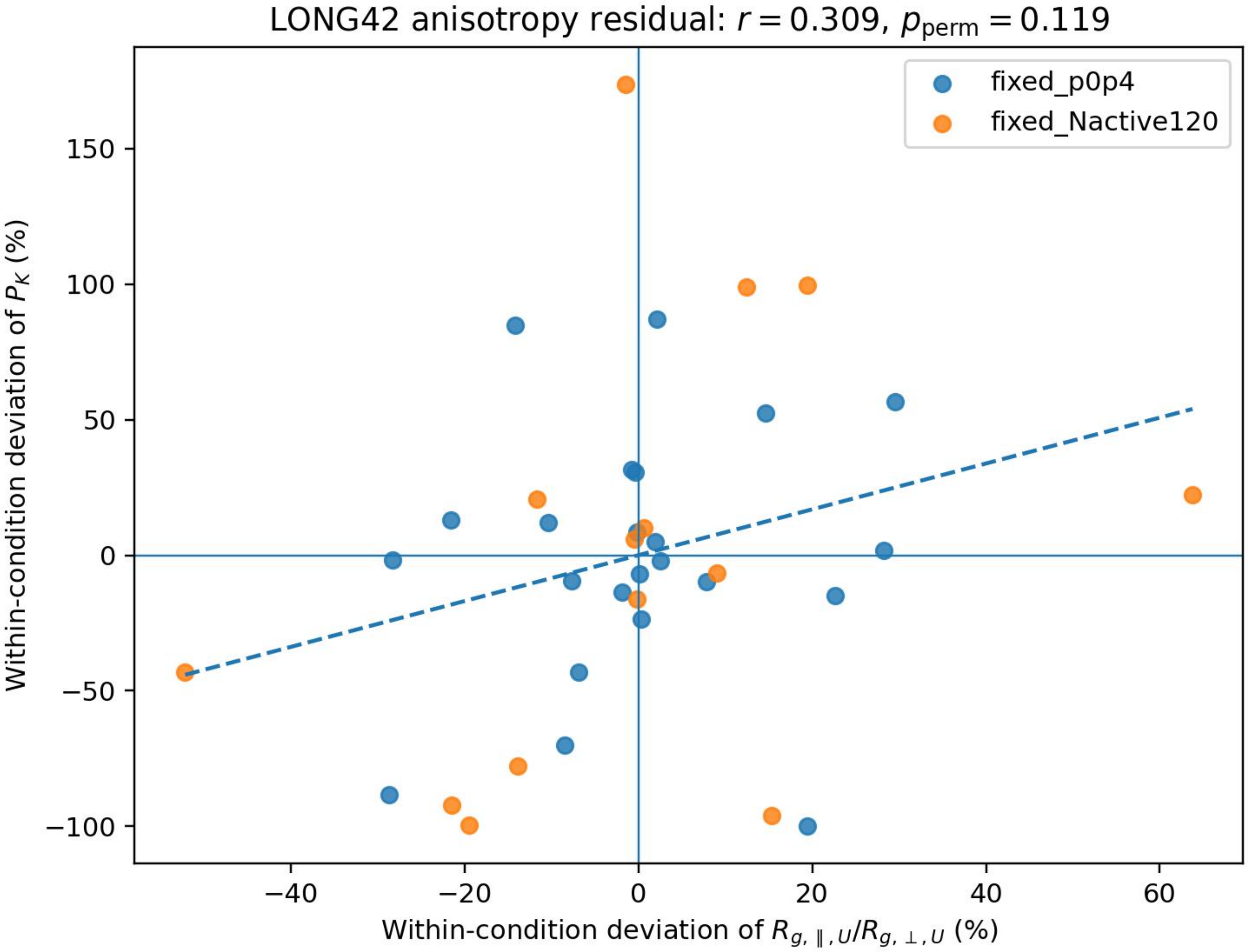

FIG. S34. Exploratory within-condition anisotropy residual for the 42 independently restarted long trajectories. The correlation is positive but not statistically robust ($r \approx 0.309$, $p_{\mathrm{perm}} \approx 0.119$), so anisotropy is not promoted to a universal susceptibility coordinate.

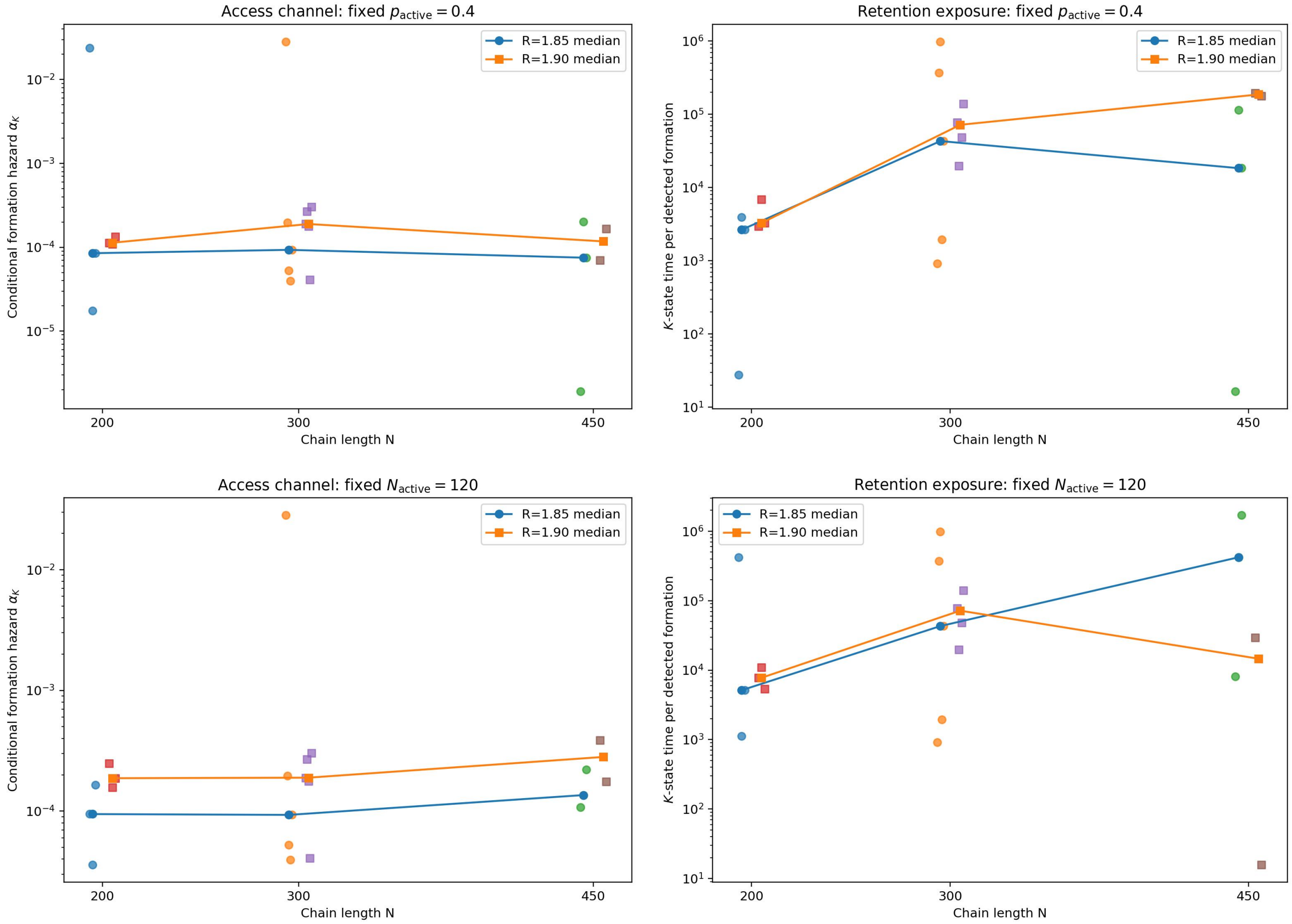

FIG. S35. Detailed chain-length access/retention accounting for both protocols at $R = 1.85$ and 1.90. Conditional formation hazard measures access, whereas K-state time per detected formation is an occupancy-budget accounting quantity and is not treated as an independent causal lifetime measurement. This is an expanded accounting view: the fixed-active-fraction access/retention panels replot the corresponding components introduced in Fig. S16, while the fixed-active-block panels add the complementary protocol.

Across the two chain-length protocols, the combined message is therefore qualitative but restrictive: neither total contour length, active contour length, transverse size, nor a single access/escape rate is sufficient by itself to organize all observed occupancies. The controls do not identify a universal chain-length scaling function, and we do not claim one. Instead they show that geometry and activity architecture reshape the set of long-lived mechanical organizations available to the chain. This provides an independent motivation for the state-resolved formulation tested prospectively in S13.

## S12. Protocol-resolved chain-length mechanics

Tables S4–S9 summarize the protocol-resolved chain-length design, common-time sensitivity, replica splitting, and mechanical/kinetic residual tests. The corresponding graphical diagnostics are distributed across Secs. S6–S11, including the replica-resolved and topology-conditioned structural controls; redundant archive-only renderings remain omitted.

Table S4. Radius-integral fit summary for the $N = 300$ state-selection analysis.

| Pe | $R_0$ | $\chi$ | RMSE | n points |
|---|---|---|---|---|
| 1 | 2.1052 | 7.35 | 0.050 | 11 |
| 3 | 1.8438 | 50.41 | 0.127 | 5 |
| 5 | 1.8214 | 38.61 | 0.009 | 5 |

Table S5. Extended-duration chain-length design. The $N = 300$ reference belongs to both protocol comparisons because $p_{active} = 0.4$ implies $N_{active} = 120$ at $N = 300$.

| Protocol | N | R values in long set | Seeds / condition |
|---|---|---|---|
| fixed $p_{active} = 0.4$ | 200 | 1.80,1.85,1.90,1.95,2.00,2.05 | 3 |
| fixed $p_{active} = 0.4$ | 450 | 1.80,1.85,1.90 | 3 |
| fixed $N_{active} = 120$ | 200 | 1.80,1.85,1.90 | 3 |
| fixed $N_{active} = 120$ | 450 | 1.85,1.90 | 3 |
| both protocols (long reference) | 300 | 1.825,1.850,1.875,1.900,1.925 | 5 |

**Table S6.** $R = 1.90$ **protocol contrast in** $N = 450$ **minus** $N = 200$ **knot occupancy for full trajectories and the common** $1.03413 \times 10^{10}$**-step horizon.**

| Protocol | Full $\Delta P_K$ | Common-horizon $\Delta P_K$ |
|---|---|---|
| fixed $p_{\text{active}} = 0.4$ | 0.63881 | 0.64770 |
| fixed $N_{\text{active}} = 120$ | 0.04243 | 0.10280 |

Table S7. Within-condition mechanical-state residual tests for the 42 independently restarted trajectories. The transverse-size association is sampling-sensitive and is treated as exploratory.

| Subset | Observable | r | $p_{\text{perm}}$ | n | conditions |
|---|---|---|---|---|---|
| all | U transverse RMS | -0.396 | 0.0364 | 37 | 13 |
| all | U axial RMS | 0.318 | 0.1296 | 37 | 13 |
| all | U anisotropy | 0.309 | 0.1194 | 37 | 13 |
| fixed p | U transverse RMS | -0.371 | 0.2408 | 23 | 8 |
| fixed $N_{\text{active}}$ | U transverse RMS | -0.417 | 0.0624 | 14 | 5 |

**Table S8. Within-condition kinetic residual tests. Formation hazard is supportive in the combined set and strongest in the fixed-p subset; complete-episode lifetime is not a universal predictor.**

| Subset | Observable | r | $p_{\text{perm}}$ | n | conditions |
|---|---|---|---|---|---|
| all | log10($\alpha_K$) | 0.415 | 0.0471 | 37 | 13 |
| all | log10(mean complete K lifetime) | -0.125 | 0.5960 | 37 | 13 |
| fixed p | log10($\alpha_K$) | 0.611 | 0.0376 | 23 | 8 |
| fixed $N_{\text{active}}$ | log10($\alpha_K$) | 0.062 | 0.8459 | 14 | 5 |

**Table S9.** $N = 450$**, fixed** $N_{\text{active}} = 120$ **replica splitting at** $R = 1.90$**. Seed 5 is fully knotted and therefore has no defined unknotted geometry or conditional formation hazard.**

| Seed | $P_K$ | U $R_{g,\perp}$ | U $R_{g,\parallel}$ | $\alpha_K$ | K-time/formation |
|---|---|---|---|---|---|
| 4 | 0.9186 | 0.9554 | 52.45 | 3.86e-4 | 2.92e4 |
| 5 | 1.000 | undefined | undefined | undefined | undefined |
| 6 | 0.00278 | 0.9567 | 35.44 | 1.76e-4 | 15.87 |

## S13. Prospective phase-space branching experiment: primary state-dependence test

### S13.1. Prespecified design and separation from retrospective analysis

The prospective experiment asks a single primary question: at identical external controls, do mechanically distinct pre-existing unknotted phase-space states carry different probabilities of future topological switching? Independent calibration trajectories specify the classifier, normalization, selection thresholds, eligibility rules, and branch horizon before any branch outcome is available. The horizon $H$ is the elapsed simulation time measured from the parent frame to a fixed prospective endpoint; it is not inferred from the observed switching times. Sixteen new $Pe = 3$ parent-generation trajectories—eight at $R = 1.85$ and eight at $R = 1.90$—then supply unknotted parent states. The classifier $s_{\text{conf}}(t) = \left[-z\left(\overline{W(t)}\right) - z\left(\overline{R^2_{g,\perp}(t)}\right)\right]/2$ uses 1000-frame trailing averages ending at the parent frame and is intentionally distinct from the retrospective whole-replica score $m^*$.

**Operational definitions used in the frozen prospective design.**

The frozen custom post-processing analyzer (version 2.3) defines $R^2_{g,\perp} = N^{-1}\sum_i \{|\mathbf{r}_i - \mathbf{r}_{\text{cm}}|^2 - [(\mathbf{r}_i - \mathbf{r}_{\text{cm}}) \cdot \mathbf{e}_a]^2\}$, where $\mathbf{e}_a$ is the cylinder-axis unit vector. All beads have unit mass, so this arithmetic-center expression is identical to the mass-weighted gyration definition used in the independent gyration validation. For radial gap $g = R - \rho$ inside the wall cutoff, the analyzer uses $f_\rho(g) = \epsilon\left[-\frac{6}{5}\frac{\sigma^9}{g^{10}} + 3\frac{\sigma^3}{g^4}\right]$, with the positive radial force direction defined away from the cylinder axis (increasing $\rho_i$); $f_\rho = 0$ beyond the cutoff. The wall-load proxy is $W = -\sum_i f_{\rho,i}$. Thus inward repulsion contributes positively to W.

For parent frame j, $\overline{W}_j$ and $\overline{R^2_{g,\perp}}_j$ are means over frames j−999,...,j (1000 saved frames, inclusive of the parent frame). The parent-selection reconstruction reproduced all 71 stored trailing means with a single inclusive-current-frame convention. For each radius separately, the discovery reducer first drew exactly equal numbers of eligible landmarks from each of five seeds by deterministic midpoint subsampling, pooled those landmarks, and computed $\mu_R$ and the sample standard deviation $s_R$ using denominator $n_R - 1$ (equivalent to numpy.std(..., ddof=1)), where $n_R$ is the pooled calibration count at that radius. The frozen score is $s_{\text{conf}} = \frac{-(\overline{W}-\mu_{W,R})/s_{W,R} - \left(\overline{R^2_{g,\perp}}-\mu_{g,R}\right)/s_{g,R}}{2}$.

For the 1000-frame calibration, $R = 1.85$ used 7801 landmarks per seed (39005 pooled): $\mu_g$ =0.8448825166, $s_g$ =0.1160063901, $\mu_W$ =2704.5717346, $s_W$ =1454.3940599, Q20 $= -0.7155897300$, Q80 $= 1.2209146978$. $R = 1.90$ used 4648 landmarks per seed (23240 pooled): $\mu_g$ =0.7907017011, $s_g$ =0.0995795159, $\mu_W$ =1227.6044736, $s_W$ =803.6386157, Q20 $= -0.5703719587$, Q80 $= 0.7788622118$. Quantiles were evaluated by numpy.quantile in the frozen reducer. LOW requires $s_{\text{conf}} \leq Q20$ and HIGH requires $s_{\text{conf}} \geq Q80$; the middle 60% is excluded. The thresholds were frozen before any prospective outcomes.

For a preserved parent state $X_i$, $q_H(X_i) = P(T_{\text{knot}} < H | X_i)$. Throughout the prospective figures and tables, $q_H$ is used as shorthand for the empirical 32-branch estimator $k_i/32$ unless probability notation is explicitly required. Unless otherwise stated, r denotes a Pearson product-moment correlation coefficient; exploratory Spearman rank correlations are denoted $\rho_S$.

### S13.2. Parent selection and branching protocol

The prespecified protocol uses a 1000-frame trailing window, requires a continuously unknotted eligibility history of at least 2000 saved frames ending at the current frame, defines LOW and HIGH with $Q20/Q80$ thresholds fixed from independent calibration trajectories; LOW denotes the lowest 20% and HIGH the highest 20% of the prespecified score distribution, and these labels do not use later knotting outcomes. The protocol imposes a 2000-frame (20,000,000-step) minimum separation between parents from the same source trajectory, caps each radius/class at 20 parents, launches 32 independent Langevin branches per parent, and uses $H = 20{,}000{,}000$ steps as the primary observation horizon. With dumps every 10,000 steps, this is 2000 saved intervals (2001 saved branch frames including the parent); with $\Delta t = 0.001$, it is $2 \times 10^4$ reduced time units. Class sizes were not adjusted after outcomes became available; consequently $R = 1.90$ contains 11 LOW and 20 HIGH parents.

Table S10. Prespecified prospective-branching protocol and inferential role.

| Design element | Prespecified value | Role |
|---|---|---|
| Conditions | Pe = $3$; $R = 1.85$ and 1.90 | Fixed external controls |
| Primary score | $s_{\text{conf}} = \frac{-z(\overline{W}) - z\left(\overline{R^2_{g,\perp}}\right)}{2}$ | Pre-branch trailing-history classifier |
| Trailing window | 1000 saved frames | Mechanical reservoir history |
| U-state eligibility | Current U; continuously U ≥2000 frames | Exclude immediate post-knot relaxation |
| LOW/HIGH | Discovery Q20/Q80 | No prospective-outcome tuning |
| Parent separation | ≥20,000,000 steps within a source trajectory | Reduce near-duplicate parents |
| Maximum parents | 20 per R/class | Predeclared cap |
| Branches per parent | 32 | Estimate $q_H$ for each parent |
| Primary horizon | 20,000,000 steps = 20,000 LJ time units | Frozen finite-time endpoint |
| Dump interval | 10,000 steps; 2001 frames including $t = 0$ | Topology time grid |

### S13.3. Phase-space initialization and numerical consistency

The selection yielded 71 parents: 20 LOW and 20 HIGH at $R = 1.85$ and 11 LOW and 20 HIGH at $R = 1.90$, drawn from 13 of the 16 parent-generation trajectories. The saved parent states retained coordinates, image flags, and velocities, so all descendants of a parent began from the same phase-space state; branches differed only in the Langevin random seed. No velocity recreation, pre-relaxation, or active re-equilibration was applied before branching.

The 1000-frame histories used for parent classification were independently reconstructed from the mechanism time series. All 71,000 history frames were unknotted, and recomputed trailing transverse-size and wall-load means agreed with the selection values to relative errors below $9 \times 10^{-13}$. These checks establish that class labels were determined entirely from the intended pre-branch trailing history ending at the parent frame and that the same archived phase-space states seeded the prospective descendants.

### S13.4. Primary endpoint and statistical inference

For parent i, $q_H(i) = \frac{k_i}{32}$, where $k_i$ is the number of its 32 branches whose first detected knot occurs before H. The parent configuration is therefore the primary statistical unit; the 2272 descendants are not treated as independent mechanical states.

A single parent-generation trajectory can contribute several parents, so those parents share a source history and are not treated as fully independent statistical units. HIGH-versus-LOW comparisons are performed separately at each radius using exact label permutation within source trajectory while preserving the observed class counts. Confidence intervals resample whole source-trajectory clusters. The radius×state interaction is the difference between the two HIGH−LOW effects. Branch-level time-to-first-knot analyses are secondary and use parent-clustered covariance.

**Table S11.** Parent-level finite-horizon outcomes.

| **R** | **Class** | **Parents** | **Events/branches** | **Mean $q_H$** | **95% source-trajectory-cluster bootstrap CI** |
|---|---|---|---|---|---|
| 1.85 | LOW | 20 | 68/640 | 0.10625 | [0.072, 0.144] |
| 1.85 | HIGH | 20 | 308/640 | 0.48125 | [0.350, 0.596] |
| 1.90 | LOW | 11 | 265/352 | 0.75284 | [0.684, 0.817] |
| 1.90 | HIGH | 20 | 411/640 | 0.64219 | [0.580, 0.702] |

Table S12. HIGH−LOW contrasts and the radius×state interaction.

| **Contrast** | **Estimate** | **95% source-trajectory-cluster bootstrap CI** | **Exact conditional permutation p** |
|---|---|---|---|
| $R = 1.85$: HIGH−LOW | +0.37500 | [0.23849, 0.49361] | $2.05 \times 10^{-7}$ |
| $R = 1.90$: HIGH−LOW | -0.11065 | [-0.22514, -0.03125] | 0.01539 |
| ($R = 1.85$ effect) − ($R = 1.90$ effect) | +0.48565 | [0.32601, 0.64884] | $7.88 \times 10^{-9}$ |

### S13.5. Prospective switching at $R$ = 1.85

At $R = 1.85$, HIGH parents produce 308 knotting events in 640 branches, compared with 68 in 640 for LOW parents, giving mean $q_H = 0.48125$ and 0.10625. The parent-level difference is 0.37500, with a source-trajectory-cluster bootstrap interval [0.23849,0.49361] and exact within-source-trajectory permutation $p = 2.05 \times 10^{-7}$. The prespecified mechanical classifier therefore carries substantial predictive information in branch trajectories generated with independent Langevin noise at fixed R and Pe. The parent-level result is shown in Fig. S36.

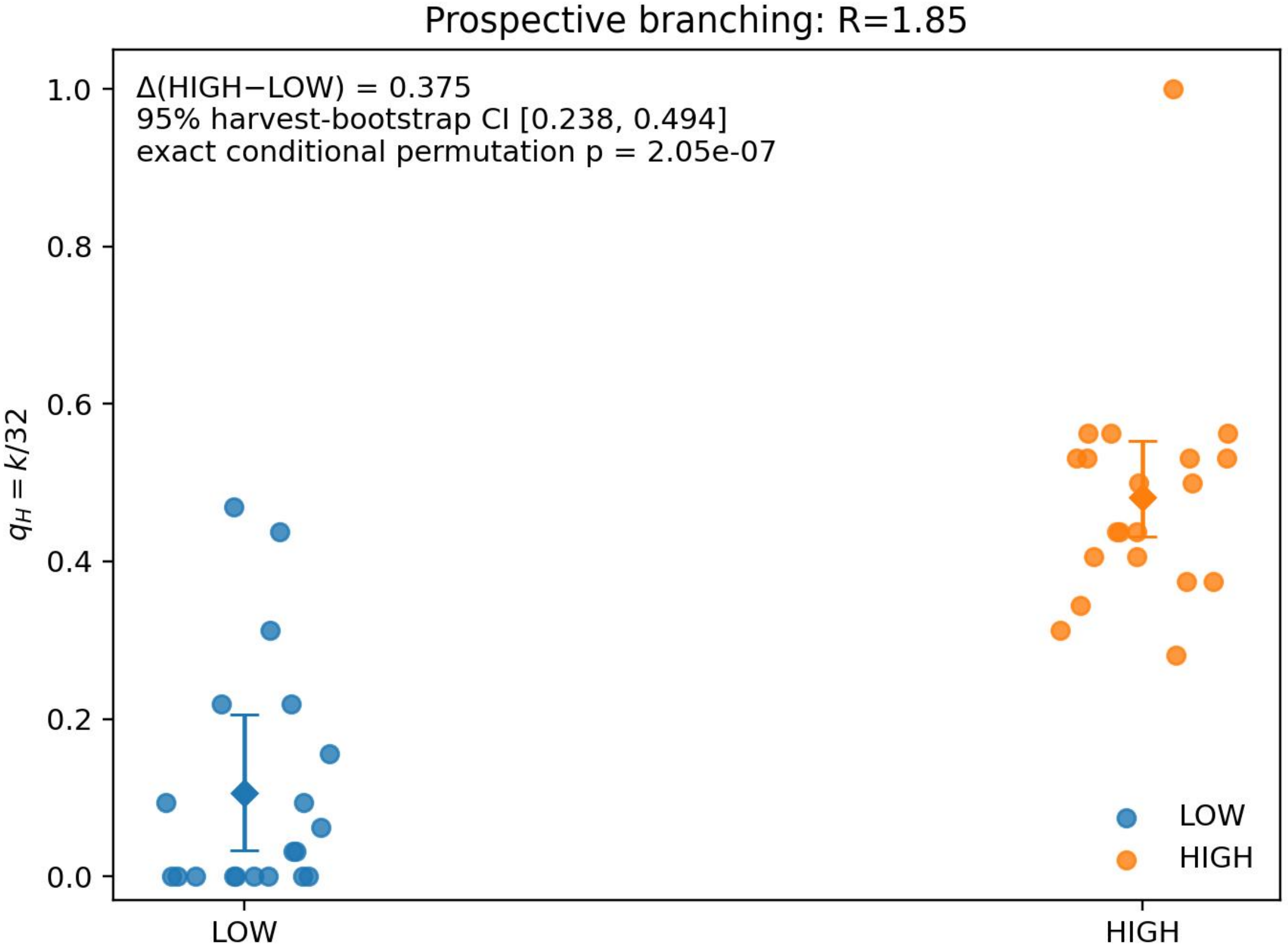

FIG. S36. Parent-level prospective knotting probability at $R$ = $1.85$. Each point is one parent state and $q_H = \frac{k}{32}$. HIGH parents are strongly shifted upward relative to LOW; the diamond/error bar reports the group mean and source-trajectory-cluster bootstrap interval.

### S13.6. Reversal at $R$ = 1.90

At $R = 1.90$, the same classifier does not become merely uninformative; its ordering reverses. HIGH parents produce 411 events in 640 branches (mean $q_H = 0.64219$), whereas the 11 LOW parents produce 265 events in 352 branches (mean $q_H = 0.75284$). The HIGH−LOW difference is -0.11065 with source-trajectory-cluster interval [-0.22514,-0.03125] and two-sided exact $p = 0.01539$. The 11-versus-20 class imbalance is included because it follows directly from the prespecified Q20/Q80 selection rule. The reversed parent-level result is shown in Fig. S37.

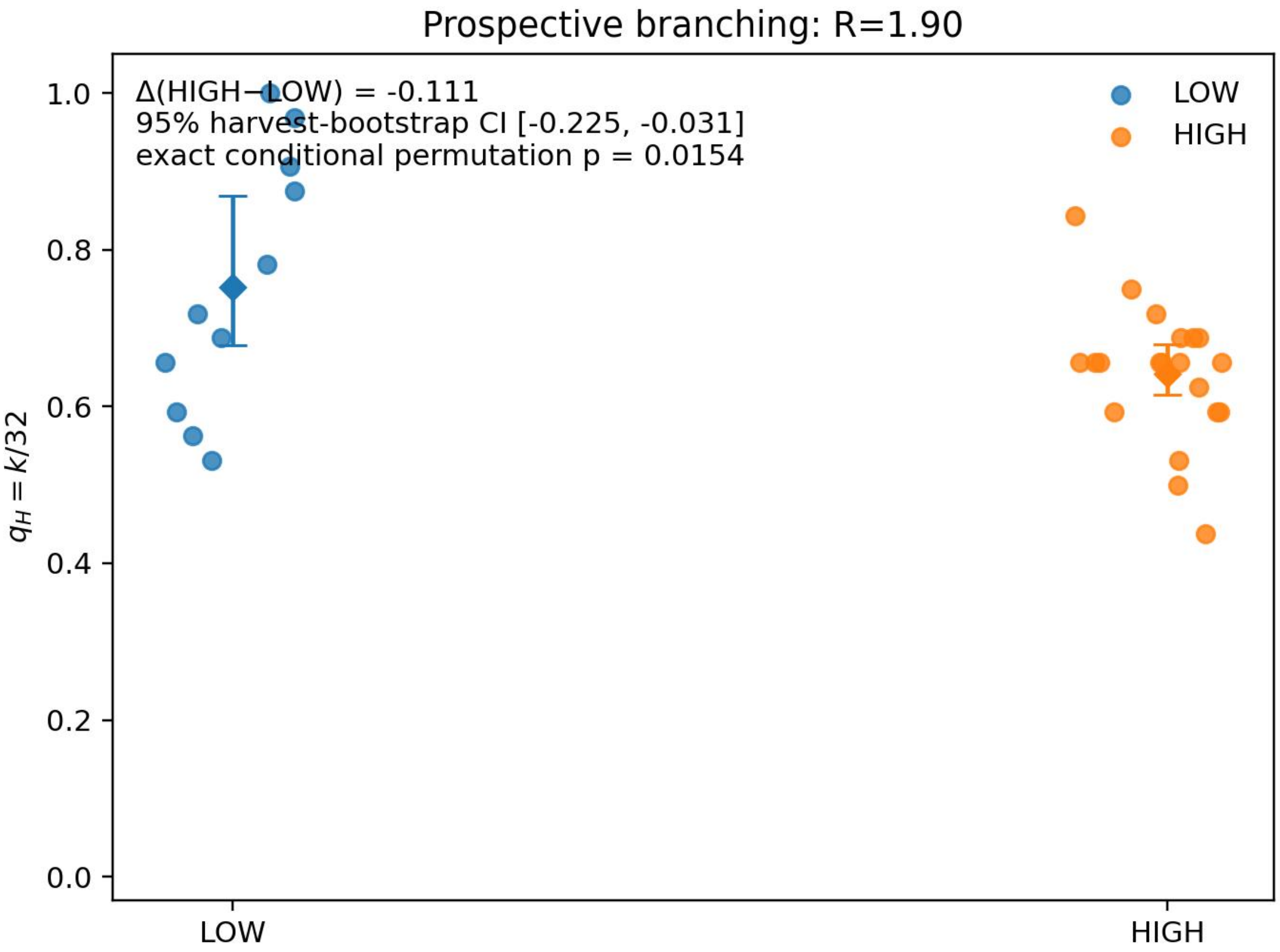

FIG. S37. Parent-level prospective knotting probability at $R$ = $1.90$. LOW parents have the larger finite-horizon switching probability. The 11-versus-20 parent count follows from the prespecified selection thresholds and is not rebalanced after observing outcomes.

### S13.7. Time dependence of first-knot probability

The class separation is not generated only by the $H = 20$-million-step endpoint. Parent-averaged cumulative incidence $P(T_{\text{knot}} \leq t)$ shows the $R = 1.85$ HIGH>LOW separation from early times: at 2, 5, and 10 million steps the means are 0.108 versus 0.009, 0.192 versus 0.027, and 0.303 versus 0.058. At $R = 1.90$, the reversed LOW>HIGH ordering is already clear at the same times: 0.318 versus 0.111, 0.469 versus 0.230, and 0.614 versus 0.406. The sign reversal is therefore a property of the transient switching dynamics rather than a late crossing at the chosen endpoint. The corresponding cumulative-incidence curves are shown in Figs. S38 and S39.

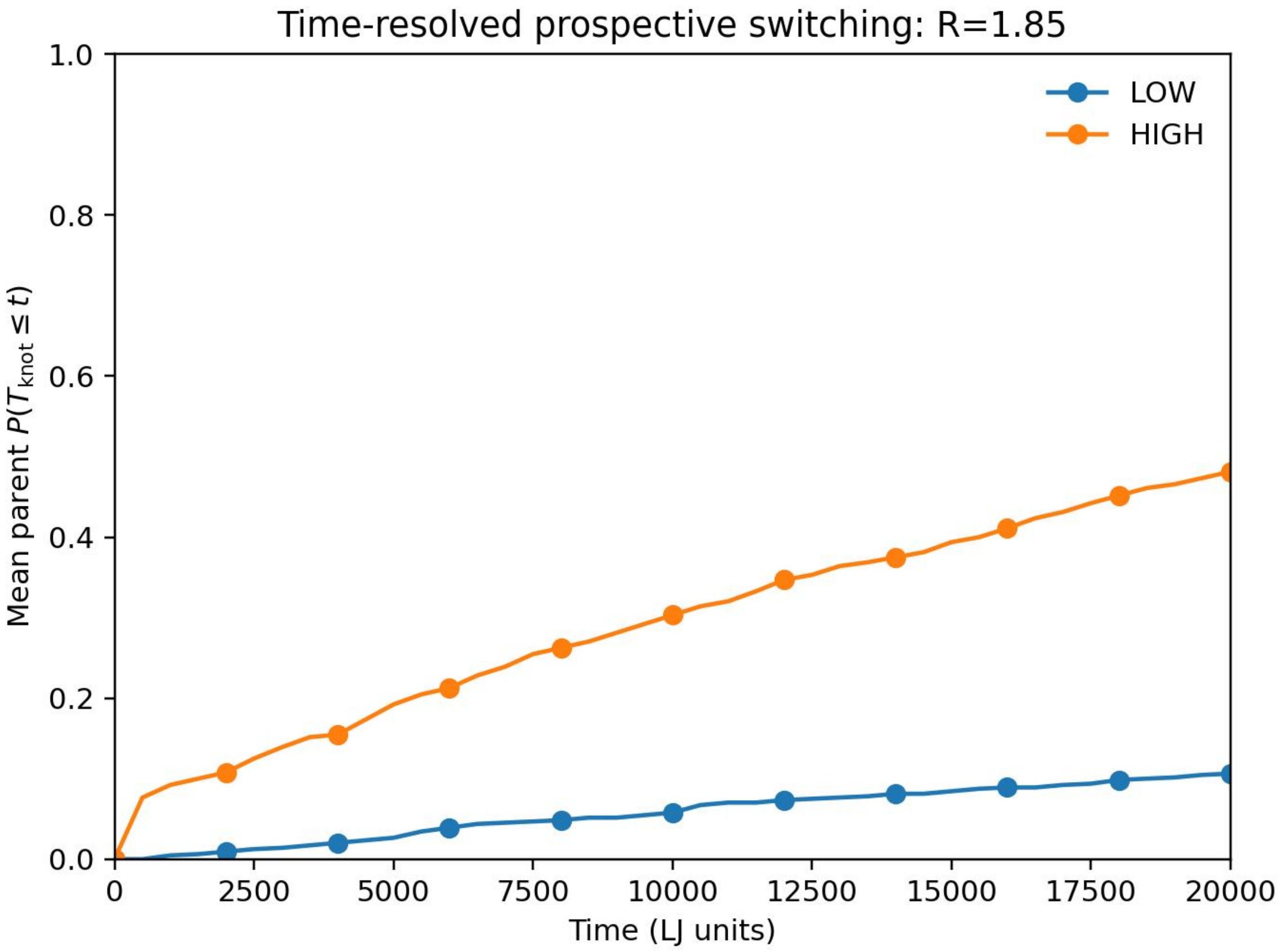

FIG. S38. Parent-averaged cumulative probability of first knot detection at *R = 1.85*. HIGH separates from LOW early and remains above it throughout the fixed observation horizon.

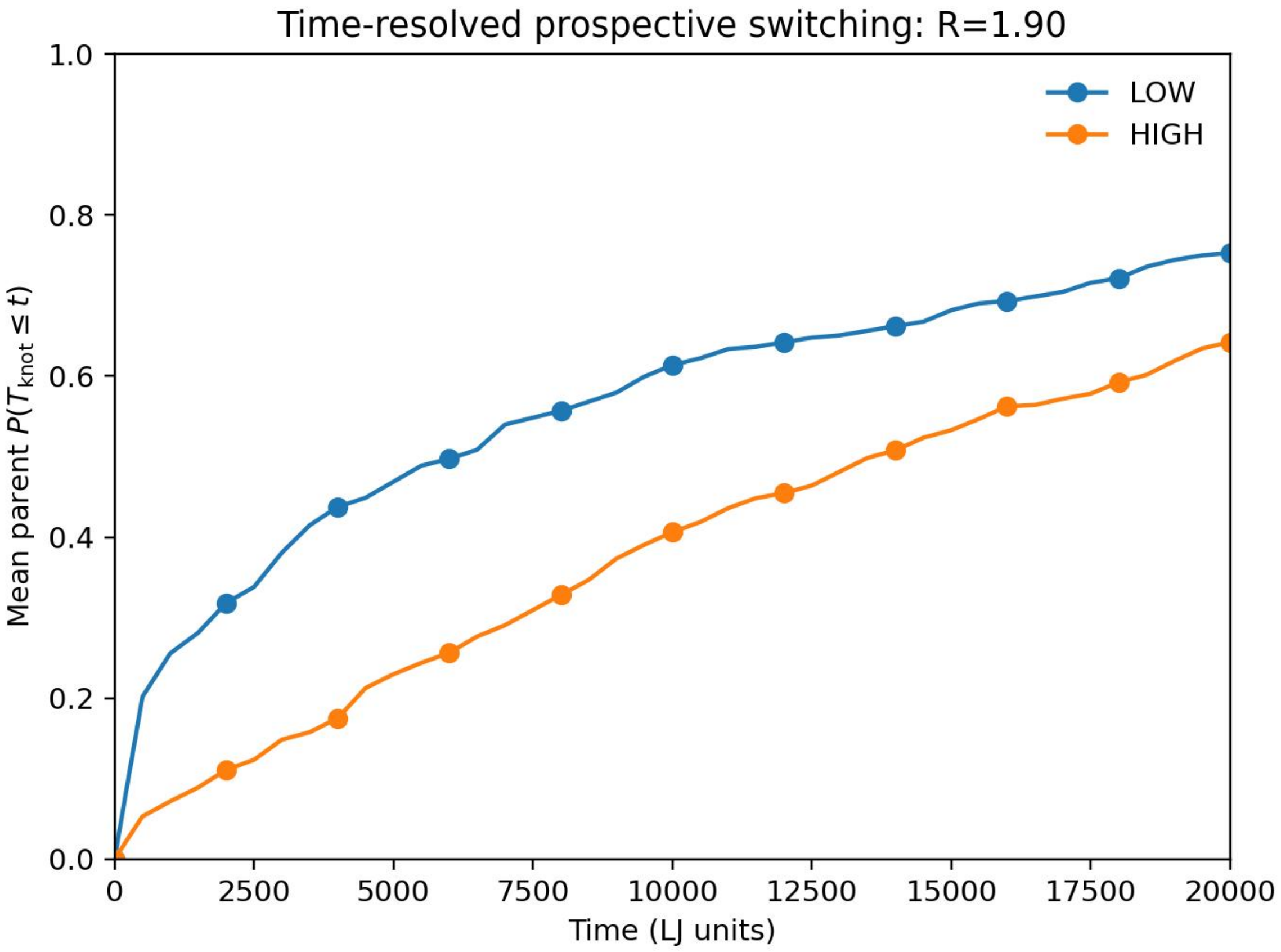

FIG. S39. Parent-averaged cumulative probability of first knot detection at $R = 1.90$. LOW is more knot-prone from early times, showing that the endpoint reversal is not created by the choice of H.

### S13.8. Continuous-score and source-trajectory robustness

The reversal is not an artifact of dichotomizing the prespecified score. Regression of $q_H$ on continuous $s_{\text{conf}}$ gives slope 0.1139 at $R = 1.85$ (source-trajectory-cluster robust $p = 1.17 \times 10^{-6}$) and -0.0523 at $R = 1.90$ ($p = 0.023$). Leave-one-source-trajectory-out HIGH−LOW estimates retain their sign for every omission: +0.336 to +0.412 at $R = 1.85$ and -0.141 to -0.083 at $R = 1.90$. Adding source-parent step and source-trajectory fixed effects leaves the adjusted HIGH coefficients positive at $R = 1.85$ (+0.356) and negative at $R = 1.90$ (-0.128); source step itself is not significant.

**Table S13.** Robustness and secondary time-to-event sensitivities.

| **Check** | $R = 1.85$ | $R = 1.90$ | **Interpretation** |
|---|---|---|---|
| Continuous score slope $\frac{dq_H}{ds_{\text{conf}}}$ | +0.1139 ($p = 1.17 \times 10^{-6}$) | -0.0523 ($p = 0.023$) | Reversal not caused by thresholding |
| Leave-one-source-trajectory-out HIGH−LOW | +0.336 to +0.412 | -0.141 to -0.083 | Sign invariant to removing one source trajectory |
| Adjusted HIGH coefficient with source-trajectory effects + source step | +0.356 | -0.128 | Not explained by parent position along the source trajectory |
| Secondary Cox HIGH/LOW hazard ratio | 5.876 | 0.623 | Time-to-event sensitivity agrees with $q_H$ sign |

### S13.9. Primary inference: geometry-dependent failure of susceptibility ordering along $s_{\text{conf}}$

The prospective experiment establishes two points without using the exploratory candidate search. First, preserved unknotted phase-space states at identical external controls carry reproducible information about future topological switching. Second, the ordering supplied by the prespecified static configuration score is not geometry independent: HIGH>LOW is strongly confirmed at $R = 1.85$ and reverses at $R = 1.90$. Continuous-score and leave-one-source-trajectory-out checks show that the reversal is not created by dichotomization or one source trajectory. Thus no single monotonic function $q_H = f(s_{\text{conf}})$ preserves susceptibility ordering across the two neighboring radii. This is more than a classifier failure: it shows that geometry can reorganize the mapping between an internal mechanical coordinate and future fate. The conclusion is specific to the independently fixed $s_{\text{conf}}$ and does not exclude every possible scalar coordinate.

Section S14 asks what additional structure in the pre-branch trailing history might organize the reversal. Because the $R = 1.90$ outcome was already known when candidate variables were assessed, every result in S14 is exploratory. No candidate is used to relabel parents, alter thresholds, redefine H, or modify the primary $q_H$ contrasts or their p values.

### S13.10. Causal scope and relation to the retrospective score

The branching design is prospective but not interventional. A parent is selected from its observed mechanical history rather than forced to a prescribed wall load, transverse size, or lability. The 32 branches therefore estimate the finite-horizon conditional probability $q_H(X)$ for a preserved phase-space state $X$: the probability that first detected knotting occurs before the fixed endpoint $H$. This is stronger than retrospective association because classification is completed before any branch outcomes are generated, but it does not prove that manipulating $s_{\text{conf}}$ alone would change the true committor. The finite-horizon quantity $q_H$ should therefore be read as a prospective susceptibility at a prespecified observation time, not as an equilibrium state probability or an infinite-time reaction probability.

The retrospective $m^*$ and prospective $s_{\text{conf}}$ have distinct roles. $m^* = \frac{-z(W_U)-z\left(R^2_{g,\perp,U}\right)+z(P_{\parallel,U})}{3}$ is a whole-replica summary used to organize occupancy and formation-hazard associations. $s_{\text{conf}} = \frac{-z(\overline{W})-z\left(\overline{R^2_{g,\perp}}\right)}{2}$ is a trailing-history classifier used to select individual parent states. The branching experiment tests the broader mechanical-state-selection hypothesis and the predictive content of $s_{\text{conf}}$, while the $R = 1.90$ reversal shows why neither score should be treated as a geometry-independent one-dimensional committor.

## S14. Exploratory second-coordinate analysis after the prospective reversal

### S14.1. Candidate variables, multiplicity control, and exploratory status

After the $R = 1.90$ reversal was observed, we examined the same 1000-frame pre-branch trailing histories ending at each of the 71 parent frames. The data set contains 71,000 unknotted frames. Feature extraction uses only quantities already present in the pre-branch mechanism and micro-metric time series; branch outcomes enter only at the subsequent association stage. Because the reversal motivated this search, all results in S14 are hypothesis-generating. They do not alter parent labels, primary endpoints, prespecified contrasts, or prospective p values.

Candidate variables were limited to pre-existing measures of axial/backfold organization, temporal fluctuations, end accessibility and nonlocal contacts, active-power/force channels, and stress anisotropy. For each candidate, both $q_H$ and the candidate are residualized against the prespecified continuous score and source-trajectory fixed effects. Residual associations are assessed by two-sided permutation within source trajectory, with Benjamini–Hochberg correction applied separately at each radius. This analysis asks whether a candidate contributes structure beyond the static score; it is not a confirmatory feature-selection test.

**Table S14.** Candidate second-coordinate families and their role.

| **Family** | Representative data setd variables | **Question addressed** |
|---|---|---|
| Static axial organization | axial span; bond-axis nematic $P_2$ ; mean hairpin count; backfold fraction | Does a second static conformation variable distinguish the reservoirs? |
| Dynamic lability | Cov( $R^2_{g,\perp}$ ,W); Corr( $R^2_{g,\perp}$ ,W); SD( $R^2_{g,\perp}$ ); $\text{SD}(W)$ ; SD(hairpin count) | Does fluctuation structure add information beyond trailing means? |
| End accessibility | end-1/end-N nonlocal contact count and distance; end wall gaps | Is one free end pre-positioned for nonlocal approach? |
| Active drive / stress | axial/radial/azimuthal active power and fluctuations; stress anisotropy | Does drive partitioning or stress provide a residual route variable? |

### S14.2. The score reversal is shared by both static components

The two components of $s_{\text{conf}}$ reverse together. At $R = 1.85$, $q_H$ is negatively associated with both trailing transverse size and wall load (Spearman $\rho_S = -0.622$ and -0.625; within-source-trajectory $r = -0.751$ and -0.780). At $R = 1.90$, both associations become positive ($\rho_S = +0.309$ and +0.312; within-source-trajectory $r = +0.512$ and +0.519). The reversal therefore reflects a change in the static mean-state mapping itself rather than cancellation between two oppositely predictive terms.

### S14.3. Dynamical lability as a candidate second coordinate

After accounting for source-trajectory identity, two history descriptors remain positively associated with $q_H$ at both radii and have a direct interpretation as reservoir lability: $\text{Cov}_{1000}\left(R^2_{g,\perp}, W\right)$, which measures coupled transverse-size and wall-load fluctuations, and the 1000-frame standard deviation of axial hairpin count. Frame by frame, the axial hairpin count is the number of robust sign reversals of successive bond projections along the pore axis after excluding nearly transverse bonds with |cos(theta_axis)|<0.15. Its 1000-frame standard deviation therefore measures fluctuations in forward/backward contour reorganization. For visualization we define $s_{\text{lab}} = \frac{z_R[\text{Cov}_{1000}(R^2_{g,\perp},W)]+z_R[\sigma_{1000}(N_{\text{hairpin}})]}{2}$, with within-radius standardization.

Because this coordinate was defined after the prospective reversal was known, it is an exploratory summary rather than an independently validated classifier.

At $R = 1.85$, the coefficient of $s_{\mathrm{lab}}$ in $q_H \sim z_R(s_{\mathrm{conf}}) + s_{\mathrm{lab}}$ is +0.0589 with source-trajectory-cluster bootstrap 95% interval [0.0076,0.1059]. At $R = 1.90$ it is +0.1072 with interval [0.0743,0.1764]. Removing any one source trajectory leaves both coefficients positive. Thus the lability direction retains the same association with future switching across the two neighboring geometries, unlike the static score ordering.

**Table S15. Key group-level pre-branch trailing-history descriptors.**

| R | class | n | $q_H$ | $s_{\mathrm{lab}}$ | Cov(Rg,W) | SD hairpin | mean hairpin | SD wall | $P_2$ | axial span | passive-end contacts | SD active-end contacts | axial active power |
|---|---|---|---|---|---|---|---|---|---|---|---|---|---|
| R1p85 | HIGH | 20 | 0.481 | +0.627 | 14.73 | 5.09 | 13.3 | 253.1 | 0.629 | 190.9 | 0.000 | 2.65 | 19.31 |
| R1p85 | LOW | 20 | 0.106 | -0.627 | 1.20 | 4.34 | 46.3 | 262.2 | 0.419 | 62.2 | 0.078 | 1.05 | 0.68 |
| R1p9 | HIGH | 20 | 0.642 | -0.355 | 12.26 | 6.09 | 15.2 | 207.8 | 0.574 | 178.0 | 0.010 | 2.73 | 19.57 |
| R1p9 | LOW | 11 | 0.753 | +0.645 | 25.81 | 7.93 | 26.7 | 508.9 | 0.434 | 84.0 | 1.022 | 2.75 | 16.88 |

### S14.4. Geometry-dependent coupling of static score and lability

At $R = 1.85$, $z_R$ ($s_{\mathrm{conf}}$) and $s_{\mathrm{lab}}$ are positively correlated ($r =+ 0.718$; within-source-trajectory $r =+ 0.766$). At $R = 1.90$ they are negatively correlated ($r =- 0.661$; within-source-trajectory $r =- 0.676$). Mean $s_{\mathrm{lab}}$ is +0.627 for $R = 1.85$ HIGH and -0.627 for LOW, whereas at $R = 1.90$ HIGH=-0.355 and LOW=+0.645. Because $s_{\mathrm{lab}}$ retains a positive association with $q_H$ at both radii, the coupling reversal provides a parsimonious organization of the primary sign reversal: the compact/unloaded and dynamically labile directions are aligned at $R = 1.85$ but anticorrelated at $R = 1.90$. The reversal of HIGH/LOW lability ordering is shown in Fig. S40.

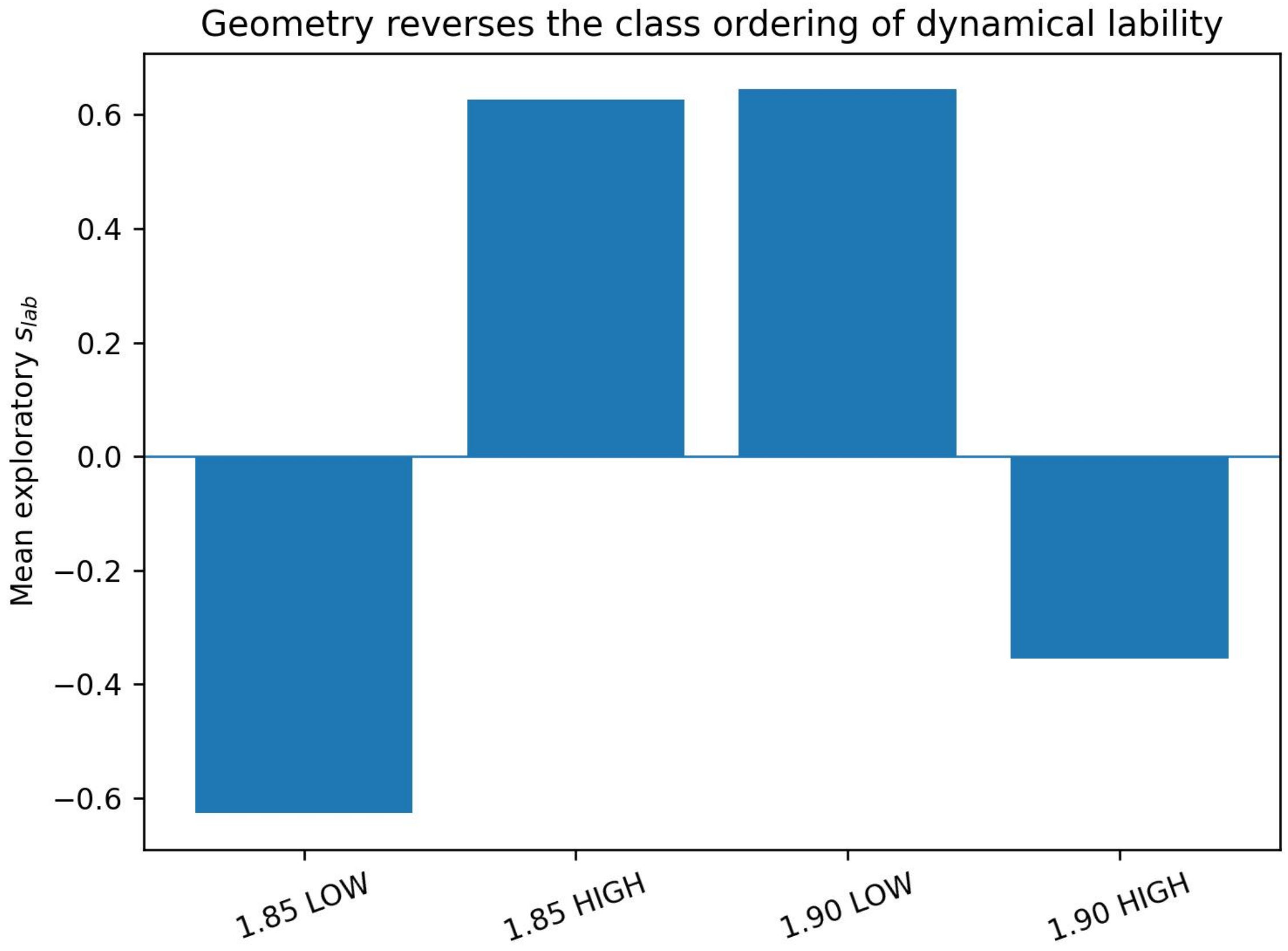

FIG. S40. Geometry reverses the HIGH/LOW ordering of the exploratory lability coordinate. The class label selected by the prespecified static score corresponds to higher lability at $R = 1.85$ but lower lability at $R = 1.90$.

### S14.5. Radius-dependent residual candidate structure

At $R = 1.85$, after removing the prespecified score and source-trajectory effects, the strongest residual candidate is the standard deviation of active-end (end N, bead 300) nonlocal-contact count (partial $r = 0.532$, permutation $p = 4.0 \times 10^{-4}$, BH $q = 0.010$). Mean axial active power ($r = 0.571$, $q = 0.041$) and wall-load fluctuation ($r = 0.329$, $q = 0.041$) are also included, while $R^2_{g,\perp}$-wall-load covariance is suggestive ($r = 0.354$, $q = 0.061$). This pattern connects the pre-branch reservoir history to the independent observation that active-end nonlocal contacts rise before knot formation. The adjusted candidate panel is shown in Fig. S41.

At $R = 1.90$, residual structure shifts toward lability. Transverse-size fluctuation has partial $r = 0.585$ (BH $q = 0.026$), $R^2_{g,\perp}$-wall-load correlation $r = 0.574$ ($q = 0.026$), and their covariance $r = 0.504$ ($q = 0.032$). Mean passive-end nonlocal-contact count is next ($r = 0.498$, $q = 0.059$). LOW parents are more wall-load intermittent (mean SD 509 versus 208), more hairpin rich (26.7 versus 15.2), shorter axially (84.0 versus 178.0), less nematically aligned ($P_2 = 0.434$ versus 0.574), and more frequently in passive-end nonlocal contact (1.02 versus 0.010 contacts per frame). These differences identify a folded, dynamically accessible branch that is not represented by the static score alone. The corresponding adjusted candidate panel is shown in Fig. S42.

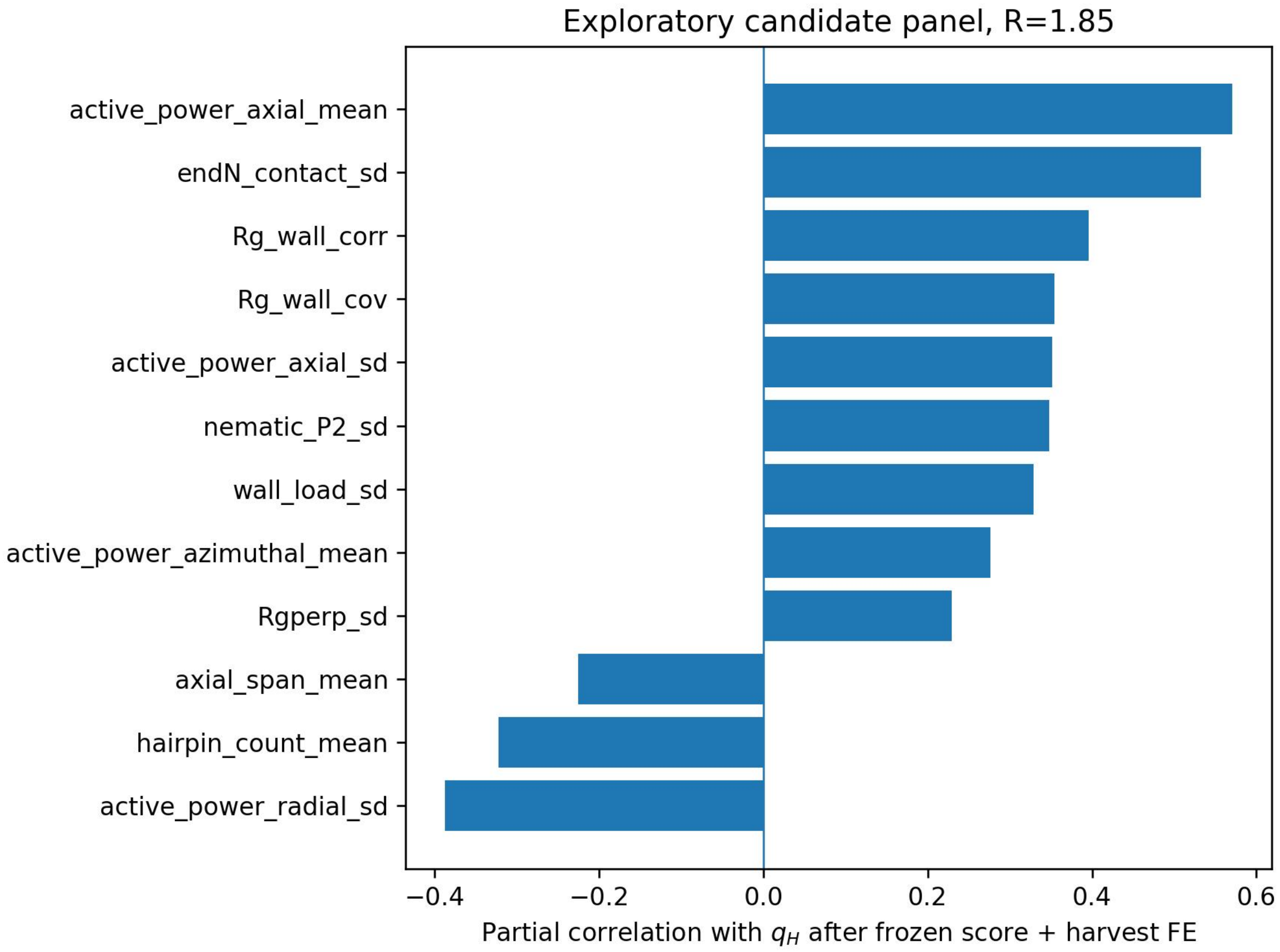

FIG. S41. Exploratory candidate associations at $R = 1.85$ after removing the prespecified score and source-trajectory effects. Bars show partial correlations; inference and multiplicity-adjusted values are listed in Table S16.

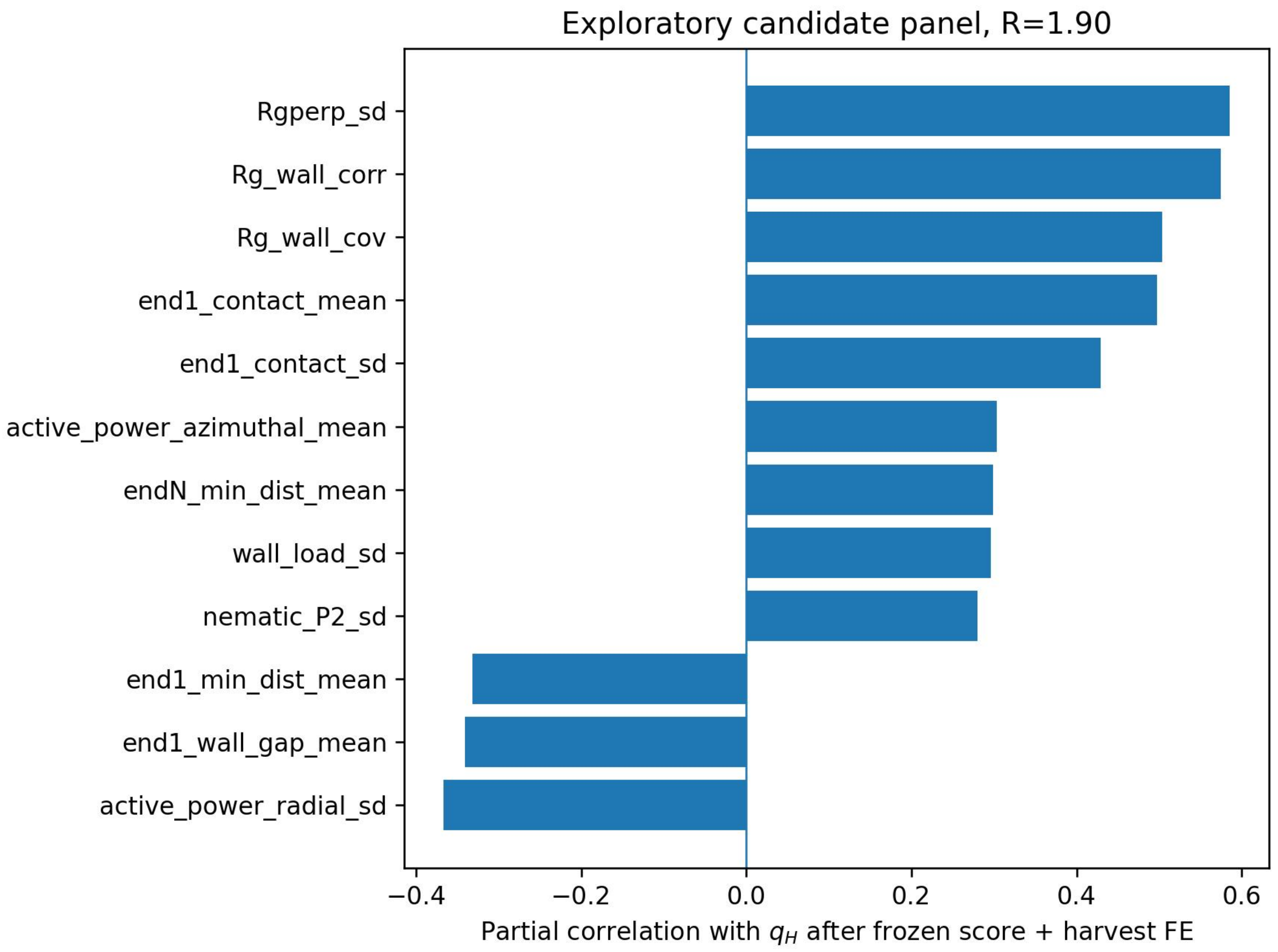

FIG. S42. Exploratory candidate associations at *R = 1.90*. Dynamic transverse/wall lability dominates the residual candidate panel, with passive-end accessibility providing the next signal.

Table S16. Exploratory candidate panel after adjustment for the prespecified score and source-trajectory effects.

| **R** | **candidate** | **partial r** | **within-source-trajectory permutation p** | **BH q** |
|---|---|---|---|---|
| R1p85 | endN_contact_sd | +0.532 | 0.0004 | 0.009999 |
| R1p85 | active_power_axial_mean | +0.571 | 0.0036 | 0.04083 |
| R1p85 | wall_load_sd | +0.329 | 0.0049 | 0.04083 |
| R1p85 | Rg_wall_cov | +0.354 | 0.009799 | 0.06124 |
| R1p85 | Rg_wall_corr | +0.396 | 0.0476 | 0.238 |
| R1p85 | active_power_axial_sd | +0.352 | 0.1076 | 0.2445 |
| R1p85 | active_power_azimuthal_mean | +0.276 | 0.1065 | 0.2445 |
| R1p85 | active_power_radial_sd | -0.387 | 0.08079 | 0.2445 |
| R1p85 | axial_span_mean | -0.226 | 0.08339 | 0.2445 |
| R1p85 | hairpin_count_mean | -0.323 | 0.09809 | 0.2445 |
| R1p85 | nematic_P2_sd | +0.348 | 0.06179 | 0.2445 |
| R1p85 | Rgperp_sd | +0.229 | 0.2287 | 0.4398 |
| R1p85 | end1_min_dist_mean | -0.188 | 0.2251 | 0.4398 |
| R1p85 | end1_wall_gap_mean | -0.148 | 0.3217 | 0.5744 |
| R1p85 | active_power_azimuthal_sd | -0.177 | 0.4693 | 0.6949 |
| R1p85 | endN_contact_mean | +0.136 | 0.4246 | 0.6949 |
| R1p85 | nematic_P2_mean | -0.127 | 0.4726 | 0.6949 |
| R1p85 | backfold_fraction_mean | -0.081 | 0.6453 | 0.8363 |
| R1p85 | end1_contact_mean | +0.064 | 0.7123 | 0.8363 |
| R1p85 | end1_contact_sd | +0.049 | 0.7359 | 0.8363 |
| R1p85 | endN_wall_gap_mean | +0.048 | 0.6865 | 0.8363 |
| R1p85 | hairpin_count_sd | +0.074 | 0.6627 | 0.8363 |
| R1p85 | endN_min_dist_mean | +0.042 | 0.7745 | 0.8419 |
| R1p85 | active_power_radial_mean | +0.036 | 0.8223 | 0.8566 |
| R1p85 | end_to_end_abs_axial_mean | -0.010 | 0.9521 | 0.9521 |

| R | candidate | partial r | within-source-trajectory permutation p | BH q |
|---|---|---|---|---|
| R1p9 | Rg_wall_corr | +0.574 | 0.0021 | 0.02625 |
| R1p9 | Rgperp_sd | +0.585 | 0.0016 | 0.02625 |
| R1p9 | Rg_wall_cov | +0.504 | 0.0038 | 0.03166 |
| R1p9 | end1_contact_mean | +0.498 | 0.009499 | 0.05937 |
| R1p9 | end1_contact_sd | +0.429 | 0.0246 | 0.123 |
| R1p9 | active_power_radial_sd | -0.366 | 0.06159 | 0.2566 |
| R1p9 | end1_min_dist_mean | -0.332 | 0.08359 | 0.2712 |
| R1p9 | end1_wall_gap_mean | -0.340 | 0.08679 | 0.2712 |
| R1p9 | wall_load_sd | +0.296 | 0.1173 | 0.3258 |
| R1p9 | active_power_azimuthal_mean | +0.303 | 0.1318 | 0.3295 |
| R1p9 | nematic_P2_sd | +0.280 | 0.1675 | 0.3806 |
| R1p9 | nematic_P2_mean | +0.252 | 0.1943 | 0.4048 |
| R1p9 | endN_min_dist_mean | +0.299 | 0.23 | 0.4423 |
| R1p9 | hairpin_count_sd | +0.218 | 0.2948 | 0.5264 |
| R1p9 | active_power_azimuthal_sd | -0.198 | 0.3401 | 0.5668 |
| R1p9 | end_to_end_abs_axial_mean | -0.128 | 0.5066 | 0.7916 |
| R1p9 | active_power_axial_sd | +0.100 | 0.6088 | 0.8456 |
| R1p9 | hairpin_count_mean | -0.103 | 0.5897 | 0.8456 |
| R1p9 | axial_span_mean | +0.087 | 0.646 | 0.85 |
| R1p9 | endN_contact_mean | +0.045 | 0.8318 | 0.9903 |
| R1p9 | endN_contact_sd | -0.048 | 0.8225 | 0.9903 |
| R1p9 | active_power_axial_mean | -0.002 | 0.9942 | 0.9942 |
| R1p9 | active_power_radial_mean | -0.013 | 0.9497 | 0.9942 |
| R1p9 | backfold_fraction_mean | -0.004 | 0.9856 | 0.9942 |
| R1p9 | endN_wall_gap_mean | +0.013 | 0.9542 | 0.9942 |

### S14.6. Incremental prediction and leave-one-source-trajectory-out analysis

A pooled parent-level model retaining the score×radius interaction gives an additive $s_{\mathrm{lab}}$ coefficient of 0.0787, with source-trajectory-cluster bootstrap 95% interval [0.0499,0.1090]. Allowing a separate lability slope at $R = 1.90$ does not materially improve out-of-source-trajectory prediction, and the corresponding interaction interval includes zero. In leave-one-source-trajectory-out prediction, adding $s_{\mathrm{lab}}$ reduces RMSE from 0.1486 to 0.1401, MAE from 0.1099 to 0.1019, and branch-count binomial log loss from 0.5754 to 0.5680. These modest improvements are consistent with an additional predictive dimension but remain exploratory. Direct lability-versus-future-susceptibility plots are shown in Figs. S43 and S44.

**Table S17.** Exploratory lability-model comparison.

| model | in-sample R2 | LOHO RMSE | LOHO MAE | LOHO binomial log loss |
|---|---|---|---|---|
| static_score_only | 0.751 | 0.1486 | 0.1099 | 0.5754 |
| static_score_plus_exploratory_lability | 0.780 | 0.1401 | 0.1019 | 0.5680 |

**Table S18.** Lability effect and score-lability coupling by radius.

| R | lability coeff. | source-trajectory-cluster bootstrap 95% CI | LOHO coeff. range | r(score,lability) | within-source-trajectory r | score coeff. after lability |
|---|---|---|---|---|---|---|
| R1p85 | 0.0589 | [0.0076,0.1059] | 0.0436–0.0662 | +0.718 | +0.766 | 0.1480 |
| R1p9 | 0.1072 | [0.0743,0.1764] | 0.0976–0.1223 | -0.661 | -0.676 | 0.0014 |

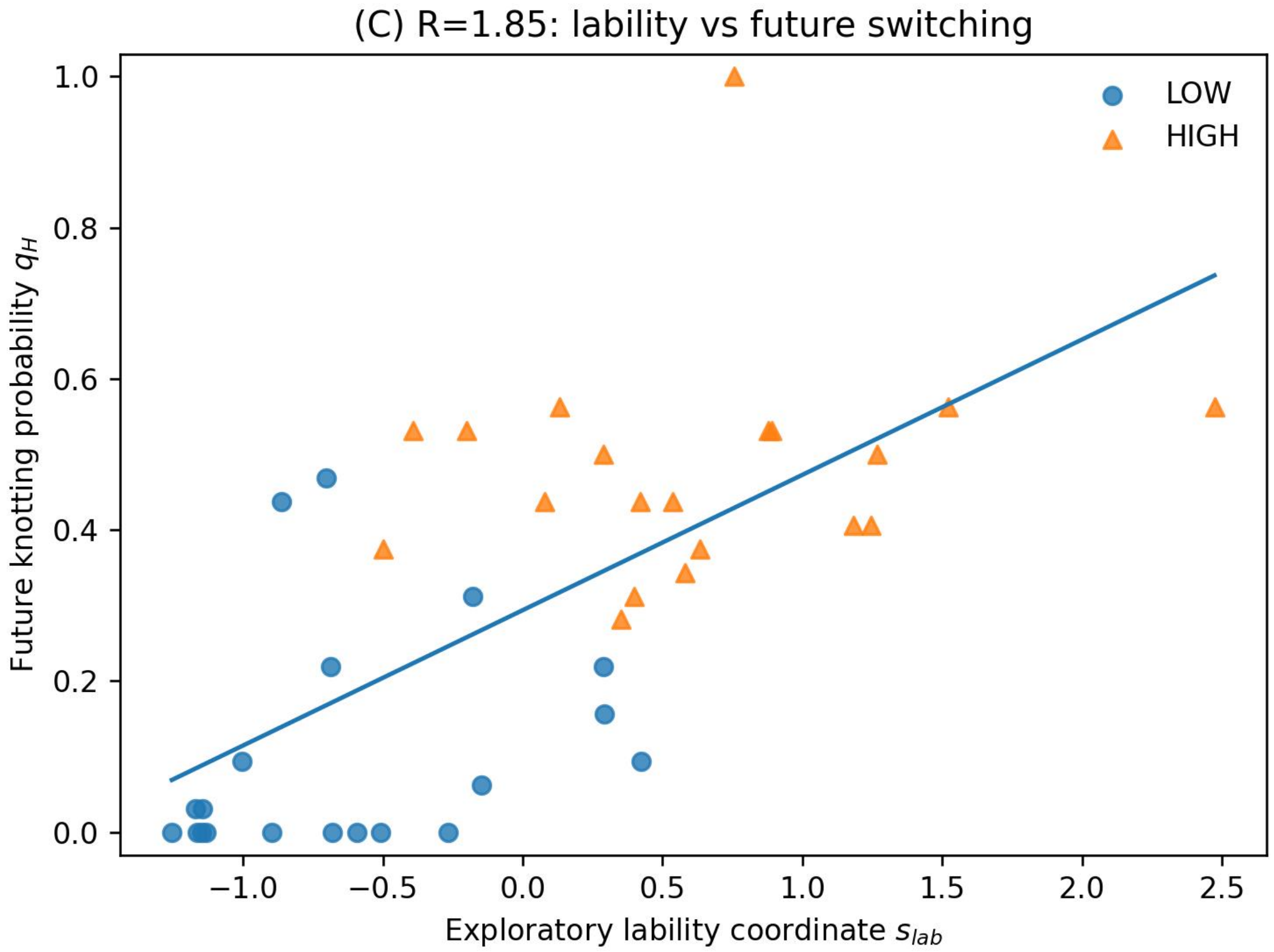

FIG. S43. Exploratory lability coordinate versus future $q_H$ at $R = 1.85$. The positive association persists when the prespecified score is included in the parent-level model.

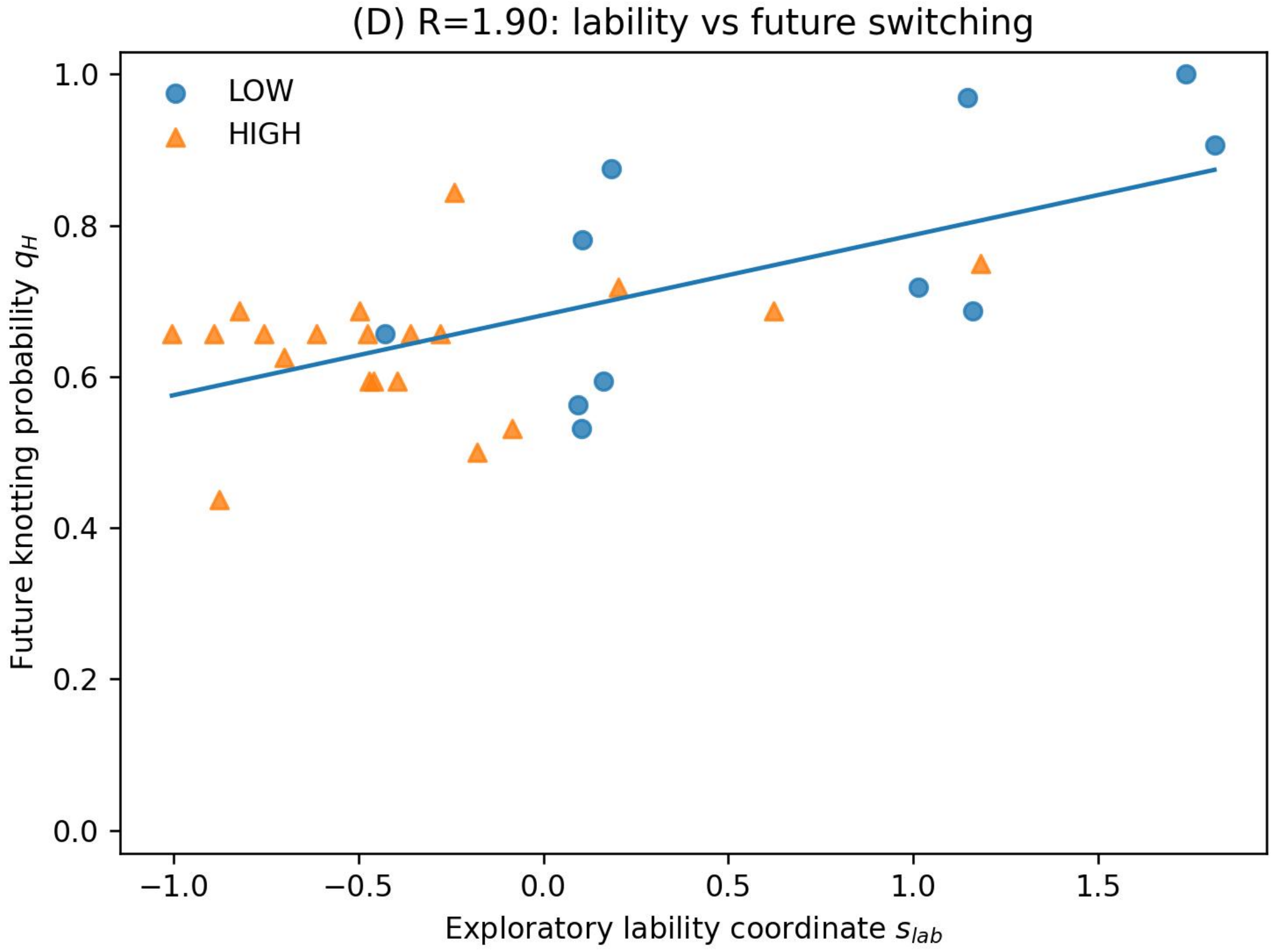

FIG. S44. Exploratory lability coordinate versus future $q_H$ at $R = 1.90$. Unlike $s_{\mathrm{conf}}$, the lability direction retains a positive association with switching in the geometry where the prespecified class ordering reverses.

### S14.7. Mechanistic interpretation

The second-coordinate analysis supports a multidimensional reservoir hypothesis rather than motivating a post hoc repair of the one-dimensional classifier. The static score measures a mean-state direction dominated by wall unloading and transverse compaction, whereas $s_{\mathrm{lab}}$ summarizes coupled wall/transverse fluctuations and axial contour reorganization. At $R = 1.85$ these directions are aligned, and active-end contact fluctuations and power provide a route consistent with the event-aligned active-end approach. At $R = 1.90$ the directions are anticorrelated: HIGH parents are compact/unloaded but comparatively extended and ordered, whereas LOW parents are more folded, intermittent, and contact rich. A parsimonious interpretation is that geometry changes the orientation of the occupied nonequilibrium state space relative to the static score.

This interpretation is stronger than attributing the $R = 1.90$ result to noise but weaker than identifying a unique microscopic committor. The candidate variables are correlated, and $s_{\mathrm{lab}}$ was defined after the prospective reversal was known. The analysis therefore does not establish that covariance, hairpin fluctuations, passive-end contacts, or any single variable is causal. It identifies a concrete two-coordinate hypothesis for an independent held-out test.

### S14.8. Implications for a future held-out test

A confirmatory follow-up should specify both coordinates before launching new independent branch trajectories. One design would retain $s_{\mathrm{conf}}$, prespecify the present $s_{\mathrm{lab}}$ definition, stratify new parent states jointly in the $(s_{\mathrm{conf}}, s_{\mathrm{lab}})$ plane at each radius, and branch from independent parent-generation trajectories. The key prediction is that $q_H$ should vary consistently along the lability direction after conditioning on static score, while geometry changes the correlation between the two coordinates. This proposed test is separate from the prospective result in S13 and the exploratory mechanism analysis reported here.

## S15. Integrated interpretation of existing controls

This section synthesizes controls already introduced in the preceding sections rather than adding a new simulation campaign. Figure S45 summarizes the protocol-resolved chain-length study and its common-seed/common-horizon checks; Fig. S46 combines the baseline broad confinement scan with the crossover-region long trajectories; and Fig. S47 combines kinetic summaries from the broad scan with the event-location classification. The purpose is to place these existing controls next to the interpretation they delimit, not to introduce new data or a new primary analysis.

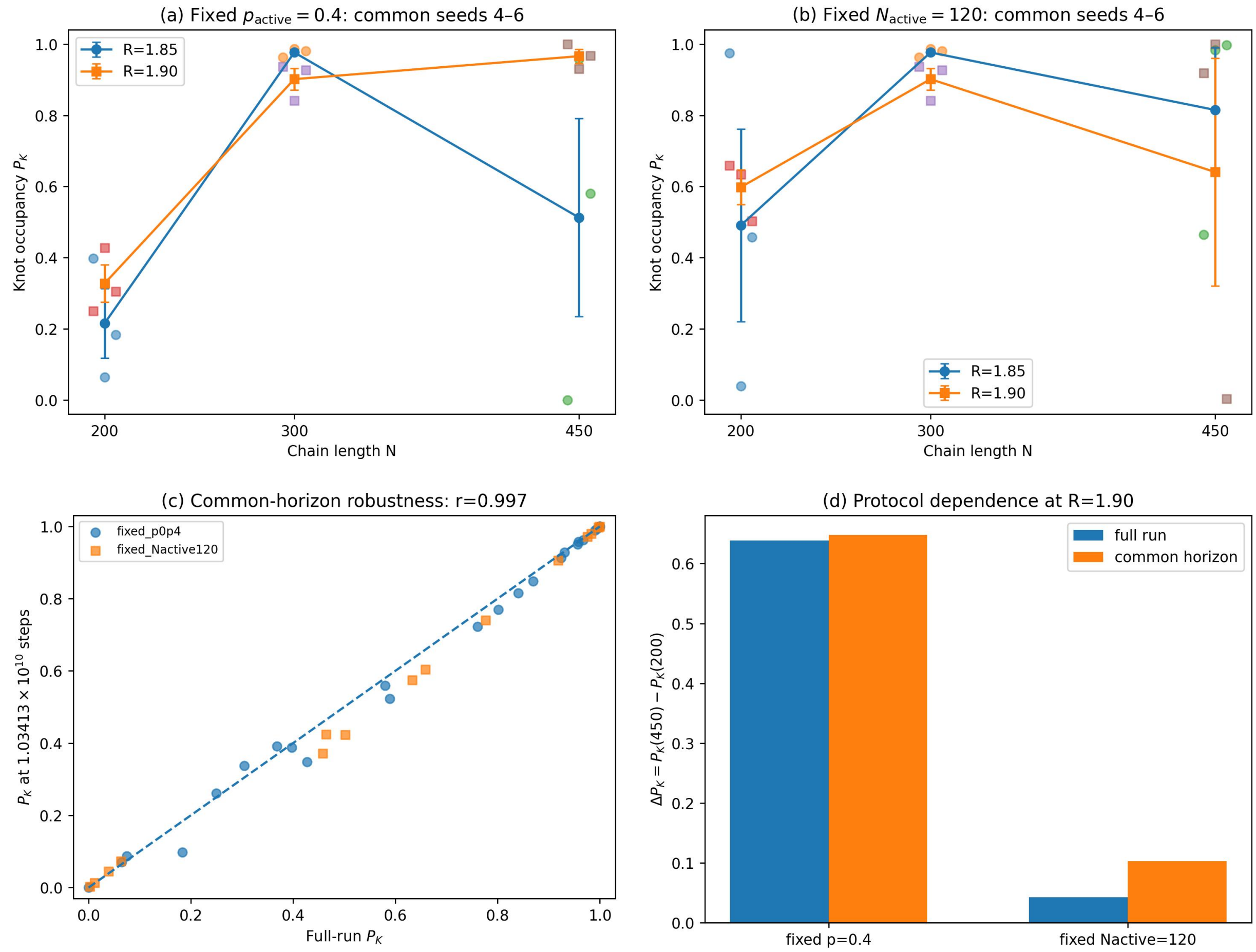

FIG. S45. Protocol-resolved chain-length test at Pe = *3*. (a,b) Knot occupancy for $N = 200$,300,450 at matched $R = 1.85$ and 1.90 under fixed active fraction and fixed active-block length, using common seeds 4–6. (c) Full-run and common-horizon occupancies for all 42 independently restarted trajectories agree closely. (d) At $R = 1.90$ the $N = 450$ minus $N = 200$ contrast is large at fixed active fraction but small and replica-heterogeneous at fixed active-block length. This is a synthesis replot of protocol-controlled results shown separately in Figs. S15 and S29-S31, assembled here to compare the two chain-length protocols in one place.

The chain-length comparison shows that adding contour is not a one-parameter perturbation. At fixed active fraction, increasing N simultaneously enlarges the driven and passive portions of the polymer; at fixed active-block length, the same increase adds passive contour while leaving the driven segment unchanged. The resulting occupancy contrasts remain qualitatively different after both common-seed and common-horizon restrictions, so they cannot be attributed to seed composition or unequal run duration. This protocol dependence is important for the Letter's interpretation: a geometry- or length-dependent change in P_K need not correspond to a unique change in 'compactness' or in one kinetic rate, because the architecture controls which mechanical organizations and residence modes are available.

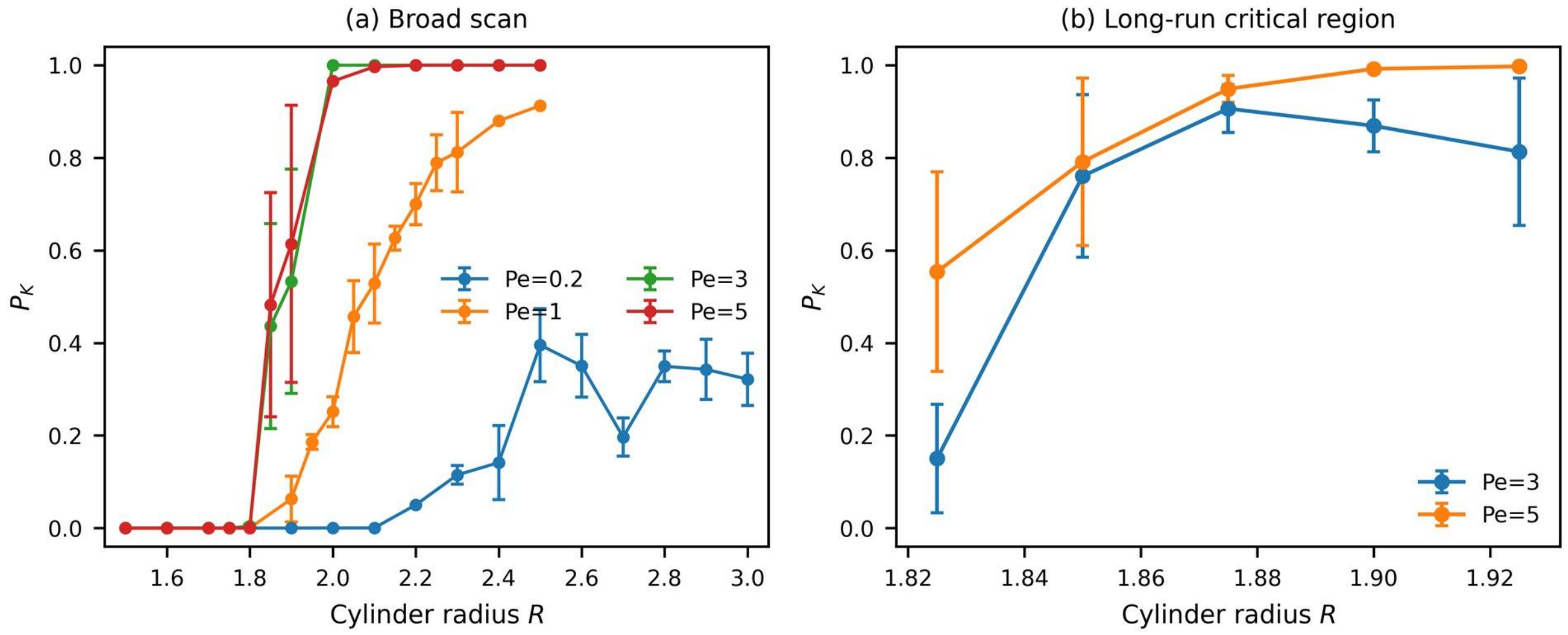

FIG. S46. External-geometry context. (a) Broad knot occupancy versus cylinder radius for Pe = *0.2*,1,3,5. (b) Long-trajectory crossover-region replicas for Pe = *3* and 5. Error bars show variability across independent trajectories. This is a compact synthesis replot of the same underlying data shown separately in Figs. S1 and S2; it is included here only to place the prospective radii in their common confinement context.

The broad and long-run confinement scans place the two prospective radii within a narrow region of pronounced replica heterogeneity rather than at unrelated points of parameter space. This matters for interpreting the sign reversal: it is not a comparison between a deeply unknotted regime and a saturated knotted regime where different physics would be expected trivially. Instead, neighboring radii sample the same crossover neighborhood while changing the organization and coupling of occupied states. The mean P_K(R) curves therefore provide environmental context, whereas the source-resolved branching experiment asks the more discriminating question of how a preserved state maps to future fate. We do not interpret the fitted crossover location as a thermodynamic critical point.

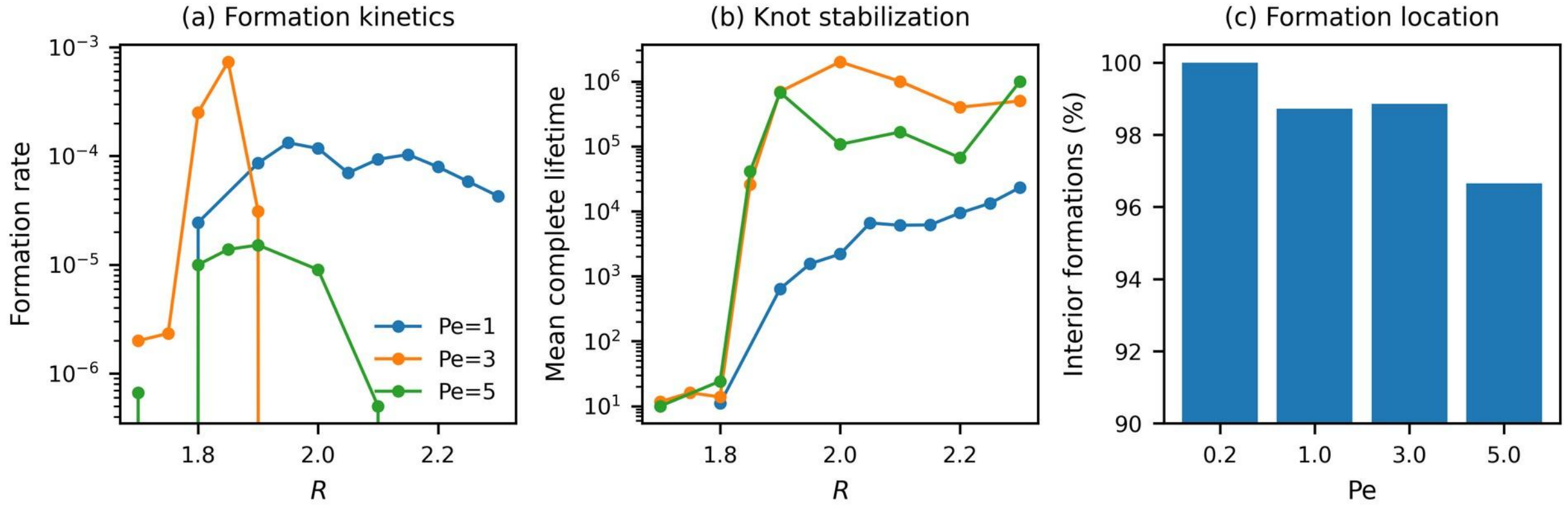

FIG. S47. Kinetic context for the geometry-dependent topology response. (a) Formation rate, (b) mean complete knot lifetime, and (c) fraction of detected formation events in the pore interior. Across the broad data set, 14,145 of 14,291 detected formation events (98.98%) are interior. Panels (a) and (b) compactly replot the broad-scan formation-rate and complete-lifetime data shown in Figs. S4 and S6; panel (c) adds the event-location summary. No panel is a new independent data set.

The kinetic context further separates access from residence. Formation rate and complete-episode lifetime both vary across confinement, but the censored survival curves in Fig. S13 are strongly nonexponential, so no single memoryless escape rate captures knot persistence. The 98.98% interior-formation fraction simultaneously shows that the dominant detected topology changes occur away from the wall, even though the wall controls the global ensemble and can produce transient mechanical loading during the approach to formation. These observations support a three-level interpretation: confinement selects a long-lived mechanical reservoir; a distinct, short interior compaction/threading path carries the system toward first knot detection; and the preserved pre-branch state controls finite-time future susceptibility in a geometry-dependent way. None of these controls alone proves the prospective result, but together they remove simpler rate-only, size-only, and direct-wall-collision explanations.

Taken as a whole, the supplemental controls support the main-text general-physics argument without adding a second novelty claim. External geometry first changes the distribution of occupied nonequilibrium states; those states are not interchangeable even at identical nominal controls; the short transition path is mechanically distinct from the long-lived susceptible reservoir; and the prospective experiment shows that the mapping from a pre-existing state coordinate to future topology can itself change with geometry. This is why the Letter emphasizes state selection and coordinate transferability rather than a universal knotting rate, size law, or equilibrium reaction coordinate.

S16. Energetic context, state-manifold synthesis, and scope

The energetic diagnostics and the state-manifold schematic play different inferential roles. Figures S10 and S11 are quantitative consistency checks: they show that activity-sensitive work injection and conservative-energy response do not reduce the transition to an equilibrium barrier picture. The prospective sign reversal in Sec. S13 remains the confirmatory result, while Sec. S14 supplies only an exploratory dynamical-lability direction. The final schematic below combines these established and exploratory layers without adding a fitted coordinate or a new statistical test. The final state-manifold synthesis is shown in Fig. S48.

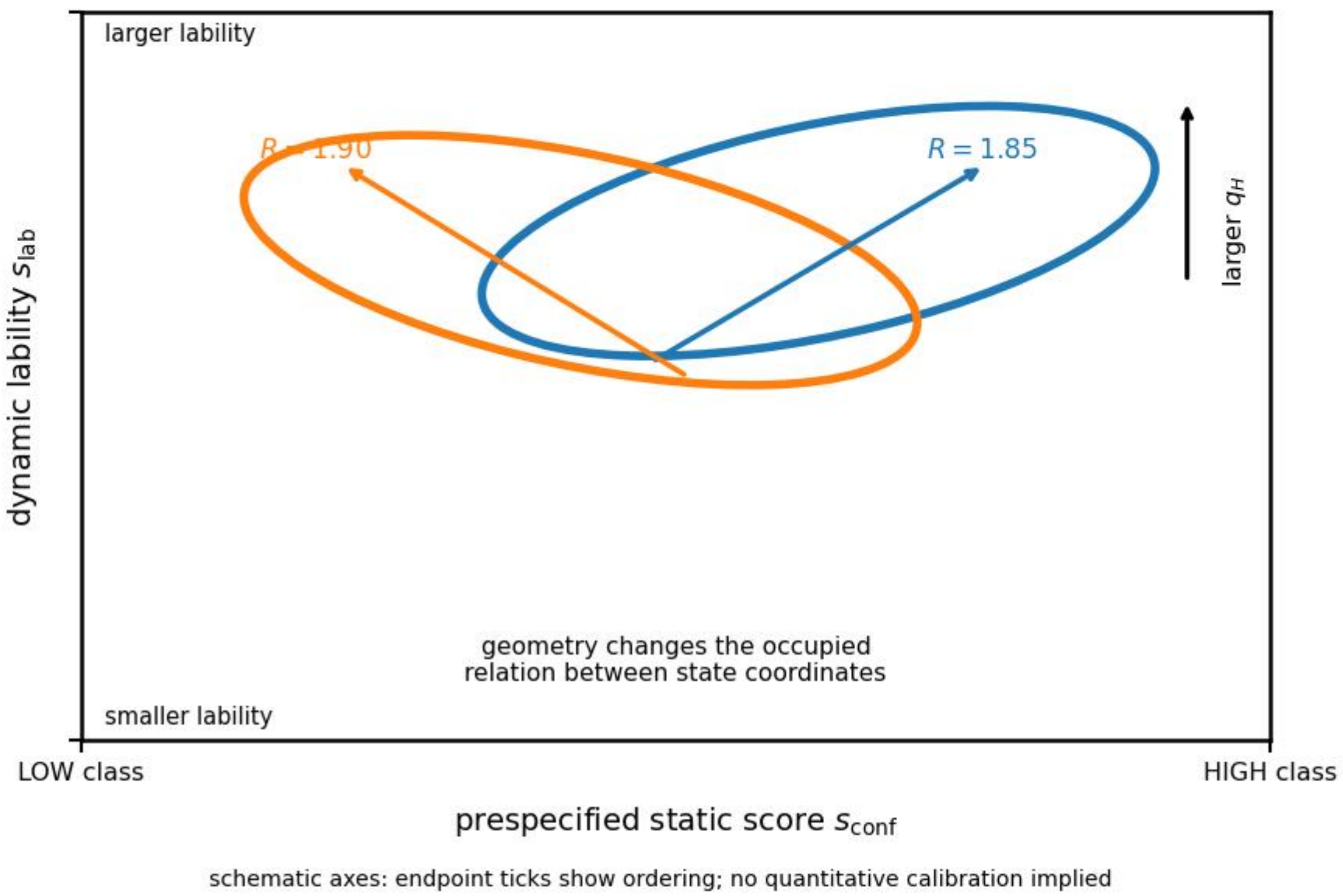

FIG. S48. Schematic multidimensional state-manifold synthesis. Geometry changes the occupied relation between the prespecified static score s_conf and exploratory dynamical lability s_lab, so a one-dimensional LOW/HIGH projection can reverse between R=1.85 and R=1.90. Ellipses and arrows are schematic and encode ordering only; they are not fitted probability contours, free-energy basins, or quantitatively calibrated reaction coordinates.

Figure S48 is deliberately a synthesis rather than additional evidence. The branching experiment establishes that s_conf is prospectively predictive at R=1.85 but reverses its susceptibility ordering at R=1.90; the post-outcome analysis then identifies a lability direction that remains positively associated with finite-horizon switching at both radii, while the coupling between s_conf and that lability changes sign. The differently oriented ellipses summarize this geometry-dependent relation between occupied state coordinates. They should not be read as probability contours or as evidence for two literal metastable basins.

Taken together with the energetic checks in Figs. S10 and S11, the synthesis separates three physical roles. Activity supplies strongly varying work injection during the short approach to threading; confinement reorganizes the population and internal organization of long-lived mechanical states; and the mapping from a pre-existing coordinate to future topological fate need not transfer unchanged between neighboring geometries. This is the intended scope of the state-manifold picture: a testable multidimensional hypothesis motivated by the prospective reversal, not a claim that a unique two-dimensional committor has already been identified.

This energy analysis is deliberately kept separate from the confirmatory claim in S13. It neither determines the sign-reversing classifier nor proves causality of any energy variable. Its value is diagnostic: it demonstrates that topology, conservative energetic cost, and injected active work need not vary in the same way. That separation is precisely what is expected when rare switching occurs on a driven multidimensional state manifold rather than along a single equilibrium landscape coordinate.